\documentclass[twoside,11pt]{article}

\usepackage{jmlr2e}
\usepackage{amsmath}

\usepackage{lastpage}

\usepackage{algpseudocode}
\usepackage{algorithm}

\usepackage{paralist}

\usepackage{amsbsy}

\usepackage[mathscr]{eucal}

\usepackage{bm}

\usepackage{tikz}
\usetikzlibrary{arrows, graphs}

\usepackage{xspace}

\usepackage{subfig}

\usepackage{booktabs}

\mathchardef\mhyphen="2D

\ifpdf
\newcommand{\Sec}[1]{\hyperref[sec:#1]{Section~\ref*{sec:#1}}} 
\newcommand{\App}[1]{\hyperref[sec:#1]{Appendix~\ref*{sec:#1}}} 
\newcommand{\Supp}[1]{\hyperref[sec:#1]{Supplement~\ref*{sec:#1}}} 
\newcommand{\Eqn}[1]{\hyperref[eq:#1]{{\rm (\ref*{eq:#1})}}} 
\newcommand{\Part}[1]{\hyperref[part:#1]{(\ref*{part:#1})}} 
\newcommand{\Fig}[1]{\hyperref[fig:#1]{Figure~\ref*{fig:#1}}} 
\newcommand{\Tab}[1]{\hyperref[tab:#1]{Table~\ref*{tab:#1}}} 
\newcommand{\Thm}[1]{\hyperref[thm:#1]{Theorem~\ref*{thm:#1}}} 
\newcommand{\Lem}[1]{\hyperref[lem:#1]{Lemma~\ref*{lem:#1}}} 
\newcommand{\Prop}[1]{\hyperref[prop:#1]{Proposition~\ref*{prop:#1}}} 
\newcommand{\Cor}[1]{\hyperref[cor:#1]{Corollary~\ref*{cor:#1}}} 
\newcommand{\Def}[1]{\hyperref[def:#1]{Definition~\ref*{def:#1}}} 
\newcommand{\Alg}[1]{\hyperref[alg:#1]{Algorithm~\ref*{alg:#1}}} 
\newcommand{\Ex}[1]{\hyperref[ex:#1]{Example~\ref*{ex:#1}}} 
\newcommand{\As}[1]{\hyperref[as:#1]{Assumption~{\rm\ref*{as:#1}}}} 
\newcommand{\Reg}[1]{\hyperref[as:#1]{Condition~\ref*{reg:#1}}} 
\newcommand{\AlgLine}[2]{\hyperref[alg:#1]{line~\ref*{line:#2} of Algorithm~\ref*{alg:#1}}}
\newcommand{\AlgLines}[3]{\hyperref[alg:#1]{lines~\ref*{line:#2}--\ref*{line:#3} of Algorithm~\ref*{alg:#1}}}
\else
\newcommand{\Sec}[1]{{Section~\ref{sec:#1}}} 
\newcommand{\App}[1]{{Appendix~\ref{sec:#1}}} 
\newcommand{\Supp}[1]{{Supplement~\ref{sec:#1}}} 
\newcommand{\Eqn}[1]{{(\ref{eq:#1})}} 
\newcommand{\Part}[1]{{(\ref{part:#1})}} 
\newcommand{\Fig}[1]{{Figure~\ref{fig:#1}}} 
\newcommand{\Tab}[1]{{Table~\ref{tab:#1}}} 
\newcommand{\Thm}[1]{{Theorem~\ref{thm:#1}}} 
\newcommand{\Lem}[1]{{Lemma~\ref{lem:#1}}} 
\newcommand{\Prop}[1]{{Proposition~\ref{prop:#1}}} 
\newcommand{\Cor}[1]{{Corollary~\ref{cor:#1}}} 
\newcommand{\Def}[1]{{Definition~\ref{def:#1}}} 
\newcommand{\Alg}[1]{{Algorithm~\ref{alg:#1}}} 
\newcommand{\Ex}[1]{{Example~\ref{ex:#1}}} 
\newcommand{\Reg}[1]{{R~\ref*{reg:#1}}} 
\fi

\newcommand{\Real}{\mathbb{R}}

\newcommand{\Tra}{^{\sf T}} 
\def\vec{\mathop{\rm vec}\nolimits}
\newcommand{\tr}{\operatorname{tr}} 

\newcommand{\V}[1]{{\bm{\mathbf{\MakeLowercase{#1}}}}} 
\newcommand{\VE}[2]{\MakeLowercase{#1}_{#2}} 
\newcommand{\Vbar}[1]{{\bm{\bar \mathbf{\MakeLowercase{#1}}}}} 
\newcommand{\Vtilde}[1]{{\bm{\tilde \mathbf{\MakeLowercase{#1}}}}} 
\newcommand{\Vhat}[1]{{\bm{\hat \mathbf{\MakeLowercase{#1}}}}} 
\newcommand{\Vn}[2]{\V{#1}^{(#2)}} 

\newcommand{\M}[1]{{\bm{\mathbf{\MakeUppercase{#1}}}}} 
\newcommand{\ME}[2]{\MakeLowercase{#1}_{#2}} 

\newcommand{\Mn}[2]{\M{#1}^{(#2)}} 

\newcommand{\T}[1]{\boldsymbol{\mathscr{\MakeUppercase{#1}}}} 
\newcommand{\Tbar}[1]{\boldsymbol{\bar \mathscr{\MakeUppercase{#1}}}} 
\newcommand{\That}[1]{\boldsymbol{\hat \mathscr{\MakeUppercase{#1}}}} 
\newcommand{\Ttilde}[1]{\boldsymbol{\tilde \mathscr{\MakeUppercase{#1}}}} 
\newcommand{\Tn}[2]{\T{#1}^{(#2)}} 

\newcommand{\Kron}{\otimes} 

\newcommand{\Mz}[2]{\M{#1}_{(#2)}} 

\newcommand{\E}{\mathcal{E}}

\newcommand{\prox}{\mathop{\rm prox}\nolimits}

\newcommand{\amp}{\mathop{\:\:\,}\nolimits}

\newtheorem{assumption}{Assumption}[section]

\jmlrheading{21}{2020}{1-\pageref{LastPage}}{3/18; Revised
2/20}{10/20}{18-155}{Eric C. Chi, Brian R. Gaines, Will Wei Sun, Hua Zhou, and Jian Yang}

\ShortHeadings{Provable Convex Co-clustering of Tensors}{Chi, Gaines, Sun, Zhou, and Yang}

\firstpageno{1}

\begin{document}

\title{Provable Convex Co-clustering of Tensors}

\author{\name Eric C.\ Chi \email eric\_chi@ncsu.edu \\
       \addr Department of Statistics\\
       North Carolina State University\\
       Raleigh, NC 27695, USA
       \AND
       \name Brian R.\ Gaines \email brian.gaines@sas.com \\
       \addr Advanced Analytics R\&D \\
       SAS Institute Inc.\@\\
       Cary, NC 27513, USA
       \AND
       \name Will Wei Sun \email sun244@purdue.edu  \\
       \addr Krannert School of Management \\
       Purdue University \\
       West Lafayette, IN 47907, USA
       \AND
       \name Hua Zhou \email huazhou@ucla.edu \\
       \addr Department of Biostatistics\\
       University of California\\
       Los Angeles, CA 90095, USA
       \AND
       \name Jian Yang \email jianyang@oath.com \\
       \addr Advertising Sciences\\
       Yahoo Research\\
       Sunnyvale, CA 94089, USA}

\editor{Francis Bach}

\maketitle

\begin{abstract}
Cluster analysis is a fundamental tool for pattern discovery of complex heterogeneous data. Prevalent clustering methods mainly focus on vector or matrix-variate data and are not applicable to general-order tensors, which arise frequently in modern scientific and business applications.  Moreover, there is a gap between statistical guarantees and computational efficiency for existing tensor clustering solutions due to the nature of their non-convex formulations. In this work, we bridge this gap by developing a provable convex formulation of tensor co-clustering. Our convex co-clustering (CoCo) estimator enjoys stability guarantees and its computational and storage costs are polynomial in the size of the data. We further establish a non-asymptotic error bound for the CoCo estimator, which reveals a surprising ``blessing of dimensionality" phenomenon that does not exist in vector or matrix-variate cluster analysis. Our theoretical findings are supported by extensive simulated studies. Finally, we apply the CoCo estimator to the cluster analysis of advertisement click tensor data from a major online company. Our clustering results provide meaningful business insights to improve advertising effectiveness.
\end{abstract}

\begin{keywords}
  Clustering, Fused lasso, High-dimensional Statistical Learning, Multiway Data, Non-asymptotic Error
\end{keywords}

\section{Introduction}

In this work, we study the problem of finding structure in multiway data, or tensors, via clustering. Tensors appear frequently in modern scientific and business applications involving complex heterogeneous data. For example, data in a neurogenomics study of brain development consists of a 3-way array of expression level measurements indexed by gene, space, and time \citep{LiuYuanZhao2017}. Other examples of 3-way data arrays consisting of matrices collected over time include email communications (sender, recipient, time) \citep{Papalexakis2013}, online chatroom communications (user, keyword, time) \citep{AcarCamtepeYener2006}, bike rentals (source station, destination station, time) \citep{GuigouresBoulleRossi2015}, and internet network traffic (source IP, destination IP, time) \citep{SunTaoFaloutsos2006}. The rise in tensor data has created new challenges in making predictions, such as in recommender systems for example \citep{Zheng2016, Symeonidis2016, SymeonidisZioupos2016, FrolovOseledets2017, BiQuShen2017+} as well as inferring latent structure in multiway data \citep{acar2009unsupervised, Anandkumar2014, cichocki2015tensor, Sidiropoulos2017}.

As tensors become increasingly more common, the need for a reliable co-clustering method grows increasingly more urgent.
Prevalent clustering methods, however, mainly focus on vector or matrix-variate data. The goal of vector clustering is to identify subgroups within the vector-variate observations \citep{ma2008, ShenHuang2010, shen2012clustering, wang2013provable}. Biclustering is the extension of clustering to two-way data where both the observations (rows) and the features (columns) of a data matrix are simultaneously grouped together \citep{Hartigan1972, MadOli2004, BusPro2008}. In spite of their prevalence, these approaches are not directly applicable to the cluster analysis of general-order (general-way) tensors. On the other hand, existing methods for co-clustering general $D$-way arrays, for $D \geq 3$, employ one of three strategies: (i) extensions of spectral clustering to tensors \citep{WuBensonGleich2016}, (ii) directly clustering the subarrays along each dimension, or way, of the tensor using either $k$-means or variants on it \citep{JegelkaSraBanerjee2009}, and (iii) low rank tensor decompositions  \citep{SunPapadimitriouLinEtAl2009, Papalexakis2013, ZhaoWang2016}. While all these existing approaches may demonstrate good empirical performance, they have limitations. For instance, the spectral co-clustering method proposed by \cite{WuBensonGleich2016} is limited to nonnegative tensors and the CoTeC method proposed by \cite{JegelkaSraBanerjee2009}, like $k$-means, requires specifying the number of clusters along each dimension as a tuning parameter. Most importantly, none of the existing methods provide statistical guarantees for recovering an underlying co-clustering structure. There is a conspicuous gap between statistical guarantees and computational efficiency for existing tensor clustering solutions due to the nature of the non-convex formulations of the previously mentioned works.

In this paper, we propose a Convex Co-clustering (CoCo) procedure that solves a convex formulation of the problem of co-clustering a $D$-way array for $D \geq 3$. Our proposed CoCo estimator affords the following advantages over existing tensor co-clustering methods.
\begin{itemize}
\item[(i)] Under modest assumptions on the data generating process, the CoCo estimator is guaranteed to recover an underlying co-clustering structure with high probability. In particular, we establish a non-asymptotic error bound for the CoCo estimator, which reveals a surprising ``blessing of dimensionality" phenomenon: As the dimensions of the array increase, the CoCo estimator is {\em still} consistent even if the number of underlying co-clusters grows as a function of the number of elements in the tensor sample. More importantly, an underlying co-clustering structure can be consistently recovered with even a single tensor sample, which is a typical case in real applications. This phenomenon does not exist in vector or matrix-variate cluster analysis.
\item[(ii)] The CoCo estimator possesses stability guarantees. In particular, the CoCo estimator is Lipschitz continuous in the data and jointly continuous in the data and its tuning parameter. 
We emphasize that Lipschitz continuity in the data guarantees that perturbations in the data lead to graceful and commensurate variations in the cluster assignments, and the continuity in the tuning parameter can be leveraged to expedite computation through warm starts.
\item[(iii)] The CoCo estimator can be iteratively computed with convergence guarantees via an accelerated first order method with storage and per-iteration cost that is linear in the size of the data. \end{itemize}
In short, the CoCo estimator comes with (i) statistical guarantees, (ii) practically relevant stability guarantees at all sample sizes, and (iii) an algorithm with polynomial complexity. The theoretical properties of our CoCo estimator are supported by extensive simulation studies. To demonstrate its business impact, we apply the CoCo estimator to the cluster analysis of advertisement click tensor data from a major online company. Our clustering results provide meaningful business insights to help advertising planning.

Our work is related to, but also clearly distinct from, a number of recent developments in cluster analysis. The first related line of research tackles convex clustering \citep{HockingJoulinBachEtAl2011, Zhu2014, Chi2015, Chen2015, Tan2015, WanZhaSun2016, Radchenko2017} and convex biclustering \citep{ChiAllenBaraniuk2017}. These existing methods are not directly applicable to general-order tensors, however. Importantly, our CoCo estimator enjoys a unique ``blessing of dimensionality" phenomenon that has not been established in the aforementioned approaches. Moreover, the CoCo estimator is similar in spirit to a recent series of work approximating a noisy observed array with an array that is smooth with respect to some latent organization associated with each dimension of the array \citep{Gavish2012, Ankenman2014, Mishne2016, Yair2017}. Our proposed CoCo procedure seeks an approximating array that is smooth with respect to a latent clustering along each dimension of the array. While CoCo shares features with these array approximation techniques, namely the use of data-driven similarity graphs along tensor modes, a key distinction between our CoCo estimator and these methods is that CoCo produces an approximating array that explicitly recovers hard co-clustering assignments. As we will see shortly, focusing our attention in this work on the co-clustering model paves the way to the discovery and explicit characterization  of new and interesting fundamental behavior in finding intrinsic organization within tensors.

The rest of the paper is organized as follows. In \Sec{prelim}, we review standard facts and results about tensors that we will use. In \Sec{coco_formulation}, we introduce our convex formulation of the co-clustering problem. In \Sec{properties}, we establish the stability properties and prediction error bounds of the CoCo estimator. In \Sec{estimation_algorithm}, we describe the algorithm used to compute the CoCo estimator. In \Sec{weights}, we discuss how to specify weights used in our CoCo estimator, and in \Sec{practical_issues} we give guidance on how to set and select tuning parameters used in the CoCo estimator in practice. In \Sec{simulations}, we present simulation results. In \Sec{real_data}, we discuss the results of applying the CoCo estimator to co-cluster a real data tensor from online advertising. In \Sec{discussion}, we close with a discussion. The Appendix contains a brief review of the two main tensor decompositions that are discussed in this paper, all technical proofs, as well as additional experiments. 

\section{Preliminaries}
\label{sec:prelim}

\subsection{Notation}
\label{sec:notation}

We adopt the terminology and notation used by \cite{KoldaBader2009}. We call the number of ways or modes of a tensor its {\em order}. Vectors are tensors of order one and denoted by boldface lowercase letters, e.g.\@ $\V{a}$. Matrices are tensors of order two and denoted by boldface capital letters, e.g.\@ $\M{A}$. Tensors of higher-order, namely order three and greater, we denote by boldface Euler script letters, e.g.\@ $\T{A}$. Thus, if $\T{A}$ represent a $D$-way data array of size $n_1 \times n_2 \times \cdots \times n_D$, we say $\T{A}$ is a tensor of order $D$. We denote scalars by lowercase letters, e.g.\@ $a$. We denote the $i$th element of a vector $\V{a}$ by $\VE{a}{i}$, the $ij$th element of a matrix $\M{A}$ by $\ME{a}{ij}$, the $ijk$th element of a third-order tensor $\T{a}$ by $\ME{a}{ijk}$, and so on.

We can extract a subarray of a tensor by fixing a subset of its indices. For example, by fixing the first index of a matrix to be $i$, we extract the $i$th row of the matrix, and by fixing the second index of a matrix to be $j$, we extract a $j$th  column of the matrix. We use a colon to indicate all elements of a mode. Consequently, we denote the $i$th row of a matrix $\M{A}$ by $\M{a}_{i:}$ and the $j$th column of a matrix $\M{A}$ by $\M{a}_{:j}$. {\em Fibers} are the subarrays of a tensor obtained by fixing all but one of its indices. In the case of a matrix, a mode-1 fiber is a matrix column and a mode-2 fiber is a matrix row. {\em Slices} are the two-dimensional subarrays of a tensor obtained by fixing all but two indices. For example, a third-order tensor $\M{A}$ has three sets of slices denoted by $\T{A}_{i::}, \T{A}_{:j:}$, and $\T{A}_{::k}$.

\subsection{Basic Tensor Operations}
\label{sec:tensor_ops}

It is often convenient to reorder the elements of a $D$-way array into a matrix or a vector. Reordering a tensor's elements into a matrix is referred to as {\em matricization}, while reordering its elements into a vector is referred to as {\em vectorization}. There are many ways to reorder a tensor into a matrix or vector. In this paper, we use a canonical mode-$d$ matricization, where the mode-$d$ fibers of a $D$-way tensor $\T{A} \in \Real^{n_1 \times n_2 \times \cdots \times n_D}$ become the columns of a matrix $\M{A}_{(d)} \in \Real^{n_d \times n_{-d}}$, where $n_{-d} = \prod_{j \neq d} n_j$. Recall that the column-major vectorization of a matrix maps a matrix $\M{A} \in \Real^{p \times q}$ to the vector $\V{a} \in \Real^{pq}$ by stacking the columns of $\M{A}$ on top of each other, namely $\V{a} = \begin{pmatrix} \M{A}_{:1}\Tra & \M{A}_{:2}\Tra & \cdots & \M{A}_{:q}\Tra \end{pmatrix}\Tra \in \Real^{pq}$. In this paper, we take the vectorization of a $D$-way tensor $\T{A}$, denoted $\vec(\T{A})$, to be the column-major vectorization of the mode-1 matriciziation of $\T{A}$, namely $\vec(\T{A}) = \vec(\Mz{A}{1}) \in \Real^{n}$, where $n = \prod_{d} n_d$ the total number of elements in $\T{A}$. As a shorthand, when the context leaves no ambiguity, we denote this vectorization of a tensor $\T{A}$ by its boldface lowercase version $\V{a}$.

The Frobenius norm of a $D$-way tensor $\T{A} \in \Real^{n_1 \times n_2 \times \cdots \times n_D}$ is the natural generalization of the Frobenius norm of a matrix, namely it is the square root of the sum of the squares of all its elements,
\begin{eqnarray*}
\lVert \T{A} \rVert_{\text{F}} & = & \sqrt{\sum_{i_1 = 1}^{n_1} \sum_{i_2=1}^{n_2}\cdots\sum_{i_D=1}^{n_D} \ME{A}{i_1i_2\cdots i_D}^2}.
\end{eqnarray*}
The Frobenius norm of a tensor is equivalent to the $\ell_2$-norm of the vectorization of the tensor, namely $\lVert \T{A} \rVert_{\text{F}} = \lVert \V{a} \rVert_2$.

Let $\T{A}$ be a tensor in $\Real^{n_1 \times n_2 \times \cdots \times n_D}$ and $\M{B}$ be a matrix in $\Real^{m \times n_d}$. The {\em d-mode (matrix) product} of the tensor $\T{A}$ with the  matrix $\M{B}$, denoted by $\T{A} \times_{d} \M{B}$, is the tensor of size $n_1 \times \cdots \times n_{d-1} \times m \times n_{d+1} \times \cdots \times n_D$ whose $(i_1, i_2, \cdots, i_{d-1}, j, i_{d+1}, \cdots, i_D)$th element is given by 
\begin{eqnarray*}
(\T{A} \times_d \M{B})_{i_1\ldots i_{d-1} j i_{d+1} \cdots i_D} & = & \sum_{i_d = 1}^{n_d} \ME{a}{i_1i_2\cdots i_D}\ME{b}{j i_d},
\end{eqnarray*}
for $j \in \{1, \ldots, m\}$. The vectorization of the $d$-mode product $\T{A} \times_{d} \M{B}$ can be expressed as
\begin{eqnarray}
\label{eq:vec_identity}
\vec (\T{A} \times_d \M{B}) 
& = & (\M{I}_{n_D} \Kron \cdots \Kron \M{I}_{n_{d+1}}\Kron \M{B} \Kron \M{I}_{n_{d-1}} \Kron \cdots \Kron \M{I}_{n_1} )\V{a},
\end{eqnarray}
where $\M{I}_p$ is the $p$-by-$p$ identity matrix and $\Kron$ denotes the Kronecker product between two matrices. The identity given in \Eqn{vec_identity} generalizes the well known formula for the column-major vectorization of a product of two matrices, namely $\vec(\M{B}\M{A}) = (\M{I}\Kron\M{B})\V{a}$.

\section{A Convex Formulation of Co-clustering}
\label{sec:coco_formulation}

\begin{figure}[t]
    \centering
        \includegraphics[scale = 0.5]{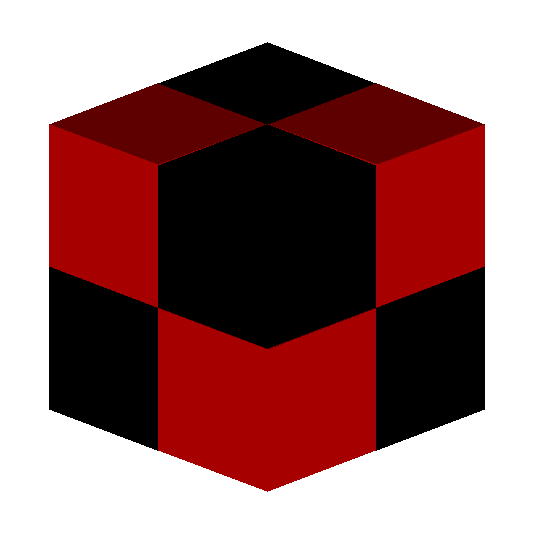} 
    \caption{A 3-way tensor with a checkerbox structure }
    \label{fig:checkerbox}
\end{figure}

We first consider a convex formulation of co-clustering problem when the data is a 3-way tensor $\T{X} \in \Real^{n_1 \times n_2 \times n_3}$ before discussing the natural generalization to $D$-way tensors. Our basic assumption is that the observed data tensor is a noisy realization of an underlying tensor that exhibits a checkerbox structure modulo some unknown reordering along each of its modes. Specifically suppose that there are $k_1, k_2,$ and $k_3$ clusters along modes 1, 2, and 3 respectively.  If the $(i_1,i_2,i_3)$-th entry in $\T{X}$ belongs to the cluster defined by the $r_1$th mode-1 group, $r_2$th mode-2 group, and $r_3$th mode-3 group, then we assume that the observed tensor element $\ME{X}{i_1i_2i_3}$ is given by
\begin{eqnarray}
\label{eq:mean_model}
\ME{X}{i_1i_2i_3} & = & \ME{c^*}{r_1r_2r_3} + \ME{\epsilon}{i_1i_2i_3},
\end{eqnarray}
where $\ME{c^*}{r_1r_2r_3}$ is the mean of the co-cluster defined by the $r_1$th mode-1 partition, $r_2$th mode-2 partition, and $r_3$th mode-3 partition, and $\ME{\epsilon}{i_1i_2i_3}$ are noise terms. We will specify a joint distribution on the noise terms later in \Sec{statistical_properties} in order to derive prediction bounds. Thus, we model the observed tensor $\T{X}$ as the sum of a mean tensor $\T{U}^* \in \Real^{n_1 \times n_2 \times n_3}$, whose elements are expanded from the co-cluster means tensor $\T{C}^* \in \Real^{k_1 \times k_2 \times k_3}$, and a noise tensor $\T{E} \in \Real^{n_1 \times n_2 \times n_3}$. We can write this expansion explicitly by introducing a membership matrix $\M{M}_d \in \{0,1\}^{n_d \times k_d}$ for the $d$th mode, where the $ik$th element of $\M{M}_d$ is one if and only if the $i$th mode-$d$ slice belongs to the $k$th mode-$d$ cluster for $k \in \{1, \ldots, k_d\}$. We require that each row of the membership matrix sum to one, namely $\M{M}_d\V{1} = 1$, to ensure that each of the mode-$d$ slices belongs to exactly one of the $k_d$ mode-$d$ clusters. Then,
\begin{eqnarray*}
\T{U}^* & = & \T{c}^* \times_1 \M{M}_1 \times_2 \M{M}_2 \times_3 \M{M}_3.
\end{eqnarray*}
\Fig{checkerbox} illustrates an underlying mean tensor $\T{U}^*$ after permuting the slices along each of the modes to reveal a checkerbox structure.

The co-clustering model in \Eqn{mean_model} is the 3-way analogue of the checkerboard mean model often employed in biclustering data matrices \citep{MadOli2004, TanWit2013, ChiAllenBaraniuk2017}. Moreover, the tensor $\T{C}^*$ of co-cluster means corresponds to the tensor of cluster ``centers" in the tensor clustering work by \cite{JegelkaSraBanerjee2009}. The model is complete and exclusive in that each tensor element is assigned to exactly one co-cluster. This is in contrast to models that allow potentially overlapping co-clusters \citep{LazOwe2002, BerIhm2003, TurBai2005, huang2008simultaneous, WitTib2009, LeeSheHua2010, SilKaiKop2011, BharHaubrockMukhopadhyayEtAl2015}.

Estimating the model in \Eqn{mean_model} consists of finding (i) the partitions along each mode and (ii) the mean values of each of the $k_1k_2k_3$ co-clusters. Estimating $\ME{c^*}{r_1r_2r_3}$, given the mode clustering assignments is trivial. Let $\mathcal{G}_1, \mathcal{G}_2,$ and $\mathcal{G}_3$ denote the indices of the $r_1$th mode-1, $r_2$th mode-2, and $r_3$th mode-3 groups respectively. If the noise terms 
$\ME{\epsilon}{i_1i_2i_3}$ are iid $N(0,\sigma^2)$ for some positive $\sigma^2$, then the maximum likelihood estimate of $\ME{c^*}{r_1r_2r_3}$ is simply the sample mean of the entries of $\T{X}$ over the indices defined by $\mathcal{G}_1, \mathcal{G}_2$, and $\mathcal{G}_3$, namely
\begin{eqnarray*}
\hat{c}^*_{r_1r_2r_3} & = & \frac{1}{\lvert \mathcal{G}_1 \rvert\lvert \mathcal{G}_2 \rvert \rvert\lvert \mathcal{G}_3 \rvert}\sum_{i_1 \in \mathcal{G}_1}\sum_{i_2 \in \mathcal{G}_2}\sum_{i_3 \in \mathcal{G}_3} \ME{X}{i_1i_2i_3}.
\end{eqnarray*}

Finding the partitions $\mathcal{G}_1, \mathcal{G}_2,$ and $\mathcal{G}_3$, on the other hand, is a combinatorially hard problem. In recent years, however, many combinatorially hard problems, that initially appear computationally intractable, have been successfully attacked by solving a convex relaxation to the original combinatorial optimization problem. Perhaps the most celebrated convex relaxations is the lasso \citep{Tibshirani1996}, which simultaneously performs variable selection and parameter estimation for fitting sparse regression models by minimizing a non-smooth convex criterion.

In light of the lasso's success, we propose to simultaneously identify partitions along the modes of $\T{X}$ and estimate the co-cluster means by minimizing the following convex objective function
\begin{eqnarray}
\label{eq:cvxtriclustr}
F_{\gamma}(\T{U})  & = & \frac{1}{2} \lVert \T{X} - \T{U} \rVert_{\text{F}}^2 + \gamma \underbrace{\bigg [R_1(\T{U}) + R_2(\T{U}) + R_3(\T{U}) \bigg ]}_{R(\T{U})},
\end{eqnarray}
where
\begin{eqnarray*}
R_1(\T{U}) &=&  \sum_{i < j} w_{1, ij}  \lVert \T{U}_{i::} - \T{U}_{j::}   \rVert_{\text{F}}  \nonumber   \\
R_2(\T{U}) &=&  \sum_{i < j} w_{2, ij}  \lVert \T{U}_{:i:} - \T{U}_{:j:}   \rVert_\text{F}   \\
R_3(\T{U}) &=&  \sum_{i < j} w_{3, ij}  \lVert \T{U}_{::i} - \T{U}_{::j}   \rVert_\text{F}  \nonumber.
\end{eqnarray*}
By seeking the minimizer $\That{U}_{\gamma}  \in \Real^{n_1 \times n_2 \times n_3}$ of \Eqn{cvxtriclustr}, we have cast co-clustering as a signal approximation problem, modeled as a penalized regression, to estimate the true co-cluster means tensor $\T{U}^*$. In the following discussion, we drop the dependence of $\gamma$ in $\That{U}_{\gamma}$ and denote our estimator as $\That{U}$ when there is no confusion. The quadratic term in \Eqn{cvxtriclustr} quantifies how well $\T{U}$ approximates $\T{X}$, while the regularization term $R(\T{U})$ in \Eqn{cvxtriclustr} penalizes deviations away from a checkerbox pattern. The nonnegative parameter $\gamma$ tunes the relative emphasis on these two terms. The parameters $w_{d, ij}$ are nonnegative weights whose purpose will be discussed shortly.

To appreciate how the regularization term $R(\T{U})$ steers the minimizer of \Eqn{cvxtriclustr} towards a checkerbox pattern, consider the effect of one of the terms $R_d(\T{U})$ in isolation. Specifically, suppose that $R(\T{U}) = R_1(\T{U})$. When $\gamma$ is zero, the minimum of \Eqn{cvxtriclustr} is attained when $\T{U} = \T{X}$. Or stated another way, $\T{U}_{i::} = \T{X}_{i::}$ for $i \in \{1, \ldots, n_1\}$.
As $\gamma$ increases, the mode-1 slices $\T{U}_{i::}$ will shrink towards each other and in fact coalesce due to the non-differentiability of the Frobenius norm at zero. In other words, as $\gamma$ gets larger, the pairwise differences of the mode-1 slices of $\That{U}$ will become increasingly sparser. Sparsity in these pairwise differences leads to a natural partitioning assignment. Two mode-1 slices $\T{X}_{i::}$ and $\T{X}_{j::}$ are assigned to the same mode-1 partition if $\T{U}_{i::} = \T{U}_{j::}$. Under mild regularity conditions, that we will spell out in \Sec{properties}, for sufficiently large $\gamma$, all mode-1 slices $\That{U}$ will be identical and therefore belong to a single cluster. Similar behavior holds if $R(\T{U}) = R_2(\T{U})$ or $R(\T{U}) = R_3(\T{U})$.

When $R(\T{U})$ includes all three terms $R_d(\T{U})$ for $d=1,2,3$, pairs of mode-1, mode-2, and mode-3 slices are {\em simultaneously} shrunk towards each other and coalesce as the parameter $\gamma$ increases. By coupling clustering along each of the modes simultaneously, our formulation explicitly seeks out a solution with a checkerbox mean structure. Moreover, we will show in \Sec{properties} that the solution $\That{U}$ produces an entire solution path of checkerbox co-clustering estimates that varies continuously in $\gamma$. The solution path spans a range of models from the least smoothed model, where $\That{U}$ is $\T{X}$ and each tensor element occupies its own co-cluster, to the most smoothed model, where all the elements of $\That{U}$ are identical and all tensor elements belong to a single co-cluster.

\begin{figure}
\centering
\begin{tikzpicture}
\node[draw=black,shape=circle] (1) at (-7.5,0)  {1};
\node[draw=black,shape=circle] (2) at (-5,0) {2};
\node[draw=black,shape=circle] (3) at (-3.75,2.25) {3};
\node[draw=black,shape=circle] (4) at (0, 0) {4};
\node[draw=black,shape=circle] (5) at (-2.5,0) {5};
\node[draw=black,shape=circle] (6) at (1.25,2.25) {6};
\node[draw=black,shape=circle] (7) at (-1.25,2.25) {7};
\draw (1) -- (2)
(2) -- (3)
(4) -- (5)
(4) -- (6)
(6) -- (7);
\node (8) at (-6.25,0.25) {$w_{1,12}$};
\node (9) at (-5, 1.125) {$w_{1,23}$};
\node (10) at (-1.25,0.25) {$w_{1,45}$};
\node (11) at (0, 1.125) {$w_{1,46}$};
\node (12) at (0, 2.5) {$w_{1,67}$};
\end{tikzpicture}
\caption{A graph that summarizes the similarities between pairs of the mode-$1$ subarrays. Only edges with positive weight are drawn.}
    \label{fig:edgeGraph}
\end{figure}
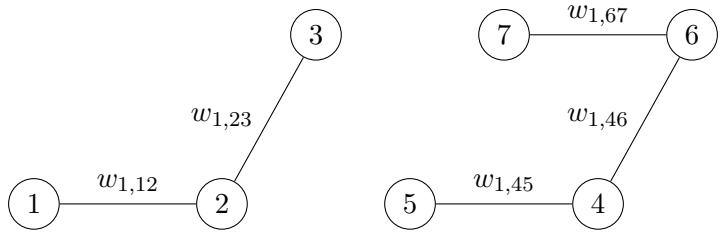

The nonnegative weights $w_{d, ij}$ fine tune the shrinkage of the slices along the $d$th mode. For example, if $w_{1,ij} > w_{1,i'j'}$, then there will be more pressure for $\T{U}_{i::}$ and $\T{U}_{j::}$ to fuse than for $\T{U}_{i'::}$ and $\T{U}_{j'::}$ to fuse as $\gamma$ increases. Thus, the weight $w_{d,ij}$ quantifies the similarity between the $i$th and $j$th mode-$d$ slices. A very large $w_{d,ij}$ indicates that the two slices are very similar, while a very small $w_{d,ij}$ indicates that they are very dissimilar. These pairwise similarities motivate a graphical view of clustering. For the $d$th mode, define the set $\E_d$ as the edge set of a similarity graph.  Each slice is a node in the graph and the set $\E_d$ contains an edge $(i, j)$ if and only if $w_{d, ij} > 0$.  \Fig{edgeGraph} shows an example of a mode-1 similarity graph, which corresponds to a tensor with seven mode-$1$ slices and positive weights that define the edge set
\begin{eqnarray*}
\E_1 & = & \{ (1,2), (2,3), (4,5), (4,6), (6,7) \}.
\end{eqnarray*}
Given the connectivity of the graph, as $\gamma$ increases, the slices $\T{U}_{1::}, \T{U}_{2::}$, and $\T{U}_{3::}$ will be shrunk towards each other while the slices $\T{U}_{4::}, \T{U}_{5::}, \T{U}_{6::}$ and $\T{U}_{7::}$ shrunk towards each other. Since $w_{d, ij} = 0$ for any $(i, j) \notin \E_d$, we can express the penalty terms for the $d$th mode as 
\begin{eqnarray*}
R_d(\T{U}) & = &  \sum_{(i, j) \in \E_d} w_{d, ij}  \lVert \T{U}_{i::} - \T{U}_{j::}   \rVert_{\text{F}}.
\end{eqnarray*}

The graph in \Fig{edgeGraph} makes readily apparent that the convex objective in \Eqn{cvxtriclustr} separates over the connected components of the similarity graph for the mode-$d$ slices. Consequently, one can solve for the optimal $\T{U}$ component by component. Without loss of generality, we assume that the weights are such that all the similarity graphs are connected. Before leaving this preliminary description of the weights, however, we want to emphasize that in practice weights are set once in a data-adaptive manner and should be considered empirically chosen hyper-parameters rather than tuning parameters. Further discussion of the weights and practical recommendations for specifying them will be discussed in \Sec{weights}.

Having familiarized ourselves with the convex co-clustering of a 3-way array, we now present the natural extension of  
\Eqn{cvxtriclustr} for clustering the fibers of a general higher-order tensor $\T{X} \in \Real^{n_1 \times \cdots \times n_D}$ along all its $D$ modes. Let $\V{\Delta}_{d,ij} = \V{e}_i\Tra - \V{e}_j\Tra$ where $\V{e}_i$ is the $i$th standard basis vector in $\Real^{n_d}$. The objective function of our convex co-clustering for a general higher-order tensor is as follows.

\begin{eqnarray}
\label{eq:primal_objective}
F_\gamma(\T{U}) & = & \frac{1}{2}\lVert \T{X} - \T{u} \rVert_{\text{F}}^2 + \gamma \sum_{d=1}^D \sum_{(i,j) \in \E_d} w_{d, ij} \lVert \T{U} \times_d \V{\Delta}_{d,ij} \rVert_{\text{F}}.
\end{eqnarray}

The difference between the convex triclustering objective \Eqn{cvxtriclustr} and the general convex co-clustering objective \Eqn{primal_objective} is in the penalty terms. Previously in \Eqn{cvxtriclustr} we penalized the difference between pairs slices whereas in \Eqn{primal_objective} we penalize the differences between pairs of mode-$d$ subarrays.

Note that the function $F_{\gamma}(\T{U})$ defined in \Eqn{primal_objective} has a unique global minimizer. This follows immediately from the fact that $F_{\gamma}(\T{U})$ is strongly convex. The unique global minimizer of $F_{\gamma}(\T{U})$ is our proposed CoCo estimator, which is denoted by $\That{U}$ for the remainder of the paper.

At times it will be more convenient to work with vectors rather than tensors.  By applying the identity in \Eqn{vec_identity}, we can rewrite the objective function in \Eqn{primal_objective} in terms of the vectorizations of $\T{U}$ and $\T{X}$ as follows
\begin{eqnarray}
\label{eq:primal_objective_vec}
F_{\gamma}(\V{U}) & = & \frac{1}{2} \lVert \V{x} - \V{U} \rVert_{2}^2 + \gamma \sum_{d=1}^D\sum_{(i,j) \in \mathcal{E}_d} \VE{w}{d,ij}\lVert \M{A}_{d,ij} \V{u} \rVert_2.
\end{eqnarray}
where $\M{A}_{d,ij}$ is the $n_{-d}$-by-$n$ matrix
\begin{eqnarray}
\label{eq:Adij}
\M{A}_{d,ij} & = & \M{I}_{n_D} \Kron \cdots \Kron \M{I}_{n_{d+1}}\Kron \V{\Delta}_{d,ij}\Kron \M{I}_{n_{d-1}} \Kron \cdots \Kron \M{I}_{n_1}
\end{eqnarray}
where $\M{I}_{n_{d}}$ is the $n_d$-by-$n_d$ identity matrix. We will refer to the unique global minimizer of \Eqn{primal_objective_vec}, $\Vhat{u} = \arg\min_{\V{u}} F_\gamma(\V{u})$, as the vectorized version of our CoCo estimator.

\begin{remark}
The fusion penalties $R_d(\T{U})$ are a composition of the group lasso \citep{YuanYi2006} and the fused lasso \citep{TibSauRos2005}, a special case of the generalized lasso \citep{TibTay2011}. When only a single mode is being clustered and only one of the terms $R_d(\T{U})$ is employed, we recover the objective function in the convex clustering problem \citep{PelDeSuy2005, She2010, LinOhlLju2011, HockingJoulinBachEtAl2011, Sharpnack2012, Zhu2014, Chi2015, Radchenko2017}. Most prior work on convex clustering employ an element-wise $\ell_1$-norm penalty on pairwise differences, as in the original fused lasso, however, 
 $\ell_2$-norm and $\ell_\infty$-norm have also been considered \citep{HockingJoulinBachEtAl2011,Chi2015}. In this paper, we restrict ourselves to the $\ell_2$-norm for two reasons. First, the $\ell_2$-norm is rotationally invariant. In general, we are reluctant to adopt a procedure whose co-clustering output may non-trivially change when the coordinate representation of the data along one of its modes is trivially changed. Second, the $\ell_2$-norm promotes the group-wise shrinkage of pairwise differences of subarrays along each mode leading to more straightforward partitioning along each mode. Pairwise differences are either exactly zero or not.
 When the tensor is a matrix and the rows and columns are being simultaneously clustered, we recover the objective function in the convex biclustering problem \citep{ChiAllenBaraniuk2017}. In general, the fusion penalties $R_d(\T{U})$ shrink solutions to vector valued functions that are piece-wise constant over the mode-$d$ similarity graph defined by the weights $w_{d,ij}$. Viewed this way, we can see our approach as simultaneously performing the network lasso \citep{HallacLeskovecBoyd2015} on $D$ similarity graphs.
\end{remark}

\begin{remark}
The CoCo estimator is invariant to permutations in the data tensor $\T{X}$ in the following sense. Suppose $\That{U}$ and $\That{U}'$ are the CoCo estimators when the data tensors are respectively $\T{X}$ and $\T{X}' = \T{X}\times_1 \M{\Pi}_1 \times_2 \cdots \times_D \M{\Pi}_D$ where $\M{\Pi}_1 \in \{0,1\}^{n_1 \times n_1}, \ldots, \M{\Pi}_D \in \{0,1\}^{n_D \times n_D}$ are permutation matrices, namely $\M{\Pi}_d\Tra\M{\Pi}_d = \M{I}$.
In words, $\T{X}'$ can be obtained from $\T{X}$ by permuting the subarrays of $\T{X}$ along the $d$th mode according to $\M{\Pi}_d$ for $d = 1, \ldots, D$, and $\T{X}$ can be recovered from $\T{X}'$ by permuting along the $d$th mode according to $\M{\Pi}_d\Tra$ for $d = 1, \ldots, D$. Since $\lVert \T{U} \times_1 \M{\Pi}_1 \times_2 \cdots \times_D \M{\Pi}_D \rVert_{\text{F}} = \lVert \T{U} \rVert_{\text{F}}$, it follows that
\begin{eqnarray*}
    \That{U}' & = & \That{U} \times_1 \M{\Pi}_1 \times_2 \cdots \times_D \M{\Pi}_D \quad\text{and}\quad
    \That{U} \amp = \amp \That{U}'\times_1 \M{\Pi}_1\Tra \times_2 \cdots \times_D \M{\Pi}_D\Tra.
\end{eqnarray*}
Permutation invariance is important because it means that the CoCo estimator is essentially unaltered by any reshuffling along the modes of the data tensor.
\end{remark}

\begin{remark}
Given the co-clustering structure assumed in \Eqn{mean_model}, one may wonder how much is added by explicitly seeking a co-clustering over clustering along each mode independently. In other words, why not solve $D$ independent convex clustering problems with $R(\T{U}) = R_d(\T{U})$? To provide some intuition on why co-clustering should be preferred over independently clustering each mode, consider the following problem. Imagine trying to cluster row vectors $\V{x}_i \in \Real^{10,000}$ for $i = 1, \ldots, 100$ drawn from a two-component mixture of Gaussians, namely
\begin{eqnarray*}
\V{x}_i & \overset{iid}{\sim} & \frac{1}{2} N(\V{\mu}, \sigma^2\M{I}) + \frac{1}{2} N(\V{\nu}, \sigma^2\M{I}).
\end{eqnarray*}
This is a challenging clustering problem due to the disproportionately small number of observations compared to the number of features. If, however, we were told that $\VE{\mu}{j} = \mu_1$ and $\VE{\nu}{j} = \nu_1$ for $j = 1, \ldots, 5,000$ and $\VE{\mu}{j} = \mu_2$ and $\VE{\nu}{j} = \nu_2$ for $i = 5,001, \ldots, 10,000$, in other words that the features were clustered into two groups, our fortunes have reversed and we now have an abundance of observations compared to the number of effective features. Even if we lack a clear-cut clustering structure in the features, this example suggests that leveraging similarity structure along the columns can expedite identifying similarity structure along the rows, and vice versa. Indeed, if there is an underlying checkerbox mean tensor we may expect that simultaneously clustering along each mode should make the task of clustering along any one given mode easier. Our prediction error result presented in \Sec{statistical_properties} in fact supports this suspicion (See Remark \ref{rem:connection}).
\end{remark}

\section{Properties}
\label{sec:properties}

We first discuss how the CoCo estimator $\That{U}$ behaves as a function of the data tensor $\T{X}$, the tuning parameter $\gamma$, and the weights $w_{d,ij}$. We will then present its statistical properties under mild conditions on the data generating process. We highlight that these properties hold regardless of the algorithm used to minimize \Eqn{primal_objective}, as they are intrinsic to its convex formulation. All proofs are given in \App{proof_smoothness} and \App{proof_final_error}.

\subsection{Stability Properties}
\label{sec:algorithmic_properties}

The CoCo estimator varies smoothly with respect to $\T{X}$, $\gamma$, and $\{w_{d,ij}\}$. Let $\M{W}_d = \{  {w}_{d, ij}   \}$ denote the weights matrix for mode $d$.

\begin{proposition}
\label{prop:cont}
The minimizer $\That{U}$ of \Eqn{primal_objective} is jointly continuous in $(\T{X}, \gamma, \M{W}_1, \M{W}_2, \ldots, \M{W}_D).$
\end{proposition}
As noted earlier, in practice we will typically fix the weights $w_{d,ij}$ and compute the CoCo estimator over a grid of the penalization parameters $\gamma$ in order to select a final CoCo estimator from among the computed candidate estimators of varying levels of smoothness. Since \Eqn{primal_objective} does not admit a closed form minimizer, we resort to iterative algorithms for computing the CoCo estimator. Continuity of $\That{U}$ in $\gamma$ can be leveraged to expedite computation through warm starts, namely using the solution $\That{U}_{\gamma}$ as the initial guess for iteratively computing $\That{U}_{\gamma'}$ where $\gamma'$ is slightly larger or smaller than $\gamma$.
Due to the continuity of $\That{U}$ in $\gamma$, small changes in $\gamma$ will result in small changes in $\That{U}$.  Empirically the use of warm starts can lead to a non-trivial reduction in computation time \citep{Chi2015}.  From the continuity in $\gamma$, we also see that convex co-clustering performs continuous co-clustering just as the lasso \citep{Tibshirani1996} performs continuous variable selection.  

The penalization parameter $\gamma$ tunes the complexity of the CoCo estimator. Clearly when $\gamma = 0$, the CoCo estimator coincides with the data tensor, namely $\That{U} = \T{X}$. The key to understanding the CoCo estimator's behavior as $\gamma$ increases is to recognize that the penalty functions $R_d(\T{U})$ are semi-norms. Under suitable conditions on the weights given in \As{connectedness} below, $R_d(\T{U})$ vanishes if and only if the mode-$d$ subarrays of $\T{U}$ are identical.

\begin{assumption}
\label{as:connectedness}
For any pair of mode-$d$ subarrays, indexed by $i$ and $j$ with $i < j$, there exists a sequence of indices $i \rightarrow k \rightarrow \cdots \rightarrow l \rightarrow j$ along which the weights, $w_{d, ik}, \ldots, w_{d,lj}$ are positive.
\end{assumption}

\begin{proposition}
\label{prop:zero}
Under Assumption 4.1, $R_d(\T{U}) = 0$ if and only if $\Mz{U}{d} = \V{1}\V{c}\Tra$ for some $\V{c} \in \Real^{n_{-d}}$.
\end{proposition}

To give some intuition for \Prop{zero}, note that the term $R_d(\T{U})$ separates over the connected components of the mode-$d$ similarity graph. Therefore, the term $R_d(U)$ penalizes variation in the mode-$d$ subarrays over the connected components of the mode-$d$ similarity graph. Assumption 4.1, states that the mode-$d$ similarity graph is connected. Thus, the only way for $R_d(U)$ to attain its minimum value and vanish under Assumption 4.1, is if there is no variation in $\T{U}$ along its mode-$d$ subarrays.

\Prop{zero} suggests that if \As{connectedness} holds for all $d = 1, \ldots, D$ then
 as $\gamma$ increases the CoCo estimator converges to the solution of the following constrained optimization problem:
\begin{eqnarray*}
\underset{\V{U}}{\min}\; \frac{1}{2} \lVert \V{X} - \V{U} \rVert_{\text{F}}^2 \quad \text{subject to $\V{U} = c\V{1}$ for some $c \in \Real$},
\end{eqnarray*}
the solution to which is just the global mean $\Vbar{X}$, whose entries are all identically the average value of $\V{X}$ over all its entries. The next result formalizes our intuition that as $\gamma$ increases, the CoCo estimator will eventually coincide with $\Vbar{X}$.

\begin{proposition}
\label{prop:coalesce}
Suppose \As{connectedness} holds for $d = 1, \ldots, D$, then $F_{\gamma}(\T{U})$ is minimized by the grand mean $\Tbar{x}$ for $\gamma$ sufficiently large.
\end{proposition}
Thus, as $\gamma$ increases from 0, the CoCo estimator $\That{U}$ traces a continuous solution path that starts from $n$ co-clusters, consisting of $\ME{U}{i_1\cdots i_D} = \ME{x}{i_1\cdots i_D}$, to a single co-cluster, where $\ME{U}{i_1\cdots i_D} = \V{x}\Tra \V{1}/n$ for all $i_1,\ldots, i_D$.

For a fixed $\gamma$, we can derive an explicit bound on sensitivity of the CoCo estimator to perturbations in the data.
\begin{proposition}
\label{prop:non_expansive}
The minimizer $\That{U}$ of \Eqn{primal_objective} is a nonexpansive or 1-Lipschitz function of the data tensor $\T{X}$, namely
\end{proposition}
\begin{eqnarray*}
\lVert \That{U}(\T{X}) - \That{U}(\Ttilde{X}) \rVert_{\text{F}} & \leq &  \lVert \T{X} - \Ttilde{X} \rVert_{\text{F}}.
\end{eqnarray*}
Nonexpansivity of $\That{U}$ in $\T{X}$ provides an attractive stability result.  Since $\That{U}$ varies smoothly with the data, small perturbations in the data are guaranteed to not lead to large variability of $\That{U}$, or consequently large variability in the cluster assignments. 
In a special case of our method, \cite{ChiAllenBaraniuk2017} showed empirically that the co-clustering assignments made by the 2-way version of the CoCo estimator was noticeably less sensitive to perturbations in the data than those made by several existing biclustering algorithms. 

\subsection{Statistical Properties}
\label{sec:statistical_properties}

We next provide a finite sample bound for the prediction error of the CoCo estimator. 
For simplicity, we consider the case where we take uniform weights within a mode in \Eqn{primal_objective_vec}, namely $w_{d,ij} = w_{d,i'j'} = 1/n_d$ for all $i,j,i',j' \in \{1, \ldots, n_d\}$. Such uniform weight assumption has also been imposed in the analysis of the vector-version of convex clustering \citep{Tan2015}.

In order to derive the estimation error of $\Vhat{u}$, we first define an important definition for the noise and introduce two regularity conditions.

\begin{definition}[\cite{VuWang2015}]
\label{def:error} 
We say a random vector $\V{y} \in \Real^n$ is $M$-concentrated if there are constants $C_1, C_2 >0$ such that for any convex, 1-Lipschitz function $\phi: \Real^n \rightarrow \Real$ and any $t>0$,
\begin{eqnarray*}
\mathbb P \bigg(\big \lvert \phi(\V{y}) - \mathbb{E}[\phi(\V{y})] \big\rvert \ge t \bigg) & \le & C_1 \exp\left(- \frac{C_2t^2}{M^2}\right).
\end{eqnarray*}
\end{definition}
The $M$-concentrated random variable is more general than the Gaussian or sub-Gaussian random variables, and it allows dependence in its coordinates. \cite{VuWang2015} provided a few examples of $M$-concentrated random variables. For instance, if the coordinates of $\V{y}$ are iid standard Gaussian, then $\V{y}$ is 1-concentrated. If the coordinates of $\V{y}$ are independent and $M$-bounded, then $\V{y}$ is $M$-concentrated. If the coordinates of $\V{y}$ come from a random walk with certain mixing properties, then $\V{y}$ is $M$-concentrated for some $M$. 

\begin{assumption}[Model]
\label{as:model} 
We assume the true cluster center $\T{C}^* \in \Real^{k_1\times \cdots \times k_D}$ has a checkerbox structure such that the mode-$d$ subarrays have $k_d$ different values (number of clusters along the $d$th mode), and each entry of $\T{C}^*$ is bounded above by a constant $C_0 > 0$. Define $\T{U}^* \in \Real^{n_1 \times \cdots \times n_D}$ as the true parameter expanded based on $\T{C}^*$, namely
\begin{eqnarray*}
\T{U}^* & = & \T{C}^* \times_1 \M{M}_1 \times_2 \M{M}_2 \times_3 \cdots \times_D \M{M}_D,
\end{eqnarray*}
where $\M{M}_d \in \{0, 1\}^{n_d \times k_d}$ are binary mode-$d$ cluster membership matrices such that $\M{M}_d\V{1} = 1$. Denote $\V{u}^*= \vec(\T{U}^*) \in \Real^n$ with $n=\prod_{d=1}^D n_d$. We assume the samples belonging to the $(r_1, \ldots, r_D)$-th cluster satisfy
\begin{eqnarray*}
\ME{X}{i_1, \ldots, i_D} & = & \ME{C^*}{r_1, \ldots, r_D} + \ME{\epsilon}{i_1, \ldots, i_D},
\end{eqnarray*}
with $i_d \in \{1,\ldots, n_d\}$ and $r_d \in \{1,\ldots, k_d\}$. Furthermore, we assume $\V{\epsilon} =  \vec(\T{E})$ is a $M$-concentrated random variable defined in $(\ref{def:error})$ with mean zero. 
\end{assumption}

The checkerbox means model in \As{model} provides the underlying cluster structure of the tensor data. As a special case, \As{model} with $D=2$ reduces to the model assumption underlying convex biclustering \citep{ChiAllenBaraniuk2017}. In contrast to the independent sub-Gaussian condition assumed in vector-version convex clustering \citep{Tan2015}, our error condition is much weaker since we allow for non-sub-Gaussian distributions as well as allow for dependence among its coordinates.

\begin{assumption}[Tuning]
\label{as:tuning} 
The tuning parameter $\gamma$ satisfies 
\begin{eqnarray*}
\frac{2 \log(n)\sqrt{n}}{D} & \le & \gamma \amp \le \amp \frac{2c_0 \log(n)\sqrt{n}}{D},
\end{eqnarray*}
for some constant $c_0 > 1$.
\end{assumption}

\begin{theorem}
\label{thm:final_error}
Suppose that \As{model} and \As{tuning} hold. The estimation error of $\Vhat{u}$ in \Eqn{primal_objective_vec} with uniform weights satisfies,
\begin{eqnarray}
\label{eq:final_error}
\frac{1}{n}\bigl\lVert \Vhat{u} - \V{u}^* \bigr\rVert_2^2 & \le & \frac{1}{D} \sum_{d=1}^D \left( \frac{1}{n_d} + \frac{\log(n)}{\sqrt{n n_d }}   \right) + \frac{C \log(n)}{D\sqrt{n}} \sum_{d=1}^D n_d \sqrt{\prod_{j\ne d} k_j},
\end{eqnarray}
with a high probability, where $C = 12c_0C_0^2$ is a positive constant, and $k_d$ is the true number of clusters in the $d$th mode. 
\end{theorem}

\Thm{final_error} provides a finite sample error bound for the proposed CoCo tensor estimator. Our theoretical bound allows the number of clusters in each mode to diverge, which reflects a typical large-scale clustering scenario in big tensor data. A notable consequence of \Thm{final_error} is that, when $D \ge 3$, namely a higher-order tensor with at least 3 modes, the CoCo estimator can achieve estimation consistency along all the $D$ modes even when we only have one tensor sample. Here the sample size refers to the number of available tensor samples. In our tensor clustering problem, we only have access to one tensor sample.

This property is uniquely enjoyed by co-clustering of tensor data with $D\ge 3$, and has not been previously established in the existing literature on vector clustering or biclustering. To see this, when $n_d$ are of the same order as $n_0$, and $k_d$ are of the same order as $k_0$, a sufficient condition for the consistency is that $n_0 \rightarrow \infty$ and $k_0 = o\big (n_0^{(D-2)/(D-1)}\big)$ up to a log term. When $D=3$, the CoCo estimator is consistent so long as the number of clusters $k_0$ in each mode diverges slightly slower than $\sqrt{n_0}$. Remarkably, as we have more modes in the tensor data, this constraint on the rate of divergence of $k_0$ gets weaker. In short, we reap a unique and surprisingly welcome ``blessing of dimensionality" phenomenon in the tensor co-clustering problem. 

\begin{remark}
\label{rem:connection}
Next we discuss the connections of our bound \Eqn{final_error} with prior results in the literature. An intermediate step in the proof of \Thm{final_error} indicates that the estimation error in the $d$th mode is on the order of $1/n_d + \log(n)/\sqrt{n n_d} + \log(n)\sqrt{n_d \prod_{j \ne d} k_j / n_{-d} }$. In the clustering along the rows of a data matrix, our rate matches with that established for vector-version convex clustering \citep{Tan2015}, up to a log term $\sqrt{\log(n)}$. Such a log term is due to that fact that \cite{Tan2015} considers the error to be iid sub-Gaussian while we consider a general $M$-concentrated error. In practice, the iid assumption on the noise $\V{\epsilon} =  \vec(\T{E})$ could be restrictive. Consequently, our theoretical analysis is built upon a new concentration inequality of quadratic forms recently developed in \cite{VuWang2015}. In addition, our rate reveals an interesting theoretical property of the convex biclustering method proposed by \cite{ChiAllenBaraniuk2017}. When $D=2$, our rate indicates that the estimation error along the row and column of the data matrix is $\log(n_1 n_2)\sqrt{n_1 k_2 / n_2}$ and $\log(n_1 n_2)\sqrt{n_2 k_1 / n_1}$, respectively. Clearly, both errors can not converge to zero simultaneously. This indicates a disadvantage of matricizing a data tensor for co-clustering. 
\end{remark}

\section{Estimation Algorithm}
\label{sec:estimation_algorithm}

We next discuss a simple first order method for computing the solution to the convex co-clustering problem. The proposed algorithm generalizes the variable splitting approach introduced for convex clustering problem described in \cite{Chi2015} to the CoCo problem. The key observation is that the Lagrangian dual of an equivalent formulation of the convex co-clustering problem is a constrained least squares problem that can be iteratively solved using the classic projected gradient algorithm.

\subsection{A Lagrangian Dual of the CoCo Problem}

Recall that we seek to minimize the objective function in \Eqn{primal_objective_vec}
\begin{eqnarray*}
F_{\gamma}(\V{U}) & = & \frac{1}{2} \lVert \V{x} - \V{U} \rVert_{2}^2 + \gamma \sum_{d=1}^D\sum_{l \in \mathcal{E}_d} \VE{w}{d,l}\lVert \M{A}_{d,l} \V{u} \rVert_2.
\end{eqnarray*}
Note that we have enumerated the edge indices in $\mathcal{E}_d$ to simplify the notation for the following derivation.

We perform variable splitting and introduce the dummy variables $\V{v}_{d,l} = \M{A}_{d,l}\V{u}$. Let $\M{V}_d$ denote the $n_{-d} \times \lvert \mathcal{E}_d\rvert$ matrix whose $l$th column is $\V{v}_{d,l}$. 
Further denote the vectorization of $\M{V}_d$ by $\V{v}_d = \vec(\M{V}_d)$ and let $\V{v} = \begin{bmatrix} \V{v}_1\Tra & \V{v}_2\Tra & \cdots & \V{v}_D\Tra \end{bmatrix}\Tra$ denote the vector obtained by stacking the vectors $\V{v}_d$ on top of each other. We now solve the equivalent equality constrained minimization
\begin{eqnarray*}
\underset{\V{v}, \V{u}}{\min}\; \frac{1}{2} \lVert \V{x} - \V{U} \rVert_{2}^2 + \gamma \sum_{d=1}^D \sum_{l \in \mathcal{E}_d} \VE{w}{d,l}\lVert \V{v}_{d,l} \rVert_2 \quad\quad \text{subject to} \quad\quad \V{v}_d = \M{A}_d\V{u},
\end{eqnarray*}
where $\M{A}_{d} = (\M{I}_{n_D} \Kron \cdots \Kron \M{I}_{n_{d+1}}\Kron \M{\Phi}_d \Kron \M{I}_{n_{d-1}} \Kron \cdots \Kron \M{I}_{n_1} )$ and $\M{\Phi}_d$ is the oriented edge-vertex incidence matrix for the $d$th mode graph, namely
\begin{eqnarray*}
\ME{\Phi}{d,lv} = \begin{cases}
1 & \text{If node $v$ is the head of edge $l$} \\
-1 & \text{If node $v$ is the tail of edge $l$} \\
0 & \text{otherwise.}
\end{cases}
\end{eqnarray*}

We introduce dual variables $\V{\lambda}_d$ corresponding to the equality constraint $\V{v}_d = \M{A}_d\V{u}$. Let $\M{\Lambda}_d$ denote the 
$n_{-d} \times \lvert \mathcal{E}_d\rvert$ matrix whose $l$th column is $\V{\lambda}_{d,l}$. Further denote the vectorization of $\M{\Lambda}_d$ by $\V{\lambda}_d = \vec(\M{\Lambda}_d)$ and $\V{\lambda} = \begin{bmatrix}\V{\lambda}_1\Tra & \V{\lambda}_2\Tra & \cdots & \V{\lambda}_D\Tra \end{bmatrix}\Tra$. 
The Lagrangian dual objective is given by
\begin{eqnarray*}
G(\V{\lambda}) & = & 
\frac{1}{2} \lVert \V{x} \rVert_2^2- \frac{1}{2} \lVert \V{x} -  \M{A}\Tra\V{\lambda} \rVert_{2}^2 
- \sum_{d=1}^D \sum_{l \in \mathcal{E}_d} \iota_{C_{d,l}}(\V{\lambda}_{d,l}),
\end{eqnarray*}
where $\M{A} = \begin{bmatrix} \M{A}_1\Tra & \M{A}_2\Tra & \cdots & \M{A}_D\Tra \end{bmatrix}\Tra$ and $\iota_{C_{d,l}}$ is the indicator function of the closed convex set $C_{d,l} = \{\V{z} : \lVert \V{z} \rVert_2 \leq \gamma w_{d,l} \}$, namely $\iota_{C_{d,l}}$ is the function that vanishes on the set of $C_{d,l}$ and is infinity on the complement of $C_{d,l}$. Details on the derivation of the dual objective $G(\V{\lambda})$ are provided in \App{dual_derivation}.

Maximizing the dual objective $G(\V{\lambda})$ is equivalent to solving the following constrained least squares problem:
\begin{equation}
\label{eq:dual_problem}
\begin{split}
\underset{\V{\lambda} \in C}{\min}&\; \frac{1}{2}\lVert \V{x} - \M{A}\Tra\V{\lambda} \rVert_2^2,
\end{split}
\end{equation}
where $C = \{\V{\lambda} : \V{\lambda}_{d,l} \in C_{d,l}, l \in \mathcal{E}_d, d = 1, \ldots, D\}$. We can recover the primal solution via the relationship:
\begin{eqnarray*}
\Vhat{u} & = & \V{x} - \M{A}\Tra\Vhat{\lambda},
\end{eqnarray*}
where $\Vhat{\lambda}$ is a solution to the dual problem \Eqn{dual_problem}. The dual problem \Eqn{dual_problem} has at least one solution by the Weierstrass extreme value theorem, but the solution may not be unique since $\M{A}\Tra$ has a non-trivial kernel. Nonetheless, our CoCo estimator $\Vhat{u}$ is still unique since $\M{A}\Tra\Vhat{\lambda}_1 = \M{A}\Tra\Vhat{\lambda}_2$ for any solutions $\Vhat{\lambda}_1, \Vhat{\lambda}_2$ to the problem \Eqn{dual_problem}.

We numerically solve the constrained least squares problem in \Eqn{dual_problem} with the projected gradient algorithm, which alternates between taking a gradient step and projecting onto the set $C$. \Alg{CoCo} provides pseudocode of the projected gradient algorithm, which has several good features. The projected gradient algorithm is guaranteed to converge to a global minimizer of \Eqn{dual_problem}. Its per-iteration and storage costs using the weight choices, described in \Sec{weights}, are both $\mathcal{O}(Dn)$, namely linear in either the number of dimensions $D$ or in the number of elements $n$. For a modest additional computational and storage cost, we can accelerate the projected gradient method, for example with FISTA \citep{BecTeb2009} or SpaRSA \citep{Wright2009}. In our experiments, we use a version of the latter, namely FASTA \citep{GoldsteinStuderBaraniuk, FASTA}.
Additional details on the derivation of the algorithmic updates, convergence guarantees, computational and storage costs, as well as stopping rules can be found in \App{projected_gradient}.

\begin{algorithm}[t]
Initialize $\Vn{\lambda}{0}$; for $m=0, 1, \ldots$
\begin{algorithmic}[0]
  \caption{Convex Co-Clustering (CoCo) Estimation Algorithm}
  \label{alg:CoCo}
\Repeat
\State $\Vn{u}{m+1} = \V{x} - \M{A}\Tra\Vn{\lambda}{m}$
\Comment{Gradient Step}
\For{$d = 1, \ldots, D$}
	\For{$l \in \E_d$}
		\State $\Vn{\lambda}{m+1}_{d,l} = \mathcal{P}_{C_{d,l}} \left( \Vn{\lambda}{m}_{d,l} + \eta \M{A}_{d,l}\Vn{u}{m+1}\right)$
		\Comment{Projection Step}
	\EndFor
\EndFor
\Until{convergence}
\end{algorithmic}
\end{algorithm}

\section{Specifying Non-Uniform Weights}
\label{sec:weights}

In \Sec{statistical_properties}, we assumed uniform weights $w_{d,ij}$ in the penalty terms $R_d(\T{U}$) to establish a prediction error bound, which revealed a surprising and beneficial ``blessing of dimensionality" phenomenon. Although this simplifying assumption gives clarity and insight into how the co-clustering problem gets easier as the number of modes increases, in practice choosing non-uniform weights can substantially improve the quality of the clustering results. In the context of convex clustering, \cite{Chen2015} and \cite{Chi2015} provided empirical evidence that  convex clustering with uniform weights struggled to produce exact sparsity in the pairwise differences of smooth estimates when there was not a strong separation between groups. Indeed, similar phenomena were observed in earlier work on the related clustered lasso \citep{She2010}. Several related works \citep{She2010, HockingJoulinBachEtAl2011, Chen2015, Chi2015} recommend a weight assignment strategy described below. In addition, the use of {\em sparse} weights can also lead to non-trivial improvements in both computational time and clustering performance \citep{Chi2015, ChiAllenBaraniuk2017}. 

To illustrate the practical value of non-uniform weights, we compare CoCo's ability to recover co-clusters, using both uniform and non-uniform weights, as the size of a 3-way tensor increases when there are two clusters per mode with balanced cluster sizes along each mode. We assess the quality of the recovered clustering performance using the Adjusted Rand Index (ARI). The ARI \citep{HubertArabie1985} varies between -1 and 1, where 1 indicates a perfect match between two clustering assignments whereas a value close to zero indicates the two clustering assignments match about as might be expected if they were both randomly generated. Negative values indicate that there is less agreement between clusterings than expected from random partitions.

\begin{figure}[t]
    \centering
    \includegraphics[scale=0.45]{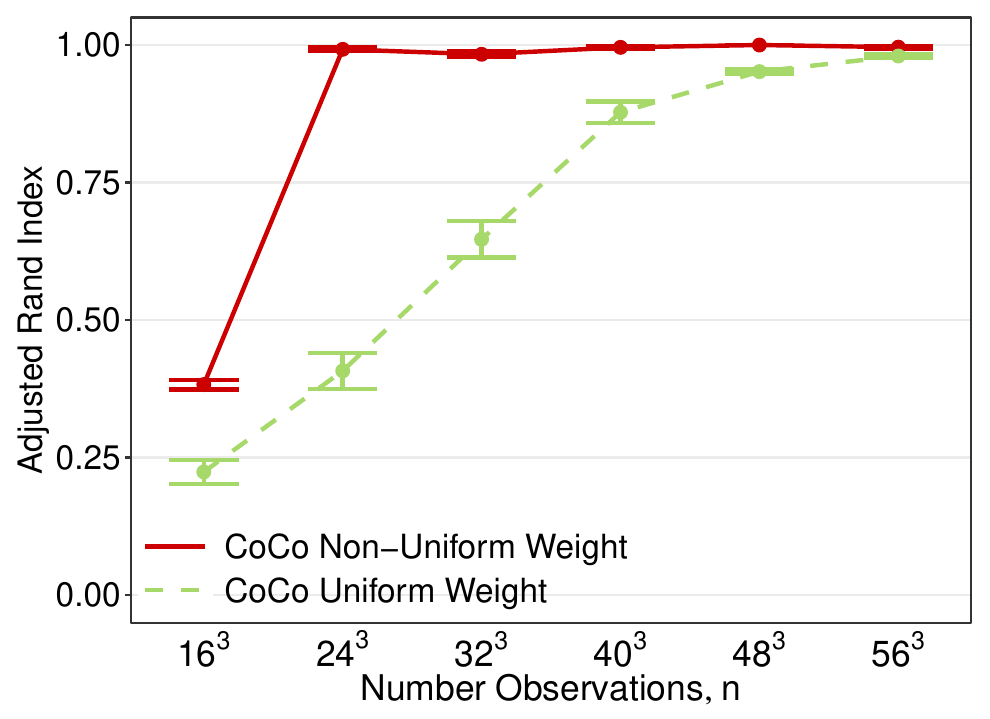}
    \caption{Uniform versus non-uniform weights:  Average Adjusted Rand Index for an increasing size. Here $n=n_0^3$ refers to a tensor of size $n_0\times n_0\times n_0$. \label{fig:checkerCube_4_sigmas_ARI_triclust}}
\end{figure}

\Fig{checkerCube_4_sigmas_ARI_triclust} shows a comparison between using non-uniform weights that are described in \Sec{tucker_weights} and uniform weights. Each plotted point in \Fig{checkerCube_4_sigmas_ARI_triclust} is the average ARI over 100 replicates. For CoCo using non-uniform weights, the smoothing parameter $\gamma$ is chosen with the data-driven extended BIC method that is detailed in \Sec{bic}. In contrast, for CoCo using uniform weights, $\gamma$ is chosen as the value that produces the estimator that minimizes the true but unknown MSE. 

We see that while using uniform weights in CoCo leads to recovering co-clusters exactly once a sufficient number of samples have been acquired, using non-uniform weights enables CoCo to recover the co-clusters exactly with notably fewer samples. The results of this experiment are especially remarkable because CoCo using non-uniform weights and a {\em data-adaptive} choice of $\gamma$ outperformed CoCo using uniform weights and an ideally chosen {\em oracle} value of $\gamma$.

As in the case of convex clustering, using non-uniform weights can lead to significantly better performance over using uniform weights in practice. We give some explanation for why this is expected in \Sec{folded_concave} but leave it to future work to develop theory proving this performance improvement. Nonetheless based on this observation, we employ non-uniform weights in CoCo for the empirical studies presented later in the paper.

\subsection{Basic Procedure for Specifying Weights}
We first describe our basic two step procedure for constructing weights before elaborating on the final refinements used in our numerical experiments.\\

\noindent {\bf Step 1:} We first calculate pre-weights $\tilde{w}_{d, ij}$ between the $i$th and $j$th mode-$d$ subarrays as 
\begin{eqnarray}
\label{eq:preweights}
\tilde{w}_{d, ij} & = & \iota^k_{\{i,j\}} \exp\left(-\tau_d \lVert \M{X}_{(d),i:} - \M{X}_{(d),j:}\rVert_{\text{F}}^2 \right).
\end{eqnarray}

The first factor on the right hand side of equation \Eqn{preweights}, $\iota^k_{\{i,j\}}$, is an indicator function that equals 1 if the $j$th slice is among the $i$th slice's $k$-nearest neighbors (or vice versa) and 0 othewise.  The purpose of this term is to control the sparsity of the weights.  The corresponding tuning parameter $k$ influences the connectivity of the mode-$d$ similarity graph.  One can explore different levels of granularity in the clustering by varying $k$ \citep{Chen2015}.   As a default, one can use the smallest $k$ such that the similarity graph is still connected. 
Note it is not necessary to calculate the exact $k$-nearest neighbors, which scales quadratically in the number of fibers in the mode.  A fast approximation to the $k$-nearest neighbors is sufficient for the sake of inducing sparsity into the weights. \cite{Chi2015} provided two reasons for using $k$-nearest neighbor weights. First, we wish to prioritize fusions between pairs of subarrays that are most similar; the subarrays that are most dissimilar should be the last pair of subarrays to fuse as the smoothing parameter $\gamma$ increases. Second, we wish to use a sparse similarity graph as the computational and storage complexity of the estimation algorithm is proportional to the number of non-zero edges in the similarity graphs (\App{projected_gradient}). Using $k$-nearest-neighbors weights accomplishes both goals.

The second factor on the right hand side of equation \Eqn{preweights} is the Gaussian kernel, which takes on larger values for pairs of mode-$d$ subarrays that are more similar to each other.  \citet{ChiSteinerberger2019} give a detailed theoretical justification for using weights like the Gaussian kernel weights in the context of convex clustering. For space considerations, we refer readers interested in these technical details to their work and give a brief intuitive rationale for the employing the Gaussian kernel here. Intuitively, the weights should be inversely proportional to the distance between the $i$th and $j$th mode-$d$ subarrays \citep{Chen2015, ChiAllenBaraniuk2017}.  The inverse of  the nonnegative parameter $\tau_d$ is a measure of scale. In practice, we can set it to be the median Euclidean distance between the $i$th and $j$th mode-$d$ subarrays that are $k$-nearest neighbors of each other. A value of $\tau_d = 0$ corresponds to uniform weights. Note that with minor modification, we can make the inverse scale parameter to be pair dependent as described in \cite{Zelnik-Manor2005}. \\

\noindent {\bf Step 2:} To obtain the mode-$d$ weights $w_{d,ij}$, we normalize the mode-$d$ pre-weights $\tilde{w}_{d,ij}$ to sum to $\sqrt{n_d/n}$.  The normalization step puts the penalty terms $R_d(\T{U})$ on the same scale and ensures that clustering along any given single mode will not dominate the entire co-clustering as $\gamma$ increases.

\subsection{Improving Weights via the Tucker Decomposition}
\label{sec:tucker_weights}

In our preliminary experiments, we found that substituting a low-rank approximation of $\T{X}$, namely a Tucker decomposition $\Ttilde{X}$, in place of $\T{X}$ in \Eqn{preweights} led to a marked improvement in co-clustering performance. To understand the boost in performance suppose that $\T{X} = \T{U}^* + \T{E}$ with $\T{U}^*$ having a checkerbox structure and the entries of $\T{E}$ are iid 
$N(0,\sigma^2)$ for simplicity. Further suppose that the $i$th and $j$th mode-$d$ subarrays of $\T{U}^\star$ belong to the same partition and $\iota^k_{\{i,j\}} = 1$. Then
\begin{eqnarray*}
\tilde{w}_{d, ij} & = &  \exp\left(-\tau_d \lVert \T{E} \times_d \V{\Delta}_{ij} \rVert_{\text{F}}^2 \right) \amp = \amp 
\exp\left(-2\tau_d \sigma^2 Z_{d,ij} \right),
\end{eqnarray*}
where $Z = \frac{\lVert \T{E} \times_d \V{\Delta}_{ij} \rVert_{\text{F}}^2}{2\sigma^2}$ is distributed as a $\chi^2$ random variable with $n_d$ degrees of freedom. If we were able to perfectly denoise the tensor $\T{X}$ so that $\sigma = 0$, then the pre-weight $\tilde{w}_{d,ij}$ would be set to its maximal value of 1, the ideal value for $\tilde{w}_{d,ij}$ since we have assumed the $i$th and $j$th mode-$d$ subarrays belong to the same partition. Thus, if we can reduce $\sigma^2$, namely denoise the observed tensor $\T{X}$, we can approach the ideal value of pre-weights. Note that we are more focused with approaching the ideal pre-weight values for pairs of subarrays that belong to the same partition and not concerned with pairs of subarrays in different partitions as the Gaussian kernel weights decay very rapidly. The Tucker decomposition is effective at reducing $\sigma^2$ when $\T{U}^*$ has a checkerbox pattern as the checkerbox pattern is a low-rank tensor that can be effectively approximated with the Tucker decomposition.

Employing the Tucker decomposition introduces another tuning parameter, namely the rank of the decomposition. In our simulation studies described in \Sec{simulations}, we use two different methods for choosing the rank as a robustness check to ensure our CoCo estimator's performance does not crucially depend on the rank selection method.
Details on these two methods can be found in \App{weight_extra_discussion}. While we found the Tucker decomposition to work well in practice, we suspect that other methods of denoising the tensor may work just as well or could possibly be more effective. We leave it to future work to explore alternatives to the Tucker decomposition. 

\subsection{Weights and Folded-Concave Penalties}
\label{sec:folded_concave}

We conclude our discussion on weights by highlighting how they provide a connection between convex clustering and other penalized regression-based clustering methods that use folded-concave penalties \citep{PanShenLiu2013, XiangShen2013, ZhuShen2013, Marchetti2014,Wu2016}. Suppose we seek to minimize the objective
\begin{eqnarray}
\label{eq:folded_concave_objective}
\tilde{f}_\gamma(\V{u}) & = & \frac{1}{2} \lVert \V{X} - \V{U} \rVert_2^2 + \gamma\sum_{d=1}^D \sum_{(i,j) \in \E_d}\varphi_d\left(\lVert \M{A}_{d,ij}\V{U} \rVert_2\right),
\end{eqnarray}
where each $\varphi_d : [0,\infty) \mapsto [0, \infty )$ has the following properties: (i) $\varphi_d$ is concave and differentiable on $(0, \infty)$, (ii) $\varphi_d$ vanishes at the origin, and (iii) the directional derivative of $\varphi_d$ exists and is positive at the origin. Such $\varphi_d$ is collectively referred to as a folded-concave penalty; prominent examples of such function include  the smoothly clipped absolute deviation \citep{FanLi2001} or minimax concave penalty \citep{Zha2010}.

Since $\varphi_d$ is concave and differentiable, for all positive $z$ and $\tilde{z}$
\begin{eqnarray}
\label{eq:folded_concave_Taylor}
	\varphi_d(z) & \leq & \varphi_d(\tilde{z}) + \varphi_d'(\tilde{z})(z - \tilde{z}).
\end{eqnarray}
The inequality \Eqn{folded_concave_Taylor} indicates that the first order Taylor expansion of a differentiable concave function $\varphi_d$ provides a tight global upper bound at the expansion point $\tilde{z}$. Thus, we can construct a function that is a tight upper bound of the function $\tilde{f}_\gamma(\V{u})$
\begin{eqnarray}
g_\gamma(\V{U} \mid \Vtilde{U}) & = & 
\frac{1}{2} \lVert \V{X} - \V{U} \rVert_2^2 + \gamma\sum_{d=1}^D \sum_{(i,j) \in \E_d}w_{d, ij} \lVert \M{A}_{d,ij}\V{U} \rVert_2 + c,
\end{eqnarray}
where the constant $c$ does not depend on $\V{U}$ and $w_{d,ij}$ are weights that depend on $\Vtilde{U}$, namely
\begin{eqnarray}
\label{eq:majorization}
w_{d, ij} & = & \varphi_d'\left(\lVert \M{A}_{d,ij} \Vtilde{U} \rVert_2 \right).
\end{eqnarray}
Note that if we take $\Vtilde{u}$ to be the vectorization of the Tucker approximation of the data, $\vec(\Ttilde{X})$, and $\varphi_d(z)$ to be  the following variation on the error function
\begin{eqnarray*}
\varphi_d(z) & = & \frac{1}{\sqrt{n_{-d} } \sum_{(i,j) \in \E_d} w_{d,ij}}\int_{0}^z e^{-\tau_d \omega^2} d\omega,
\end{eqnarray*}
then the function given in \Eqn{folded_concave_objective} coincides with the CoCo objective using the prescribed Tucker derived Gaussian kernel weights. 

The function $g_\gamma(\V{U} \mid \Vtilde{U})$ is said to majorize the function $\tilde{f}_\gamma(\V{u})$ at the point $\Vtilde{u}$  \citep{LanHunYan2000} and minimizing it corresponds to performing one-step of the local linear-approximation algorithm \citep{Zou2008, Schifano2010} which is a special case of the majorization-minimization (MM) algorithm \citep{LanHunYan2000}. The corresponding MM algorithm would consist of repeating the following two steps: (i) using a previous CoCo estimate $\Ttilde{U}$ to compute weights $w_{d,ij}$ according to \Eqn{majorization}, and (ii) computing a new CoCo estimate using the new weights. In practice, we have found one-step to be adequate, however. Indeed, \cite{Zou2008} showed that the solution to the one-step algorithm was often sufficient in terms of its statistical estimation accuracy.

\section{Other Practical Issues}
\label{sec:practical_issues}

In this section, we address other considerations for using the method in practice, namely how to choose the tuning parameter $\gamma$ and how to recover the partitions along each mode from the CoCo estimator $\That{U}$.

\subsection{Choosing $\gamma$}
\label{sec:bic}

The first major practical consideration is how to choose $\gamma$ to produce a final co-clustering result.  Since co-clustering is an exploratory method, it may be suitable for a user to manually inspect a sequence of CoCo estimators $\That{U}_\gamma$ for a range of $\gamma$ and use domain knowledge tied to a specific application to select $\gamma$ to recover a co-clustering assignment of a desired complexity.  Since this approach is time consuming and requires expert knowledge, an automated, data-driven procedure for selecting $\gamma$ is desirable.  Cross-validation \citep{stone1974cross, geisser1975predictive} and stability selection \citep{meinshausen2010stability} are popular techniques for tuning parameter selection, but since both methods are based on resampling, they are unattractive in the tensor setting due to the computational burden.  
We turn to the extended Bayesian Information Criterion (eBIC) proposed by \cite{ChenChen2008, chen2012extended}, as it does not rely on resampling and thus is not as computationally costly as cross-validation or stability selection.
\begin{eqnarray*}
\text{eBIC}({\gamma}) & = & n \log{ \left( \frac{\text{RSS}_{\gamma}}{n} \right)} + 2\text{df}_{\gamma}  \log(n),
\end{eqnarray*}
where $\text{RSS}_{\gamma}$ is the residual sum of squares $\lVert \T{X} - \That{U}_\gamma\rVert^2_{\text{F}}$ and $\text{df}_{\gamma}$ is the degrees of freedom for a particular value of $\gamma$.  We use the number of co-clusters in the CoCo estimator $\That{U}_\gamma$ as an estimate of $\text{df}_{\gamma}$, which is consistent with the spirit of degrees of freedom since each co-cluster mean is an estimated parameter. This criterion balances between model fitting and model complexity, and a similar version has been commonly employed in tuning parameter selection of tensor data analysis \citep{zhou2013, sun2015provable}.

The eBIC is calculated on a grid of values $\mathcal{S} = \{\gamma_1, \gamma_2, \ldots \gamma_s\}$, and we select the optimal $\gamma$, denoted $\gamma^\star$, which corresponds to the smallest value of the eBIC over $\mathcal{S}$, namely
\begin{eqnarray*}
\gamma^\star & = &  \underset{\gamma \in \mathcal{S}}{\arg\min}\; \text{eBIC}(\gamma).
\end{eqnarray*}

\subsection{Recovering the Partitions along Each Mode}
\label{sec:partition}

The second major practical consideration is how to extract the partitions from the CoCo estimator $\That{U}$. Recall that
the $i$th and $j$th mode-$d$ subtensors belong to the same partition if $\V{v}_{d,ij} = \T{U}\times_d \Delta_{ij} = \V{0}$. Conversely, the $i$th and $j$th mode-$d$ subtensors {\em do not} belong to the same partition if $\V{v}_{d,ij} \neq \V{0}$. Thus, a mode-$d$ partition consists of the maximal set of mode-$d$ subarrays such that for any pair $i$ and $j$ in this collection $\V{v}_{d,ij} = \V{0}$. We can automatically identify these maximal sets by extending a simple procedure employed by \cite{Chi2015} for extracting clusters in the convex clustering problem. Identifying partitions along the $d$th mode is equivalent to finding connected components of a graph, where each node corresponds to a subarray along the $d$th mode, and there is an edge between nodes $i$ and $j$ if and only if $\V{v}_{d,ij} = \V{0}$. 

We would like to read off which centroids have fused as the amount of regularization increases, namely determine partition assignments as a function of $\gamma$. Such assignments can be performed in $\mathcal{O}(n_d)$ operations, using the differences variable $\M{V}_d$. We simply apply breadth-first search to identify the connected components of the following graph induced by the $\M{V}_d$. The graph identifies a node with every data point and places an edge between the $l$th pair of points if and only if $\V{v}_l = \V{0}$. Each connected component corresponds to a partition. Note that the graph constructed to determine partitions is {\em not} the same as the graph described in \Sec{coco_formulation} with illustrative examples in \Fig{edgeGraph}.

We emphasize that the recovered partition along each mode does {\em not} depend on the ordering of the input data $\T{X}$, since it is based off of the pairwise differences along each mode, namely $\M{v}_{d}$ for $d = 1, \ldots, D$. Finally, we note that due to finite precision limitations, the difference variables $\V{v}_{d,ij}$ will likely not be exactly $\V{0}$. In \App{compute_v}, we  detail a simple and principled procedure for ensuring sparsity in these difference variables.

\section{Simulation Studies}
\label{sec:simulations}

To investigate the performance of the CoCo estimator in identifying co-clusters in tensor data, we first explore some simulated examples.  We compare our CoCo estimator to a $k$-means based approach that is representative of various tensor generalizations of the spectral clustering method common in the tensor clustering literature \citep{Kutty2011, liu2013multiview, zhang2013nntri, WuBensonGleich2016}.  We refer to this method as CPD+$k$-means. The CPD+$k$-means method \citep{Papalexakis2013, Sun2019} first performs a rank-R CP decomposition on the $D$-way tensor $\T{X}$ to reduce the dimensionality of the problem, and then independently applies $k$-means clustering to the rows of each of the $D$ factor matrix from the resulting CP decomposition.
The $k$-means algorithm has also been used to cluster the factor matrices resulting from a Tucker decomposition \citep{AcarCamtepeYener2006, SunTaoFaloutsos2006, KoldaSun2008, SunPapadimitriouLinEtAl2009, Kutty2011, liu2013multiview, zhang2013nntri, cao2015robust, oh2017shot}.  We also considered this Tucker+$k$-means method in initial experiments, but its co-clustering performance was inferior to that of CPD+$k$-means so we only report co-clustering performance results for CPD+$k$-means in the comparison experiments that follow. Note, however, that we still use the Tucker decomposition to compute CoCo weights $w_{d,ij}$ as described \Sec{weights}. Both CoCo and CPD+kmeans account for the multiway structure of the data. To assess the importance of accounting for this structure, we also include comparisons with the CoTeC method \citep{JegelkaSraBanerjee2009}, which applied $k$-means clustering along each mode and does not account for the multiway structure of the data.

All methods being compared have tuning parameters that need to be set. For the rank of the CP decomposition needed in CPD+$k$-means, we consider $R \in \{2, 3, 4, 5\}$ and use the tuning procedure in \cite{sun2015provable} to automatically select the rank.  A CP decomposition is then performed using the chosen rank, and those factor matrices are the input into the $k$-means algorithm.  A well known drawback of $k$-means is that the number of clusters $k$ needs to be specified \textit{a priori}.  Several methods for selecting $k$ have been proposed in the literature, and we use the ``gap statistic'' developed by \cite{TibshiraniWaltherHastie2001} to select an optimal $k^*$ from the specified possible values. Since CoCo estimates an entire solution path of mode-clustering results, ranging from $n_d$ clusters to a single cluster along mode $d$, we consider a rather large set of possible $k$ values to make the methods more comparable. \App{cpdkmeans} gives a more detailed description of the CPD+$k$-means procedure and the selection of its tuning parameters. CoTeC, which applies $k$-means clustering along each mode independently, also requires specifying the number of cluster along each mode. As in CPD+$k$-means, we also select this parameter along each mode using the ``gap statistic."

As described in \Sec{weights}, we employ a Tucker approximation to the data tensor in constructing weights $w_{d,ij}$. In computing the Tucker decomposition we used one of two methods for selecting the rank. In the plots within this section, TD1 denotes the results where the Tucker rank was chosen using the SCORE algorithm \citep{YokotaLee2017}, while TD2 denotes results where the rank was chosen using a heuristic. Detailed discussion on these two methods are in \App{weight_extra_discussion}.

The results presented in this section report the average CoCo estimator performance quantified by the ARI across 200 simulated replicates. All simulations were performed in \textsc{Matlab} using the Tensor Toolbox \citep{TTBSoftware}. All the following plots, except the heatmaps in \Fig{adHeatSliceM}, were made using the open source R package ggplot2 \citep{Wickham2009}.

\subsection{Cubical Tensors, Checkerbox Pattern}
\label{sec:simulations_checkerBox}

For the first and main simulation setting, we study clustering data in a cubical tensor generated by a basic checkerbox mean model according to \As{model}.  Each entry in the observed data tensor is generated according to the underlying model \Eqn{mean_model} with independent errors $\epsilon_{i_1i_2i_3} \sim N(0, \sigma^2_{r_1r_2r_3})$.
Unless specified otherwise, there are two true clusters along each mode for a total of eight underlying co-clusters.

\subsubsection{Balanced Cluster Sizes and Homoskedastic Noise}

\begin{figure}[t]
    \centering
    \label{fig:checkerCube_60_sigmas_ARI_triclust_chapter}
    \includegraphics[scale=0.45]{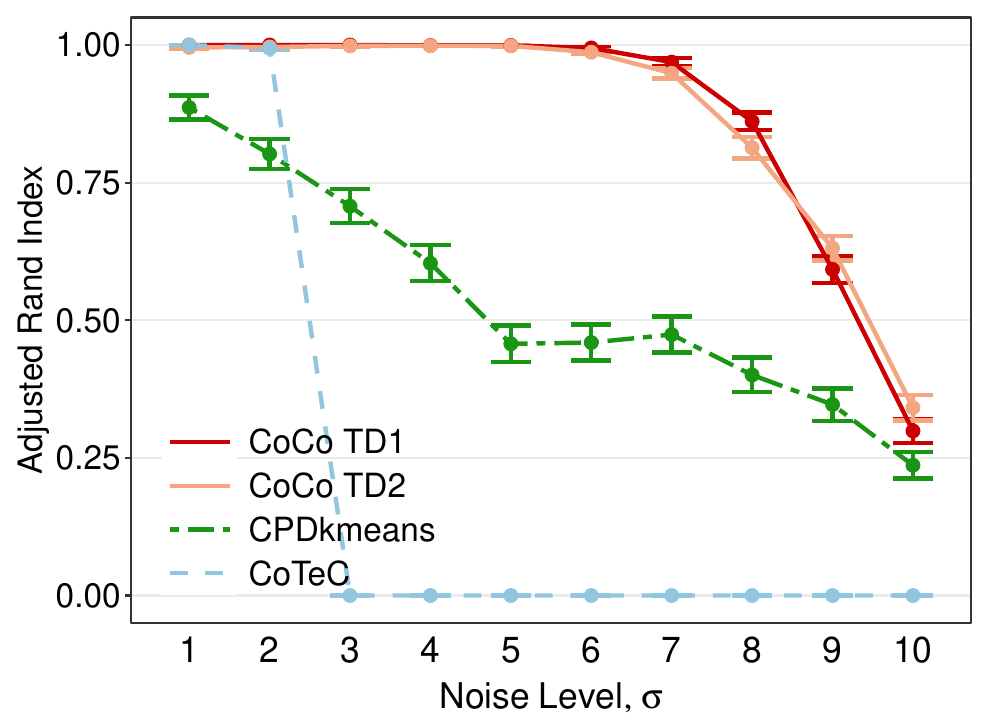}
    \caption{Checkerbox Simulation Results: Impact of Noise Level.  \small Two balanced clusters per mode across different levels of homoskedastic noise for $n_1 = n_2 = n_3 = 60$. \small For each method, the confidence interval is calculated as the mean value plus/minus one standard error.}
    \label{fig:checkerCube_60_sigmas_ARI_chapter}
\end{figure}

To get an initial feel for how the different co-clustering methods perform at recovering the true underlying checkerbox structure, we first consider a situation where the clusters corresponding to the two classes along each mode are all equally-sized, or balanced, and share the same error variance, namely $\sigma_{r_1r_2r_3} = \sigma$ for all $r_1, r_2,$ and $r_3$.   The average co-clustering performance for this setting in a tensor with dimensions $n_1 = n_2 = n_3 = 60$ are given in \Fig{checkerCube_60_sigmas_ARI_chapter} for different noise levels. \Fig{checkerCube_60_sigmas_ARI_chapter} shows that all three methods perform well when the noise level is low ($\sigma = 1$).  As the noise level increases, however, CPD+$k$-means experiences an immediate and noticeable drop off in performance.  CoTeC's performance decays even more rapidly highlighting the importance of accounting for multiway structure. The CoCo estimator, on the other hand, is able to maintain near-perfect performance until the noise level becomes rather high ($\sigma = 8$).

\begin{figure}[t]
	\centering
    \includegraphics[scale=0.45]{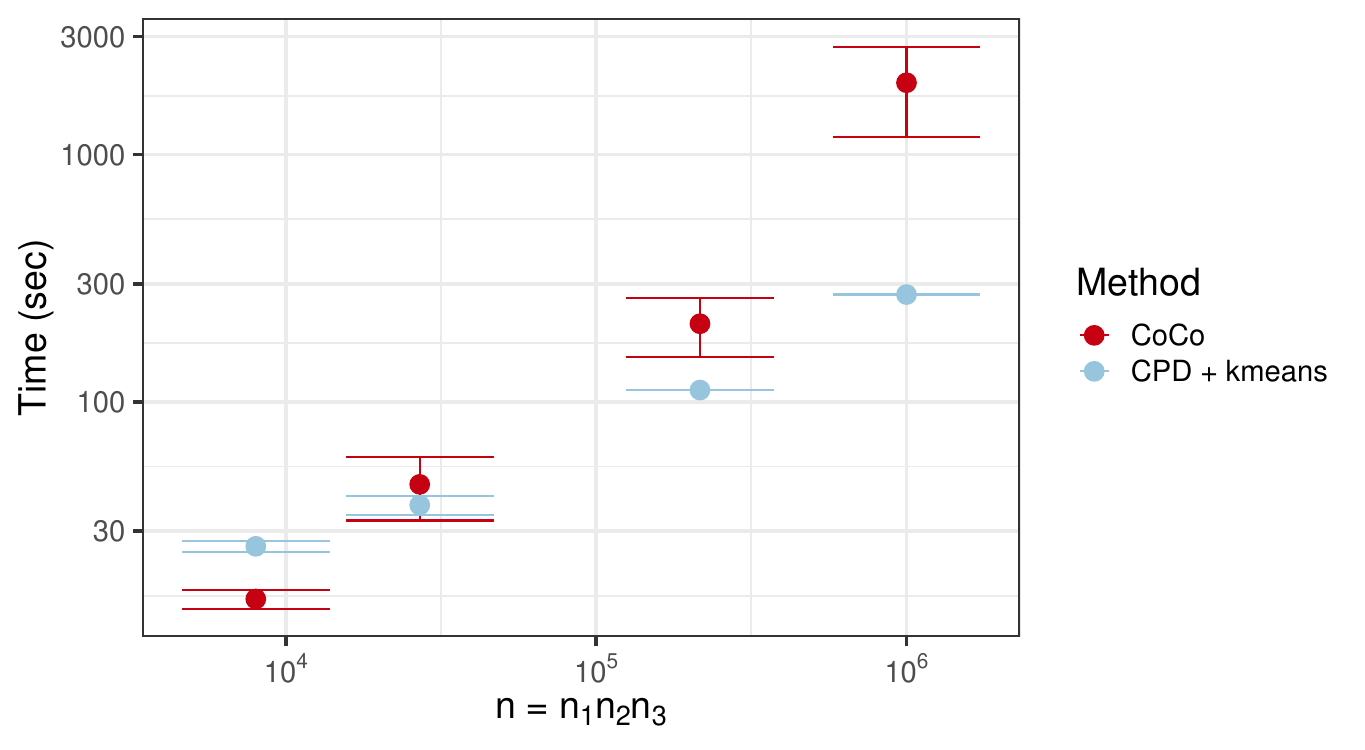}
    \caption{Timing Results: Balanced Cluster Size and Homoskedastic Noise.  \small Two balanced clusters per mode with a fixed level of homoskedastic noise for $n_1 = n_2 = n_3 = 20, 30, 60,$ and $100$. \small Vertical and horizontal axes are on a log scale.}
    \label{fig:runtime}
\end{figure}

\Fig{runtime} shows how the run times of CoCo and CPD+$k$-means vary as the size of a cubic tensor, $n = n_1n_2n_3$ with $n_1 = n_2 = n_3$ takes on the values $20^3, 30^3, 60^3,$ and $100^3$. These run times include all computations needed to fit and select a final model. For CoCo, a sequence of models were fit over a grid of $\gamma$ parameters, and a final $\gamma$ parameter was chosen using the eBIC. For CPD+$k$-means, a sequence of models were fit over a grid of possible $(k_1, k_2, k_3)$ parameters corresponding to the 3 factor matrices, and a final triple of $(k_1, k_2, k_3)$ parameters were chosen using the ``gap statistic." Timing comparisons were performed on a 3.2 GHz quad-core Intel Core i5 processor and 8 GB of RAM. The run time for CoCo scales linearly in the size of the data tensor as expected, namely proportionately with $n_1^3$. Nonetheless, as also might be expected, the clustering performance enjoyed by CoCo does not come for free, and the simpler but less reliable CPD+$k$-means algorithm enjoys a better scaling as the tensor size grows. Timing results were similar for the following experiments and are omitted for space considerations.

\subsubsection{Imbalanced Cluster Sizes}
\label{sec:simulations_imbalanced_cluster_sizes}

\begin{figure}[t]
    \centering
    \subfloat[Low Noise]{\label{fig:checkerCube_imbalance_lowNoise_60_ARI_triclust}
    \includegraphics[scale=0.45]{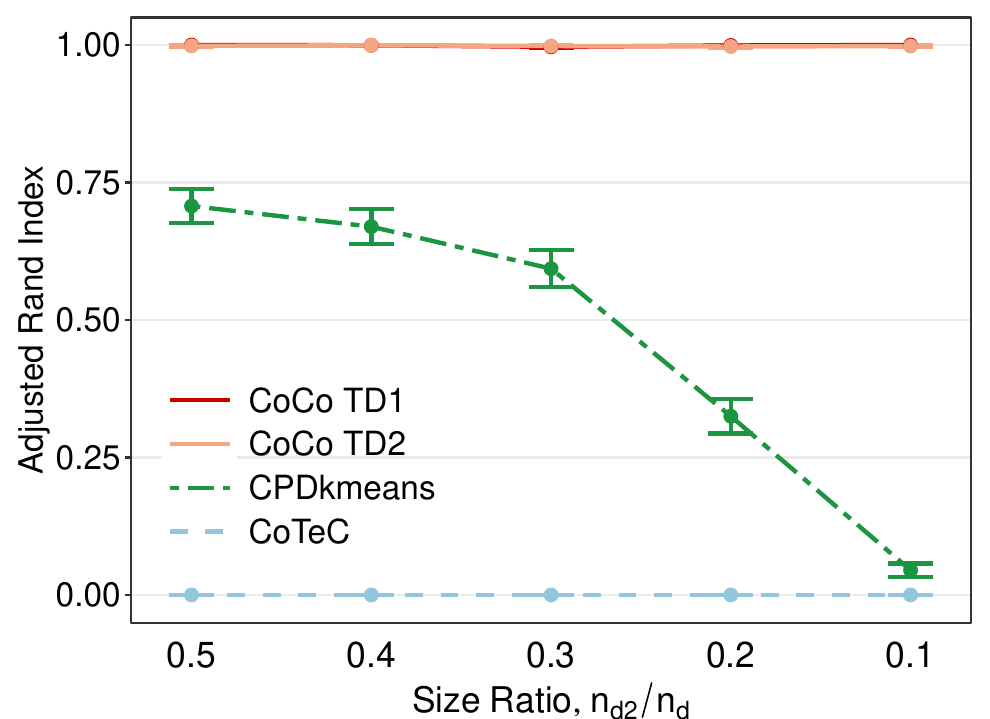}}
    \subfloat[High Noise]{\label{fig:checkerCube_imbalance_highNoise_60_ARI_triclust}
    \includegraphics[scale=0.45]{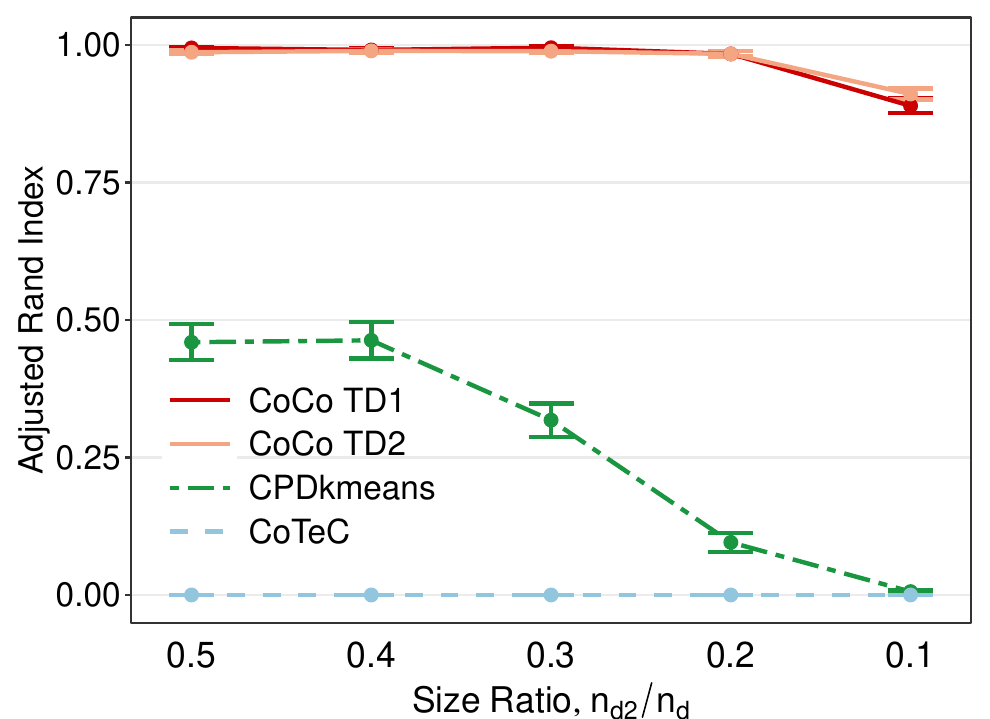}}
    \caption{Checkerbox Simulation Results:  Impact of Cluster Size Imbalance.  \small Two imbalanced clusters per mode with either low or high homoskedastic noise for $n_1 = n_2 = n_3 = 60$. \small 
    Low noise corresponds to $\sigma = 3$ while high noise refers to $\sigma = 6$.} \label{fig:checkerCube_imbalance}
\end{figure}

When comparing clustering methods, one factor of interest is the extent to which the relative sizes of the clusters impact clustering performance.  To investigate this, we again use a cubical tensor of size $n_1 = n_2 = n_3 = 60$ but introduce different levels of cluster size imbalance along each mode, which we quantify via the ratio of the number of samples in cluster 2 of mode $d$ and the total number of samples along mode $d$, for $d = 1, 2, 3$.  \Fig{checkerCube_imbalance_lowNoise_60_ARI_triclust} 
shows that when the noise level is low, CPD+$k$-means is unaffected by the imbalance until the size of cluster 2 is less than 30\% of the mode's length.  At this point, the performance of CPD+$k$-means drops off significantly and it performs as well as a random clustering assignment when the sizes are highly skewed ($n_{d2} / n_d = 0.1$).  The CoCo estimator is more or less invariant to the imbalance, and its performance is almost perfect across all levels of cluster size imbalance.  \Fig{checkerCube_imbalance_highNoise_60_ARI_triclust} shows that the CoCo estimator exhibits a slight deterioration in performance only when the cluster size ratio is $0.1$ in the high noise case. In both low and high noise scenarios, CoTeC performs poorly. 

\subsubsection{Heteroskedastic Noise}

\begin{figure}[t]
    \centering
    \subfloat[Low Noise]{\label{fig:checkerCube_hetero_lowNoise_60_ARI_triclust}
    \includegraphics[scale=0.45]{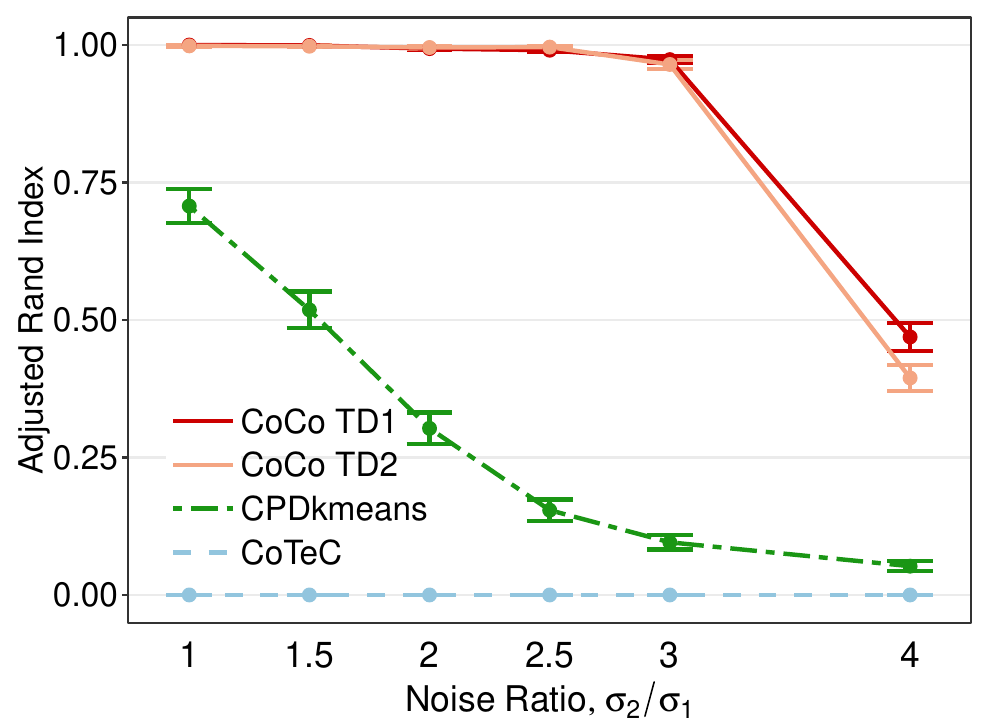}}
    \subfloat[High Noise]{\label{fig:checkerCube_hetero_highNoise_60_ARI_triclust}
    \includegraphics[scale=0.45]{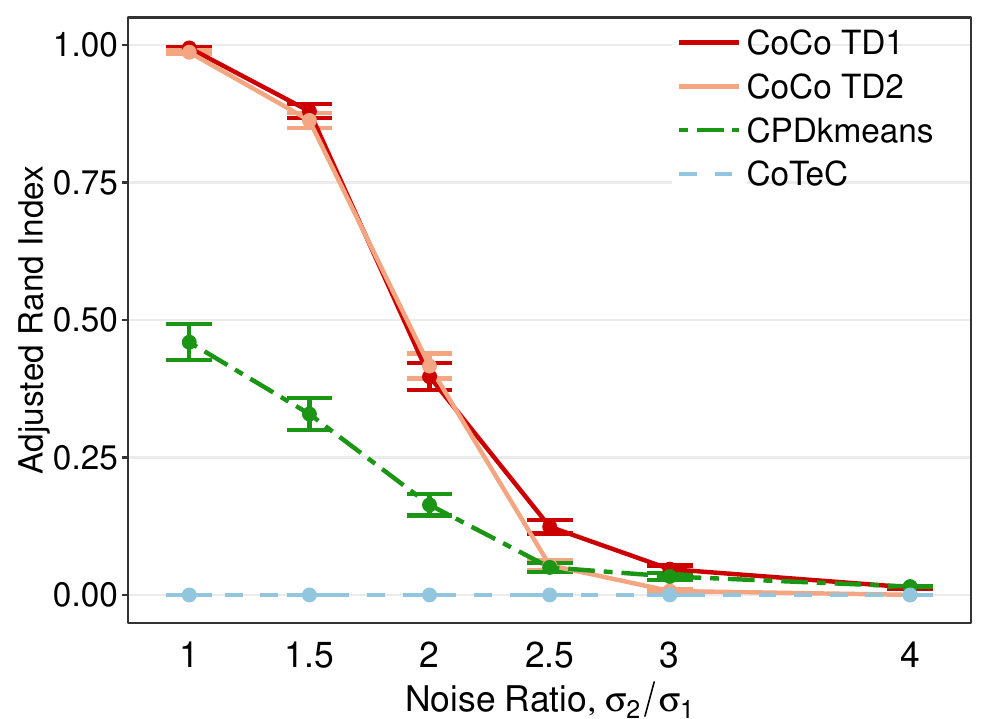}}
    \caption{Checkerbox Simulation Results: Impact of Heteroskedasticity.  \small Two balanced clusters per mode with either low or high heteroskedastic noise for $n_1 = n_2 = n_3 = 60$. \small 
    Low noise corresponds to $\sigma_1 = 3$ while high noise refers to $\sigma_1 = 6$.} \label{fig:checkerCube_hetero}
\end{figure}

Another factor of interest is how the clustering methods perform when there is heteroskedasticity in the variability of the two classes.  \Fig{checkerCube_hetero} displays the co-clustering performance for different degrees of heteroskedasticity, as measured by the standard deviation for class 2 relative to class 1's standard deviation, $\sigma_2 / \sigma_1$.  In the low noise setting, the CoCo estimator is immune to the heteroskedasticity until the noise levels differ by a factor of 4.  CPD+$k$-means in contrast is very sensitive to a deviation from homoskedasticty, experiencing a decline even when the noise ratio increases from 1 to only 1.5.  The CoCo estimator fares worse in the high noise setting and also has a drop in performance with a small deviation from homoskedasticty.  Once class 2's standard deviation is more than double the standard deviation for class 1, all three methods are essentially the same as random clustering.  This result is not terribly surprising since, in the high noise setting, this would result in one class having a very high standard deviation of $\sigma_2 = 12$.  In both low and high noise scenarios, CoTeC performs poorly. 

\subsubsection{Different Clustering Structures}
\label{sec:simulations_different_clustering_structures}

So far, we have considered only a simple situation where there are exactly two true clusters along each mode, for a total of eight triclusters.  Another factor of practical importance is how the clustering methods perform when there are more than two clusters per mode, and also when the number of clusters along each mode differs.  We investigate both of these settings in this section.  As before, the tensor is a perfect cube with $n_1 = n_2 = n_3 = 60$ observations along each mode and an underlying checkerbox pattern.   To gauge the performance, we again focus the attention on how the methods perform in the presence of both low and high noise.  

\begin{figure}[h!]
    \centering
    \includegraphics[scale=0.45]{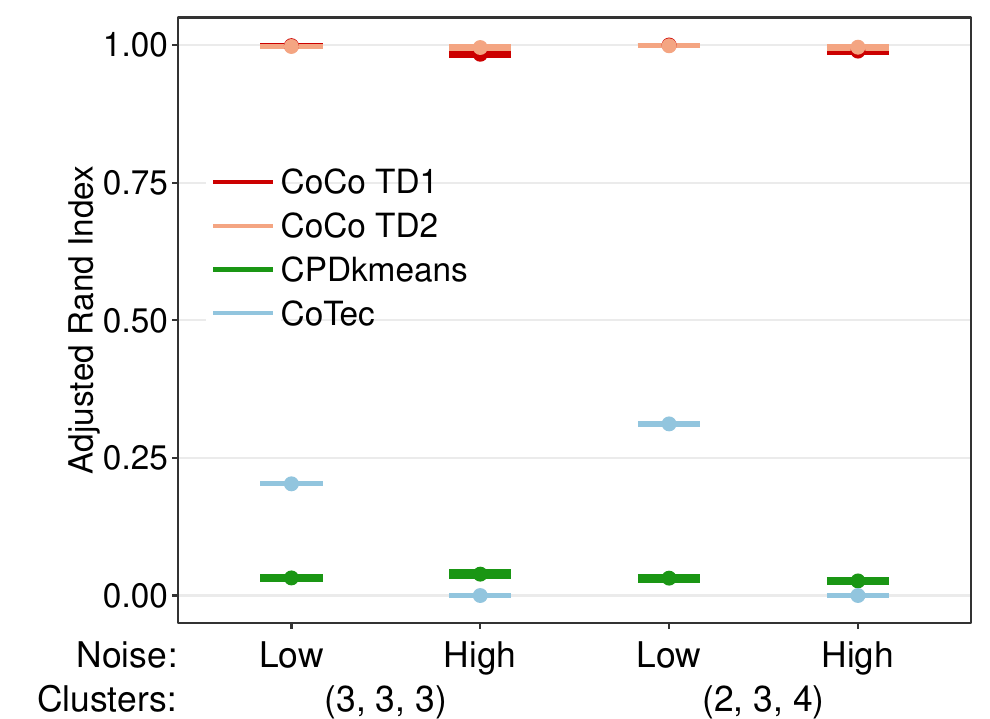}
    \caption{Checkerbox Simulation Results: Impact of Clustering Structure. \small Different balanced clusters per mode with either low or high homoskedastic noise for $n_1 = n_2 = n_3 = 60$. \small     Low noise corresponds to $\sigma = 3$ while high noise refers to $\sigma = 6$.} 
    \label{fig:checkerCube_diffClusters_60_ARI_triclust}
\end{figure}

The first situation studied is one in which there are three true clusters along each mode, resulting in a total of 27 triclusters. The left hand side of the graphs in \Fig{checkerCube_diffClusters_60_ARI_triclust} show the results from this simulation setting.  The graphs show that CoCo estimator consistently outperforms CPD+$k$-means and CoTeC in this setting across both noise levels.  The CoCo estimator is able to recover the true co-clusters almost perfectly, while CPD+$k$-means struggles to handle the increased number of clusters per mode.

We also investigated the clustering performance when the number of clusters per mode varies.  In this setting, there are two, three, and four clusters along modes one, two, and three, respectively.  From the right hand side of the graphs in \Fig{checkerCube_diffClusters_60_ARI_triclust}, we can see that the results are similar to the situation with three clusters per mode.  CPD+$k$-means again performs very poorly across both noise levels, while convex co-clustering is again able to essentially recover the true co-clustering structure.  Compared to the setting with three clusters per mode, CPD+$k$-means performs slightly worse in the face of a more complex clustering structure, while convex co-clustering is able to handle it in stride.  These results bode well for convex co-clustering as the basic clustering structure of only two clusters per mode is unlikely to be observed in practice.  

\subsection{Rectangular Tensors}
\label{sec:simulations_rectangular_tensors}

Up to this point, to get an initial feel for CoCo's performance, we restricted our attention to cubical tensors with the same number of observations per mode so as to avoid changing too many factors at once.  It is unlikely that the data tensor at hand will be a perfect cube, however, so it is important to understand the clustering performance when the methods are applied to rectangular tensors.  

\begin{figure}[t]
    \centering
    \subfloat[Adjusted Rand Index, Mode 1]{\label{fig:checkerCube_nonCube_105050_205050_ARI_mode1}
    \includegraphics[scale=0.45]{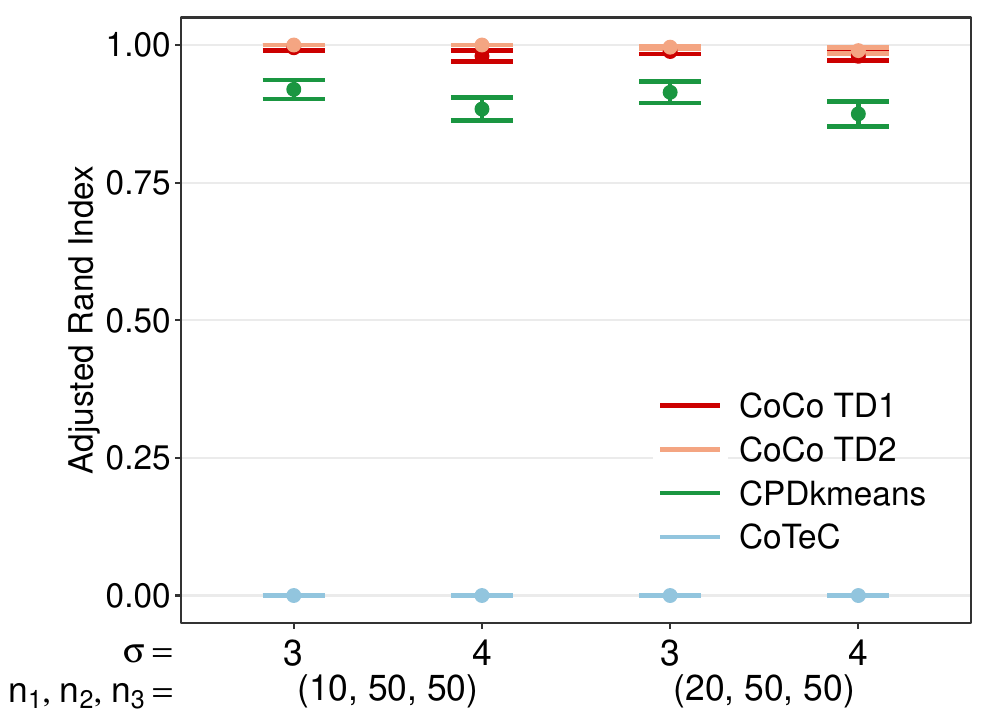}}
    \subfloat[Adjusted Rand Index, Mode 2]{\label{fig:checkerCube_nonCube_105050_205050_ARI_mode2}
    \includegraphics[scale=0.45]{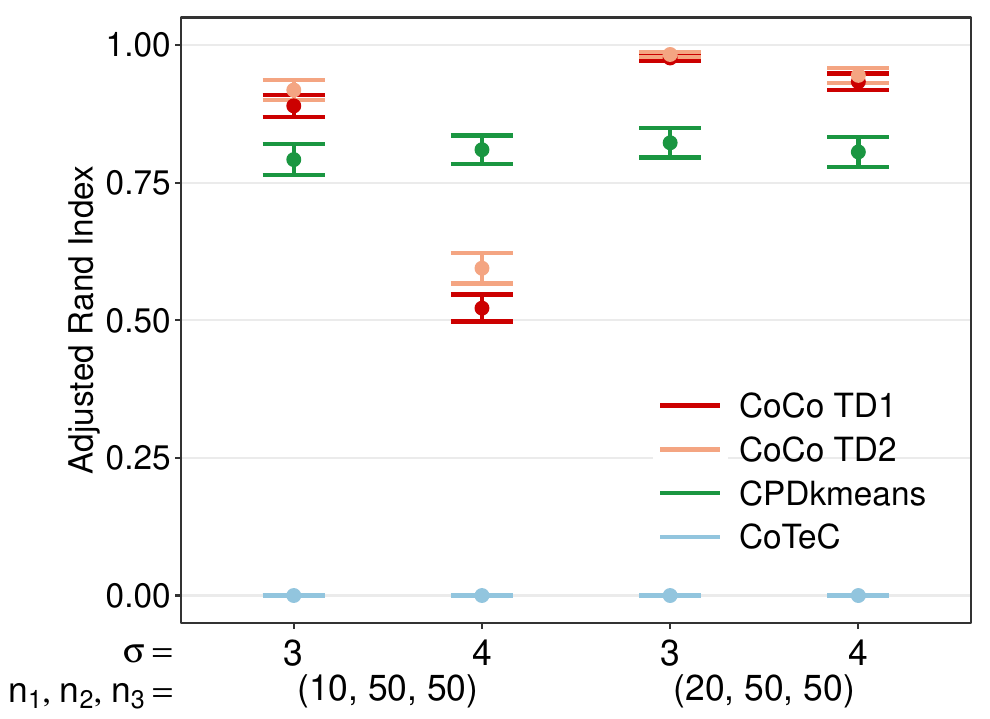}} \\
    \subfloat[Adjusted Rand Index, Mode 3]{\label{fig:checkerCube_nonCube_105050_205050_ARI_mode3}
    \includegraphics[scale=0.45]{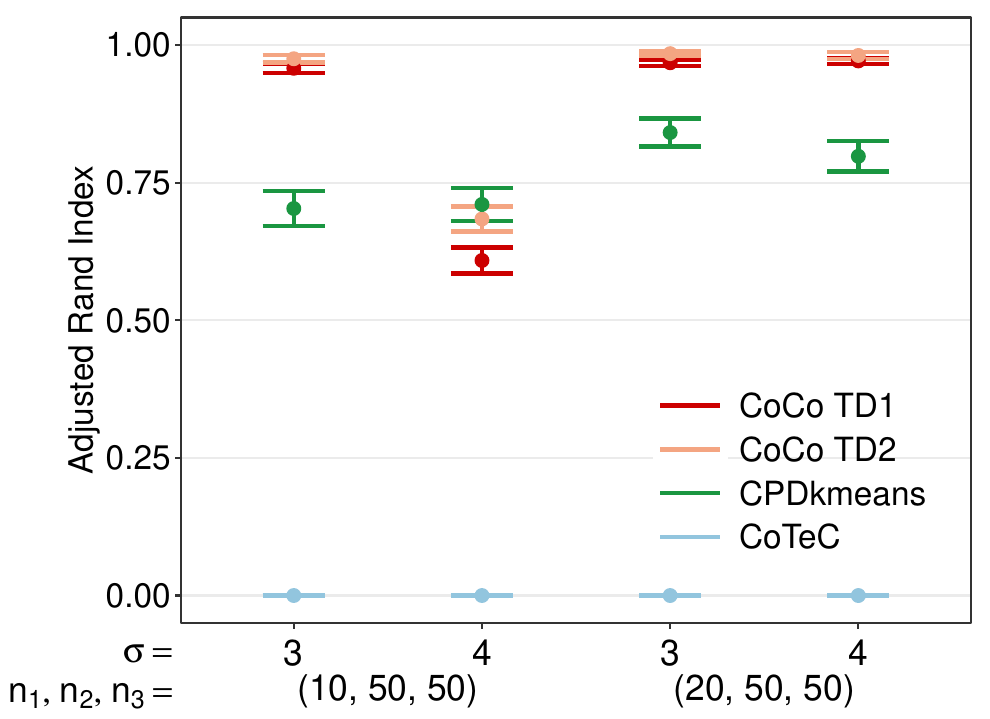}}
    \subfloat[Adjusted Rand Index, Triclusters]{\label{fig:checkerCube_nonCube_105050_205050_ARI_triclust}
    \includegraphics[scale=0.45]{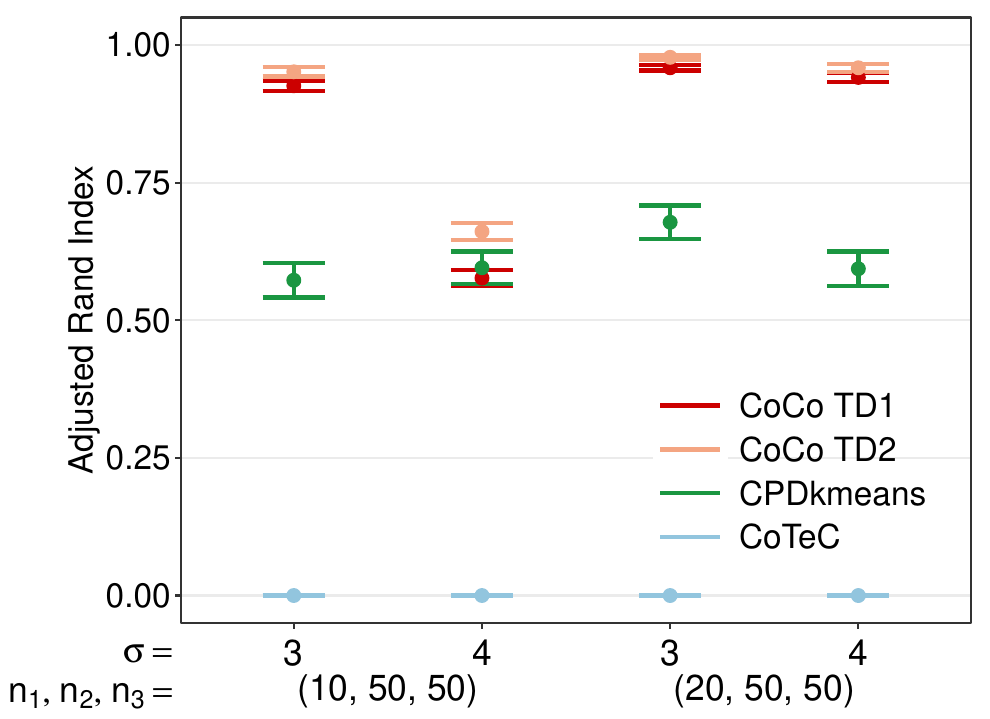}}
    \caption{Checkerbox Simulation Results:  Impact of Tensor Shape.  \small Two balanced clusters per mode with two levels of homoskedastic noise for a tensor with one short mode and two longer modes. \small Average adjusted rand index plus/minus one standard error for different noise levels and mode lengths. } \label{fig:nonCube_105050_205050}
\end{figure}

Now we turn to cluster a rectangular tensor with one short mode and two longer modes.  Two additional simulations involving rectangular tensors can be found in \App{simulations_rectangular_extra}.
\Fig{nonCube_105050_205050} shows that CoCo performs very well and better than CPD+$k$-means and CoTeC at the lower noise level ($\sigma = 3$) but has a sharp decrease in ARI at the higher noise level ($\sigma = 4$).  The decline is more pronounced for the longer modes (\Fig{checkerCube_nonCube_105050_205050_ARI_mode2} and \Fig{checkerCube_nonCube_105050_205050_ARI_mode1}) as the short mode (\Fig{checkerCube_nonCube_105050_205050_ARI_mode1}) is still able to maintain perfect performance despite the increase in noise. This is not surprising, since the shorter mode has effectively more samples.
Moreover, we see the ``blessing of dimensionality" at work when the number of samples along the short mode are doubled ($n_1 = 20$, $n_2 = n_3 = 50$), the performance along the two longer modes improves drastically in the high noise setting.

We finally note that, along the shorter mode, the use of the heuristic in determining the rank of the Tucker decomposition for calculating the weights performs better than the SCORE algorithm method along modes 1 and 2, though ultimately the co-clustering performance is comparable.  This may indicate that the SCORE algorithm struggles to correctly identify the optimal Tucker rank for short modes in the presence of relatively higher noise, while the heuristic is more immune to the noise level as it is based simply on the dimensions of the tensor. 

\subsection{CANDECOMP/PARAFAC Model}
\label{sec:simulations_CP_model}

\begin{figure}[t]
    \centering
    \subfloat[Bullseye]{\label{fig:factorMats_6_cpd}
    \includegraphics[scale=0.375]{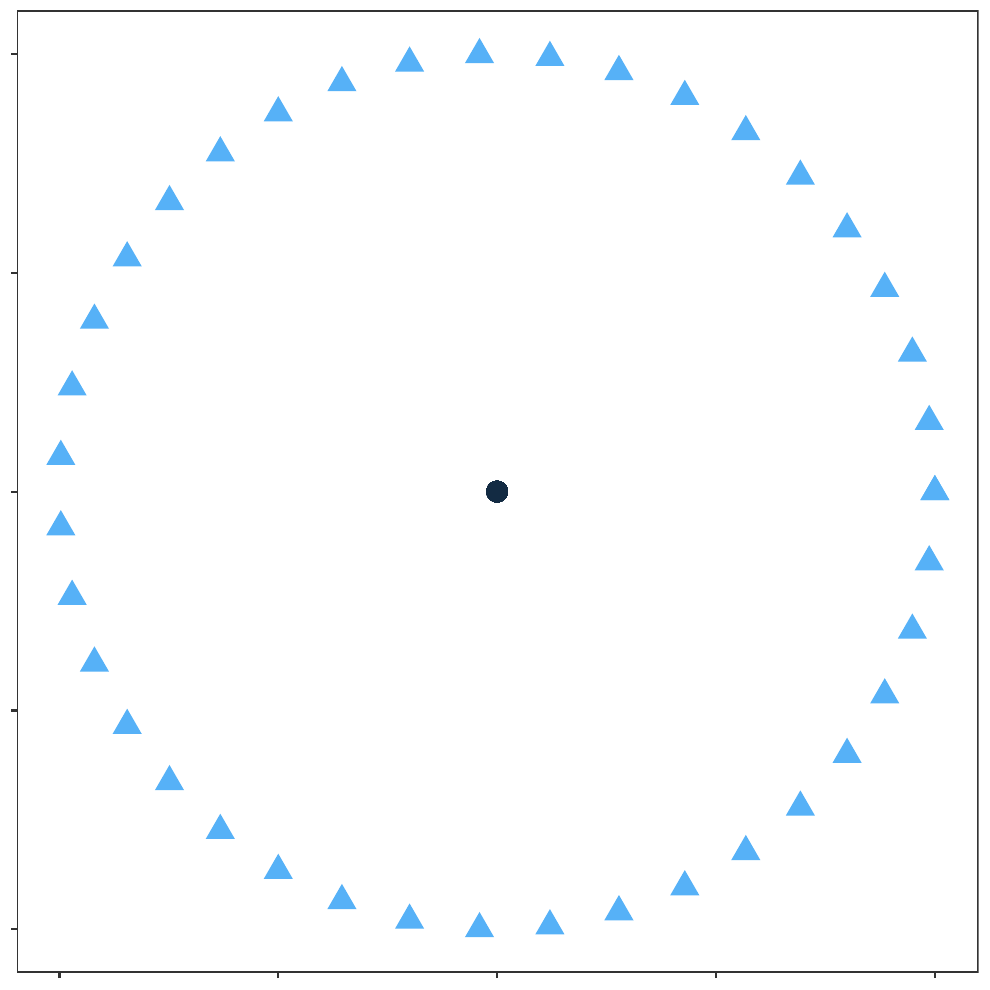}}
    \subfloat[Half Moons]{\label{fig:factorMats_9_cpd}
    \includegraphics[scale=0.375]{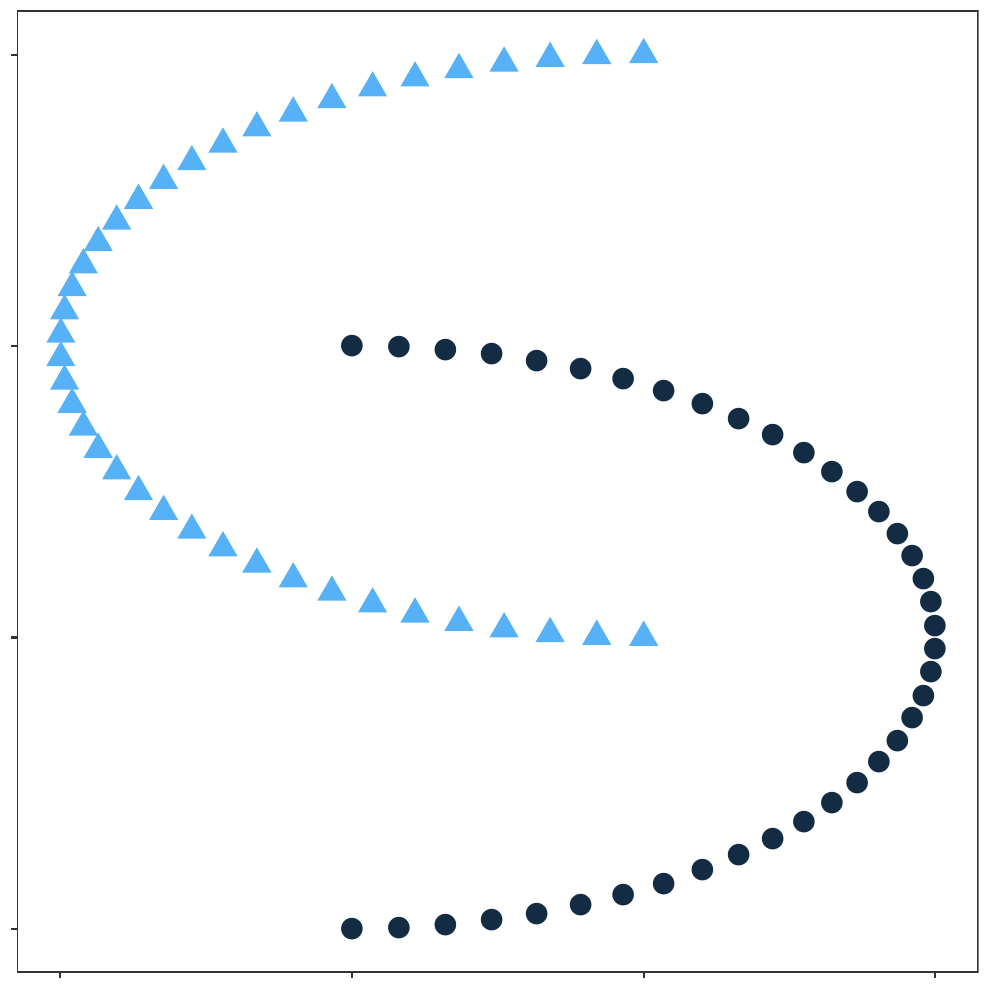}}
    \caption{Factor Matrices for the CP Models.}  \label{fig:factorMats_cpd}
\end{figure}

In \Sec{simulations_checkerBox}, we saw that the CoCo estimator performs well and typically better than CPD+$k$-means when clustering  tensors whose co-clusters have an underlying checkerbox pattern. To evaluate the performance of our CoCo estimator under model misspecification, we consider the generative model as the following CP decomposition model. We first construct the factor matrix $\M{A} \in \Real^{80 \times 2}$ and construct the following rank-2 CP means tensor 
\begin{eqnarray*}
	\T{U}^* & = & \sum_{i = 1}^2  \V{a}_i  \circ \V{a}_i  \circ \V{a}_i,
\end{eqnarray*}
where $\circ$ denotes the outer product. We then added varying levels of Gaussian noise to the $\T{U}^*$ to generate the observed data tensor.  
We consider two different types of factor matrices.  As shown in \Fig{factorMats_cpd}, one shape consists of two half-moon clusters \citep{HockingJoulinBachEtAl2011, Chi2015, Tan2015} while the other shape contains a bullseye, similar to the two-circles shape studied by \cite{ng2002spectral} and \cite{Tan2015}. In either case, the triangles in \Fig{factorMats_cpd} correspond to the first 40 rows of $\M{A}$, whereas the circles correspond to the second 40 rows of $\M{A}$. Note that this data generating mechanism should favor the CPD+$k$-means method.

\begin{figure}[t]
    \centering
    \includegraphics[scale=0.45]{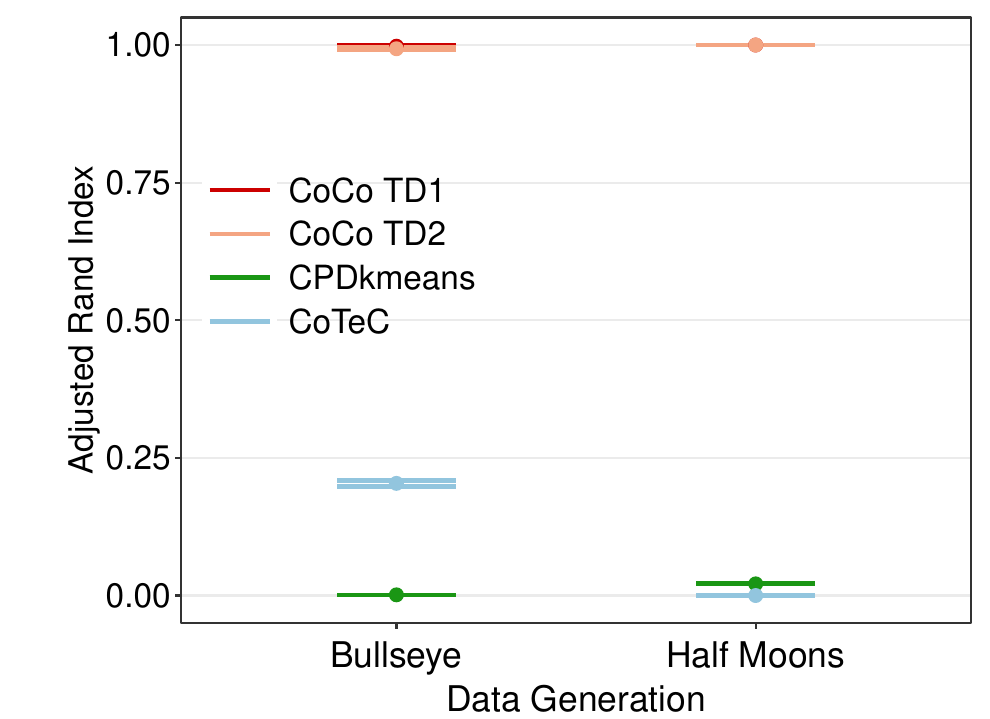}
     \caption{CP Model Simulation Results.  \small Two balanced clusters per mode with low homoskedastic noise for $n_1 = n_2 = n_3 = 40$.      ``Bullseye'' and ``Half Moons'' refer to the shape embedded in the factor matrices used to generate the true tensor.}  \label{fig:cpd_sim_results}
\end{figure}

\Fig{cpd_sim_results} shows the simulation results for using the CP model with these two non-convex shapes generating the data.  The discrepancy in performance between the CoCo estimator and the other two methods is quite large.  The CoCo estimator almost perfectly identifies the true co-clusters. In contrast, both CPD+$k$-means and CoTeC perform very poorly, even when the noise variance is small.  The poor performance of CPD+$k$-means and CoTeC are not completely surprising as other have noted the difficulty that $k$-means methods have in recovering non-convex clusters \citep{ng2002spectral, HockingJoulinBachEtAl2011, Tan2015}.  These results give us some assurances that the CoCo estimator is able to still perform well even under some model misspecification since the true co-clusters do not have a checkerbox pattern.  

\subsection{Comparison with Convex Biclustering}\phantomsection
\label{sec:simulations_biclustering}

It is natural to ask how much additional gain there is in using CoCo over convex biclustering \citep{ChiAllenBaraniuk2017} on the matricizations of a data tensor. To answer this question, we compare CoCo to the following strategy for applying convex biclustering to estimate co-clusters. We explain the strategy for a 3-way tensor; the generalization to $D$-way tensors is straightforward. We first matricize the tensor $\T{X}$ along mode-1 to obtain the matrix $\Mz{X}{1}$, apply convex biclustering on $\Mz{X}{1}$, and retain the mode-1 clustering results. Note that the mode-2 and mode-3 fibers have been mixed together through the matricization process. We then repeat the two-step procedure for mode-2 and mode-3. The final co-cluster estimates are obtained by taking the cross-products of the mode-1, mode-2, and mode-3 cluster assignments.

\begin{figure}[t]
    \centering
    \subfloat[Balanced]{\label{fig:checkerCube_60_sigmas_ARI_biclust_triclust}
    \includegraphics[scale=0.45]{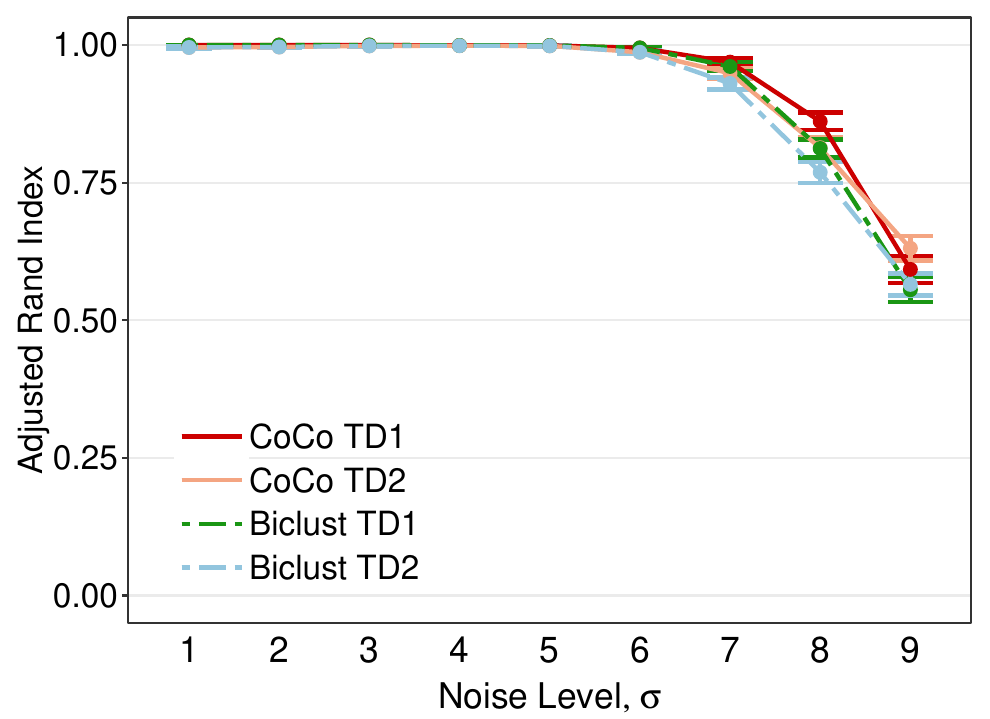}}
    \subfloat[Imbalanced]{\label{fig:checkerCube_304080_sigmas_ARI_triclust}
        \includegraphics[scale = 0.45]{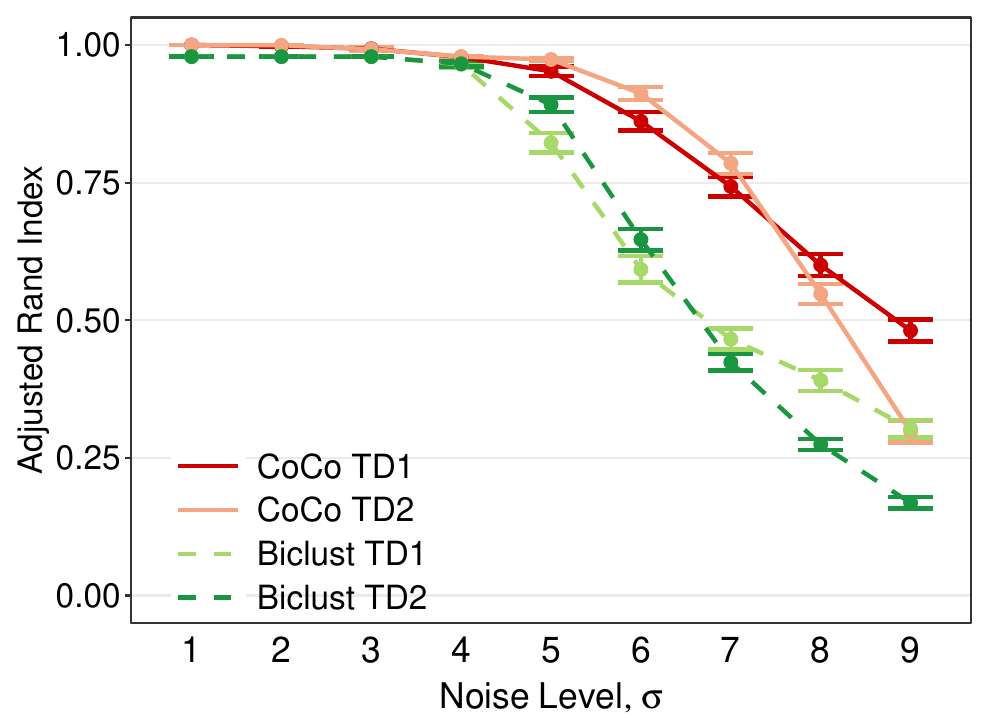}}
    \caption{A Comparison between CoCo and Convex Biclustering \small Average Adjusted Rand Index plus/minus one standard error for different noise levels. }
    \label{fig:biclustering} 
\end{figure}

We consider two illustrative scenarios to understand the value of preserving the full multiway structure with CoCo: a balanced case and imbalanced case. In the balanced case, we have a 3-way data tensor $\T{X} \in \Real^{60 \times 60 \times 60}$ with two clusters along each mode, where clusters are of equal size and homoskedastic iid Gaussian noise has been added to all elements of the tensor. This scenario is similar to the one shown in \Fig{checkerCube_60_sigmas_ARI_chapter}. In the imbalanced case, we have a 3-way data tensor $\T{X} \in \Real^{30 \times 40 \times 80}$. There are two clusters along mode-1 of sizes 10 and 20, three clusters along mode-2 of sizes 8, 12, and 20, and four clusters along mode-3 of sizes 5, 10, 20, and 45. Homoskedastic iid Gaussian noise has been added to all elements of the tensor. Finally, we note that the empirical performance of convex biclustering, like that of CoCo's, depends on choosing good weights for the rows and columns of the input data matrix \citep{ChiAllenBaraniuk2017}. To create a fair comparison, we construct convex biclustering weights based off of the same TD1 and TD2 denoising procedure used for CoCo, putting the preprocessing for both methods on equal footing.

\Fig{checkerCube_60_sigmas_ARI_biclust_triclust} and \Fig{checkerCube_304080_sigmas_ARI_triclust} show the co-clustering performance of CoCo and the convex biclustering method in the balanced and imbalanced cases respectively. We see that in the balanced case, CoCo's performance is marginally better than that of the convex biclustering method. On the other hand, we see that in the imbalanced case, CoCo's performance degrades more gracefully than that of the convex biclustering method as the noise level increases. The example illustrates that CoCo has better co-cluster recovery when there is more imbalance in the data tensor - the aspect ratios of the tensor dimensions are more skewed and the number of clusters and the cluster sizes are more heterogenous.

The key formulation difference between CoCo and the convex biclustering method that provides some insight into these two results is that CoCo imposes a finer level of smoothness that respects the multi-way structure in the data tensor. Imposing such finer level of smoothness imparts greater robustness in the presence of increasing noise to recovering the smaller co-clusters in the imbalanced scenario. An added incentive for using CoCo and preserving the multiway structure in the data is that the gains in co-cluster recovery over the convex biclustering method do not come at a greater computational cost.
Note that the computational complexity of convex biclustering is $\mathcal{O}(n)$, using sparse weights for the row and column similarity graphs. For a $D$-way tensor, the computational complexity then becomes $\mathcal{O}(Dn)$, which is the same as the computational complexity of CoCo applied directly on the $D$-way tensor. 

To summarize, in comparison to the convex biclustering method, CoCo (i) does not come at additional computational costs, (ii) can recover underlying co-clustering structure in imbalanced scenarios which are more likely to be encountered in practice, and (iii) has the ability to consistently recover an underlying co-clustering structure according to \Thm{final_error}, with even a single tensor sample, which is a typical case in real applications. 
Since this phenomenon does not exist in vector or matrix variate cluster analysis, the convex biclustering method lacks this theoretical guarantee.

\section{Real Data Application}
\label{sec:real_data}

Having studied the performance of the CoCo estimator in a variety of simulated settings, we now turn to using the CoCo estimator on a real data set.  The proprietary data set comes from a major online company and contains the click-through rates for advertisements displayed on the company's webpages from May 19, 2016 through June 15, 2016.  The click-through rate is the number of times a user clicks on a specific advertisement divided by the number of times the advertisement was displayed.  The data set contains information on 1000 users, 189 advertisements, 19 publishers, and 2 different devices, aggregated across time.  Thus, the data forms a fourth-order tensor where each entry in the tensor corresponds to the click-through rate for the given combination of user, advertisement, publisher, and device.  Here a publisher refers to a different webpage within the online company's website, such as the main home page versus a page devoted to either breaking news or sports scores.  The two device types correspond to how the user accessed the page, using either a personal computer or a mobile device such as a cell phone or tablet computer. The goal in this real application is to simultaneously cluster users, advertisements, and publishers to improve user behavior targeting and advertising planning.

In the click-through rate tensor data, over $99\%$ of the values are missing since one user likely has seen only a handful of the possible advertisements. If a specific advertisement is never seen by a user, it is considered as a missing value. Since the proposed CoCo estimator can only handle complete data, we first preprocess the data by imputing the missing values before any clustering can be done. To impute the missing entries, we use the CP-based tensor completion method \cite{jain2014provable} and tune its rank via the information criterion proposed by \cite{sun2015provable}. This tuning method chooses the optimal rank as $R=20$ from the rank list $\{1, 2, 3, 4, 5, 6, 8, 10, 12, 14, 16, 18, 20, 22\}$. Finally, the imputed values are truncated to ensure all the values of the tensor are within 0 and 1 since click-through rates are proportions.

One mode of the fourth-order tensor has only two observations and those observations already have a natural grouping (device type).  Therefore, for the sake of clustering we analyze the devices separately.  We compare our method with CPD+$k$-means.  Furthermore, the tuning parameter for convex co-clustering is automatically selected using the eBIC (\Sec{bic}) while the number of clusters in CPD+$k$-means is chosen via the gap statistic \citep{TibshiraniWaltherHastie2001}. We do not include comparisons with CoTeC given its poor performance in the simulation experiments.

\begin{figure}[t]
    \centering
    \subfloat[Personal Computers]{\label{fig:adHeatSliceM_PC_user_33}
    \includegraphics[scale=0.375]{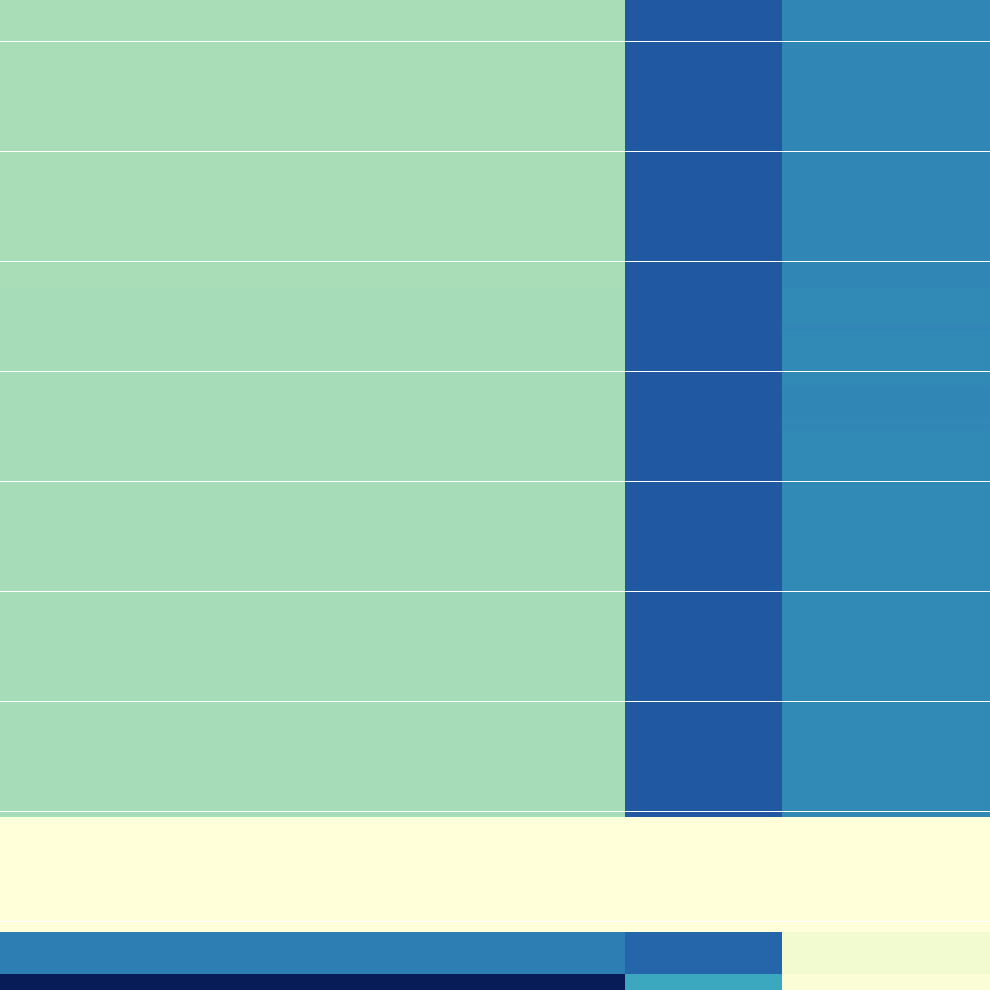}} \hspace{1.5cm}
    \subfloat[Mobile]{\label{fig:adHeatSliceM_mobile_user_33}
    \includegraphics[scale=0.375]{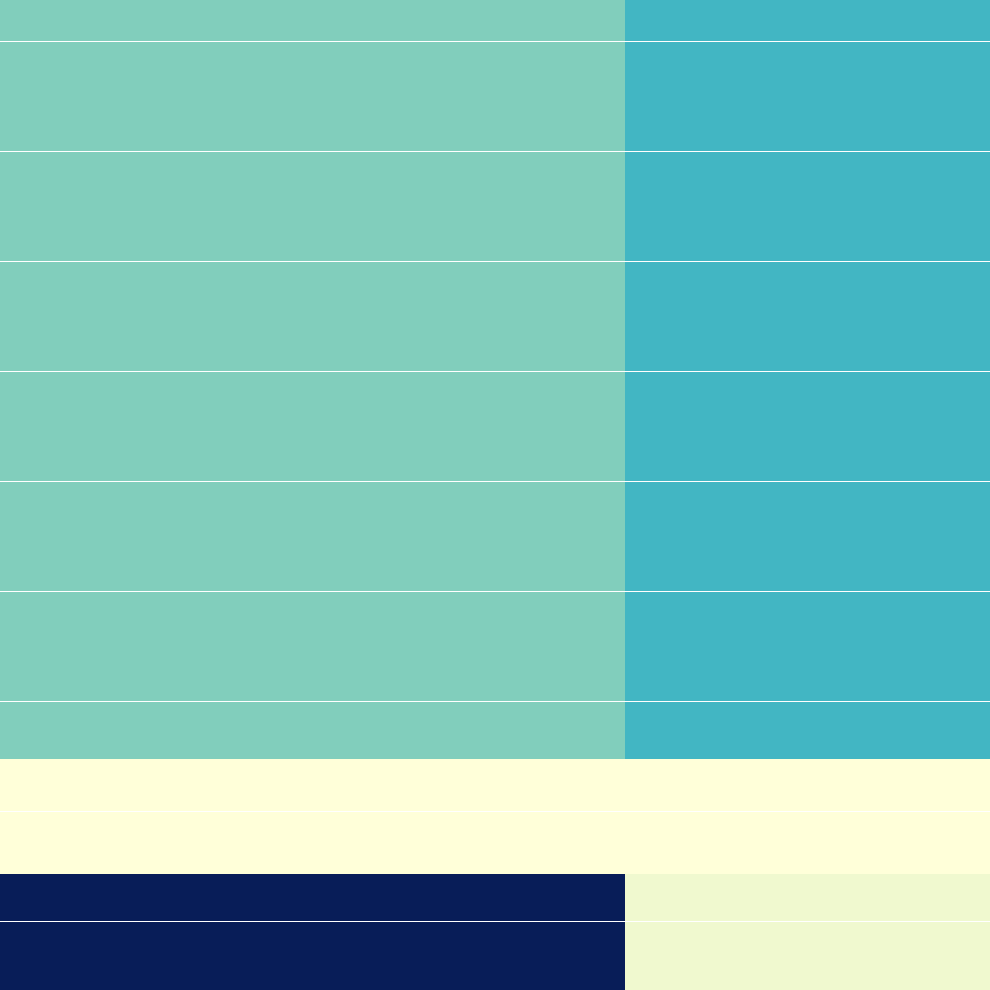}}
    \caption{Advertisement and Publisher Click-Through Rate Biclusters for a Randomly Selected User. \small The rows correspond to different advertisements and the columns correspond to different publishers.  Darker blue corresponds to higher click-through rates for a given device.}  \label{fig:adHeatSliceM}
\end{figure}

We first look at the clustering results from clustering the click-through rates for users accessing the advertisements through a personal computer (PC).  \Tab{adData} contains the number of clusters identified as well as the sizes of the clusters, while \Fig{adHeatSliceM_PC_user_33} visualizes the advertisement-by-publisher biclusters for a randomly selected user.  As to be expected,   the advertisement-by-publisher slices display a checkerbox pattern, which turns into a checkerbox pattern when the slices are meshed together.  The clustering results for the users are omitted in this paper to ensure user privacy. However, co-clustering the tensor does not result in the loss of information that would occur if the tensor was converted into a matrix by averaging across users or flattening along one of the modes.  \Tab{adData} and \Fig{adHeatSliceM_PC_user_33} show that the CoCo estimator identifies four advertisement clusters, with one cluster being much bigger than the others.  The advertisements in this large cluster have click-through rates that are close to the grand average in the data set.  One of the small clusters has very low click-through rates, while the other two clusters tend to have much higher click-through rates than the rest of the advertisements.  On the other hand, CPD+$k$-means clusters the advertisements into 57 groups, which is less-useful from a practical standpoint.  Many of the clusters are similarly-sized and contain only a few advertisements, likely due to the inability of CPD+$k$-means to handle imbalanced cluster sizes as was observed in the simulation experiments (\Sec{simulations_imbalanced_cluster_sizes}).  In terms of the publishers, the CoCo estimator identifies 3 clusters while CPD+$k$-means does not find any underlying grouping and simply identifies one big cluster, which again is not terribly useful (\Tab{adData}). We next provide some interpretations of the obtained clustering results of the publishers. One way online advertisers can reach more users is by entering agreements with other companies to route traffic to the advertiser's website.  For example, Google and Apple have a revenue-sharing agreement in which Google pays Apple a percentage of the revenue generated by searches on iPhones \citep{macworld}.  Similarly, the online company being studied partners with several internet service providers (ISPs) to host the defaut home pages for the ISP's customers.  It would make sense that these slightly different variants of the online company's main home page would have similar click-through rates, and the CoCo estimator in fact assigned these variants into the same cluster.

\begin{table}[]
\centering
\begin{tabular}{ccccccc}
\toprule
\multicolumn{1}{}{}        & \multicolumn{4}{c}{CoCo Estimator}                                                                                                                                                                                                 & \multicolumn{2}{c}{CPD+kmeans}                                                                                       \\ 
\cmidrule(lr){2-7} 
\multicolumn{1}{}{}        & \multicolumn{2}{c}{Advertisements}                                                                                & \multicolumn{2}{c}{Publisher}                                                                                       & Advertisements                                            & Publisher                                                 \\ 
\multicolumn{1}{c}{Device} & \begin{tabular}[c]{@{}c@{}}\# of\\ clusters\end{tabular} & \begin{tabular}[c]{@{}c@{}}Cluster\\ Sizes\end{tabular} & \begin{tabular}[c]{@{}c@{}}\# of \\ clusters\end{tabular} & \begin{tabular}[c]{@{}c@{}}Cluster \\ Sizes\end{tabular} & \begin{tabular}[c]{@{}c@{}}\# of \\ clusters\end{tabular} & \begin{tabular}[c]{@{}c@{}}\# of \\ clusters\end{tabular} \\ 
\cmidrule(lr){1-7} 
\multicolumn{1}{c}{PC}     & 4                                                        & (156, 22, 8, 3)                                                & 3                                                         & (4, 3, 12)                                                   & 57                                                        & 1                                                        \\ 
\multicolumn{1}{c}{Mobile} & 3                                                        & (145, 22, 22)                                                  & 2                                                         & (7, 12)                                                     & 49                                                  & 13                                                        \\
\bottomrule
\end{tabular}
\caption{Advertising Data Clustering Results}
\label{tab:adData}
\end{table}

For users accessing the advertisements through a mobile device, such as a mobile phone or tablet computer, the CoCo estimator results for the advertisements are largely similar to the results for PCs (\Tab{adData} and \Fig{adHeatSliceM_mobile_user_33}).  There is one large cluster that contains click-through rates similar to the overall average, while the two other equally-sized clusters have relatively very low or very high click-through rates, respectively.  The underlying click-through rates for the PC data have more variability than the mobile data, which is consistent with the identification of an additional cluster for the PC data.  As before, CPD+$k$-means finds a large number of advertisement clusters, most of which are roughly the same size, again likely impacted by the imbalance in the cluster sizes.  When compared to the personal computer device, one difference is that the cluster with the higher click-through rates for mobile devices is larger and has a higher average click-through rate than the similar clusters for the personal computer device.  This finding is consistent with research by the Pew Research Center that found that click-through rates for mobile devices are higher than for advertisements viewed on a personal computer or laptop \citep{mitchell2012future}.

It is also enlightening to take a closer look at the underlying advertisements clustered across the two devices.  All of the advertisements clustered in the high click-through rate cluster for the mobile devices are in the average click-through rate cluster for personal computers.  In taking a closer look at the ads in these clusters, there are several ads related to online shopping for personal goods, such as jeans, workout clothes, or neck ties.  It makes sense to shop for these types of goods using a mobile device, such as while at work when it is not appropriate to do so on a work computer.  Conversely, all of the advertisements in either of the two higher PC click-through rate clusters are in the large, average click-through rate cluster for the mobile devices.  There are several financial-related ads in these two PC clusters, such as for mortgages or general investment advice.  On the other hand, there are not many online shopping ads in those clusters, with the exception of more expensive technology-related goods that one may want to invest more time in researching before making a purchase.  

In terms of the publisher clusters on mobile device, \Tab{adData} shows that the CoCo estimator identifies two clusters of publishers while CPD+$k$-means identifies 13 small clusters.  Contrary to the advertisement clusters, the publisher clusters across both devices are very similar.  In fact, the only difference is that the smaller cluster for the mobile device, which contains seven publishers, is split into two clusters for personal computers.  This can be seen in the click-through rate heatmaps given in \Fig{adHeatSliceM} in looking at the right part of each heatmap.  The publishers in these smaller clusters have higher click-through rates on average than those in the larger cluster.  Additionally, five of the seven (71\%) publishers in the high click-through rate clusters have stand-alone apps that display ads, while only three of the twelve (25\%) publishers in the larger cluster do.  For mobile devices, it has been observed that in-app advertisements have higher click-through rates and browser-based ads \citep{forbesMobileAds}.  We conjecture that this is also true for personal computer apps, which is consistent with the clustering results.  Thus it again appears that the clusters identified by CoCo also make sense practically.    
 
\section{Discussion}
\label{sec:discussion}
 
In this paper, we formulated and studied the problem of co-clustering of tensors as a convex optimization problem. The resulting CoCo estimator enjoys features in theory and practice that are arguably lacking in existing alternatives, namely statistical consistency, stability guarantees, and an algorithm with polynomial computational complexity. Through a battery of simulations, we observed that the CoCo estimator can identify co-clustering structures under realistic scenarios such as imbalanced co-cluster sizes, imbalanced number of clusters along each mode, heteroskedasticity in the noise distribution associated with each co-cluster, and even some violation of the checkerbox mean tensor assumption.

We have leveraged the power of the convex relaxation to engineer a computationally tractable co-clustering method that comes with statistical guarantees. These benefits, however, do not come for free. The CoCo estimator incurs similar costs that using the lasso incurs as a surrogate for a cardinality constraint or penalty. It is well known that the lasso leads to parameter estimates that are shrunk towards zero. This shrinkage toward zero is the price for simultaneously estimating the support, or locations of the nonzero entries, in a sparse vector as well as the values of the nonzero entries. In the context of convex co-clustering, the CoCo estimator $\That{U}$ is shrunk towards the tensor $\Tbar{X}$, namely the tensor whose entries are all equal to the average over all entries of $\T{X}$. The weights, however, play a critical role in reducing this bias. In fact, the weights can be seen as serving the same role as weights used in the adaptive lasso \citep{Zou2006}. 

There are several possible extensions and open problems that have been left for future work. First, we note that there is a gap between what our theory predicts and what seems possible from our experiments. Specifically, \Thm{final_error} assumes uniform weights for each mode, yet simulation experiments indicate that the CoCo estimator using Tucker derived Gaussian kernel weights \Eqn{preweights} can significantly outperform the CoCo estimator using uniform weights. One open problem is to derive prediction error bounds that relax the uniform weights assumption.

Second, although we have developed automatic methods for constructing the weights that work well empirically, other approaches to constructing the weights is a direction of future search.  For example, other tensor approximation methods, such as the use of the $\ell_1$-norm to make the decomposition most robust to heavy tail noise as done by \cite{cao2015robust}, could possibly improve the quality of the weights.

Third, in this paper we have focused on additive noise that is a zero-mean $M$-concentrated random variable. Real data, however, may not follow such a distribution motivating co-clustering procedures that can handle outliers. To address potential distributional departures, the CoCo framework could be extended to handle outliers by swapping the sum of squared residuals term in \Eqn{primal_objective_vec} with an analogous Huber loss or Tukey's Biweight function.

Finally, while our first order algorithm for co-clustering tensors scales linearly in the size of the data, data tensors inevitably will only increase in size motivating the need for more scalable algorithms for computing the CoCo estimator.  A natural approach would be to adopt an existing distributed version of the proximal methods, such as one the methods proposed by \cite{combettes2011proximal}, \cite{chen2012fast}, \cite{li2013distributed}, or \cite{Eckstein2017}. Another natural approach would be to investigate if stochastic versions of the recently proposed generalized dual gradient ascent \citep{HoLinJordan2019} could be adapted to compute the CoCo estimator. Additionally, in practice many data tensors that we would like to co-cluster may be very sparse. The first order algorithm presented here assumes the data tensor is dense. Consequently, an important direction of future work is to investigate alternative optimization algorithms that could leverage the sparsity structure within a data tensor.


\acks{The authors thank Xu Han for his help with the simulation experiments during the revision of this work. The authors also thank the action editor and three reviewers for their helpful comments and suggestions which led to a much improved presentation.  Eric Chi acknowledges support from the National Science Foundation (DMS-1752692) and National Institutes of Health (R01GM135928). Will Wei Sun acknowledges support from the Office of Naval Research (ONR N00014-18-1-2759). Hua Zhou acknowledges support from the National Institutes of Health (R01GM053275 and R01HG006139). Finally, this research collaboration was partially funded by the National Science Foundation under grant DMS-1127914 (the Statistical and Applied Mathematical Sciences Institute). Any opinions, findings, and conclusions or recommendations expressed in this material are those of the authors and do not necessarily reflect the views of the National Science Foundation, the National Institutes of Health, or the Office of Naval Research.}

\appendix

\section*{Summary of Appendix}

The Appendix expands upon several topics in the main body of the paper. \App{tensor_decomp} reviews the CP and Tucker decompositions. \App{proof_smoothness} contains proofs of the stability and continuity properties of the CoCo estimator. \App{proof_final_error} contains auxiliary lemmas, their proofs, as well as the proof of our main theoretical result, the non-asymptotic error bound of the CoCo estimator. \App{dual_derivation} details the derivation of the Lagrangian dual of the CoCo estimation problem. \App{projected_gradient} provides more detail on the projected gradient algorithm given outlined in \Alg{CoCo}, namely the per-iteration and storage costs, algorithmic convergence guarantees, and stopping rule based on monitoring the duality gap. \App{weight_extra_discussion} elaborates on the connection between the weights recommended in \Sec{weights} and clustering methods that employ folded-concave penalties as well as details on the Tucker denoising methods used in our experiments. \App{cpdkmeans} provides more detail on the CPD+$k$-means procedure including selection of its tuning parameters. Finally, \App{simulations_rectangular_extra} contains two additional simulations involving rectangular tensors.

\section{Tensor Decompositions}\phantomsection
\label{sec:tensor_decomp}

We review two basic tensor decompositions that generalize the singular value decomposition (SVD) of a matrix:
(i) the CANDECOMP/PARAFAC (CP) decomposition \citep{carroll1970analysis, harshman1970foundations} and (ii) the Tucker decomposition \citep{Tucker1966}. Just as the SVD can be used to construct a lower-dimensional approximation to a data matrix, these two decompositions can be used to construct a lower dimensional approximation to a $D$-way tensor $\T{X} \in \Real^{n_1 \times n_2 \times \cdots \times n_D}$ 

The CP decomposition aims to approximate $\T{X}$ by a sum of rank-one tensors, namely
\begin{eqnarray*}
	\T{X} & \approx & \sum_{i = 1}^R \Vn{a}{1}_i  \circ \Vn{a}{2}_i \circ \cdots \circ \Vn{a}{D}_i,
\end{eqnarray*}
where $\circ$ represents the outer product and $\Vn{a}{d}_i$ is the $i$th column of the $d$th factor matrix $\Mn{A}{d} \in \Real^{n_d \times R}$.
The positive integer $R$ denotes the rank of the approximation. For sufficiently large $R$, we can exactly represent $\T{X}$ with a CP decomposition.

The Tucker decomposition aims to approximate $\T{X}$ by a core tensor $\T{H} \in \Real^{R_1 \times R_2 \times \cdots \times R_D}$ multiplied by factor matrices along each of its modes, namely
\begin{equation*}
	\T{X} \approx \T{H} \times_1 \Mn{A}{1} \times_2 \Mn{A}{2} \times_3 \cdots \times_D \Mn{A}{D}  =  \sum_{i_1=1}^{R_1} \sum_{i_2=1}^{R_2} \cdots \sum_{i_D=1}^{R_D} \ME{H}{i_1i_2\cdots i_D}\Vn{a}{1}_{i_1} \circ \Vn{a}{2}_{i_2} \circ \cdots \circ \Vn{a}{D}_{i_D},
\end{equation*}
where $\Vn{a}{d}_{i_d}$ is the $i_d$th column of the $d$th factor matrix $\Mn{A}{d} \in \Real^{n_d \times R_d}$. Typically the columns of $\Mn{A}{d}$ are computed to be orthonomal and can be interpreted as principal components or basis vectors for the $d$th mode. For sufficiently large $R_1, \ldots, R_D$, we can exactly represent $\T{X}$ with a Tucker decomposition.

\section{Proofs of Smoothness Properties}\phantomsection
\label{sec:proof_smoothness}

\subsection{Proof of \Prop{cont}}
\label{sec:proof_continuity}

Without loss of generality, we can absorb $\gamma$ into the weights matrices. Thus, we seek to show the continuity of $\That{U}$ with respect to $(\T{X}, \M{W}_1, \ldots, \M{W}_D)$. We use the following compact representation of the weights
\begin{eqnarray*}
\V{w} & = & (\vec(\M{W}_1)\Tra,\vec(\M{W}_2)\Tra, \ldots, \vec(\M{W}_D))\Tra \in \Real^{\sum_{d=1}^D {n_d \choose 2}}.
\end{eqnarray*}
We check to see if the solution $\That{U}$ is continuous in the variable $\V{\zeta} = (\V{x}\Tra,\V{w}\Tra)\Tra$. It is easy to verify that the following function is jointly continuous in $\T{U}$ and $\V{\zeta}$
\begin{eqnarray*}
f(\T{U},\V{\zeta}) & = & \frac{1}{2} \lVert \T{X} - \T{U} \rVert_{\text{F}}^2 + R(\T{U}, \V{w}),
\end{eqnarray*}
where
\begin{eqnarray*}
R(\T{U}, \V{w}) & = & \sum_{d=1}^D \sum_{i < j} w_{d,ij} \lVert \T{U} \times_d \V{\Delta}_{d,ij} \rVert_{\text{F}}
\end{eqnarray*}
is a convex function of $\T{U}$ that is continuous in $(\T{U},\V{W})$. Let
\begin{eqnarray*}
\T{U}^\star(\V{\zeta}) & = & \underset{\T{U}}{\arg\min}\; f(\T{U},\V{\zeta}).
\end{eqnarray*}
Since $f(\T{U},\V{\zeta})$ is strongly convex in $\T{U}$, the minimizer $\T{U}^\star(\V{\zeta})$ exists and is unique. 

We proceed with a proof by contradiction. Suppose $\T{U}^\star(\V{\zeta})$ is not continuous at a point $\V{\zeta}$. Then there exists an $\epsilon > 0$ and a sequence $\{\Vn{\zeta}{m}\}$ converging to a limit $\V{\zeta}$ such that $\lVert \Tn{U}{m} - \T{U}^\star(\V{\zeta}) \rVert_{\text{F}} \geq \epsilon$ for all $m$ where
\begin{eqnarray*}
\Tn{U}{m} & = & \underset{\T{U}}{\arg\min}\; f(\T{U},\Vn{\zeta}{m}).
\end{eqnarray*}
Since $f(\T{U},\V{\zeta})$ is strongly convex in $\T{U}$, the minimizer $\Tn{U}{m}$ exists and is unique. Without loss of generality, we can assume
$\lVert \Vn{\zeta}{m} - \V{\zeta} \rVert_{\text{F}} \leq 1$.
This fact will be used later in proving the boundedness of the sequence $\Tn{U}{m}$.

If $\Tn{U}{m}$ is a bounded sequence, then we can pass to a convergent subsequence with limit $\Tbar{U}$. Fix an arbitrary point $\Ttilde{U}$. Note that $f(\Tn{U}{m}, \Vn{\zeta}{m}) \leq f(\Ttilde{U}, \Vn{\zeta}{m})$ for all $m$. Since $f$ is continuous in $(\T{U},\V{\zeta})$, taking limits gives us the inequality
\begin{eqnarray*}
f(\Tbar{U}, \V{\zeta}) & \leq & f(\Ttilde{U}, \V{\zeta}).
\end{eqnarray*}
Since $\Ttilde{U}$ was selected arbitrarily, it follows that $\Tbar{U} = \T{U}^\star(\V{\zeta})$, which is a contradiction. It only remains for us to show that the sequence $\Tn{U}{m}$ is bounded.

Consider the function
\begin{eqnarray*}
g(\T{U}) & = & \underset{\Vtilde{\zeta} : \lVert \Vtilde{\zeta}- \V{\zeta} \rVert_{\text{F}} \leq 1}{\sup}\;
\frac{1}{2} \lVert \Ttilde{X} - \T{U} \rVert_{\text{F}}^2 + R_\Vtilde{W}(\T{U}).
\end{eqnarray*}
Note that $g$ is convex, since it is the point-wise supremum of a collection of convex functions. Since $f(\T{U}, \Vn{\zeta}{m}) \leq g(\T{U})$ and $f$ is strongly convex in $\T{U}$, it follows that $g(\T{U})$ is also strongly convex and therefore has a unique global minimizer $\T{U}^*$ such that $g(\T{U}^*) < \infty$. It also follows that
\begin{eqnarray}
\label{eq:ineqA}
f(\Tn{U}{m},\Vn{\zeta}{m}) & \leq & f(\T{U}^*,\Vn{\zeta}{m}) \amp \leq \amp g(\T{U}^*)
\end{eqnarray}
for all $m$. By the reverse triangle inequality it follows that
\begin{eqnarray}
\label{eq:ineqB}
\frac{1}{2} \left (
 \lVert \Tn{U}{m} \rVert_{\text{F}} - \lVert \Tn{X}{m} \rVert_{\text{F}} 
\right )^2 \amp \leq \amp \frac{1}{2} \lVert \Tn{U}{m} - \Tn{X}{m} \rVert_{\text{F}}^2 \amp \leq \amp
f(\Tn{U}{m},\Vn{\zeta}{m}).
\end{eqnarray}
Combining the inequalities in \Eqn{ineqA} and \Eqn{ineqB}, we arrive at the conclusion that
\begin{eqnarray*}
\frac{1}{2} \left (
 \lVert \Tn{U}{m} \rVert_{\text{F}} - \lVert \Tn{X}{m} \rVert_{\text{F}} 
\right )^2 \amp \leq \amp g(\T{U}^*),
\end{eqnarray*}
for all $m$. Suppose the sequence $\Tn{U}{m}$ is unbounded, namely $\lVert \Tn{U}{m} \rVert_{\text{F}} \rightarrow \infty$. 
But since $\Tn{X}{m}$ converges to $\T{X}$, the left hand side must diverge. Thus, we arrive at a contradiction if $\Tn{U}{m}$ is unbounded. \hfill $\blacksquare$

\subsection{Proof of \Prop{zero}}
\label{sec:zero}

First suppose that $\Mz{U}{d} = \V{1}\V{c}\Tra$, namely all the mode-$d$ subarrays of $\T{U}$ are identical. Recall that $\T{Z} = \T{U} \times_d \M{A}$ if and only if $\Mz{Z}{d} = \M{A}\Mz{U}{d}$. Therefore, $R_d(\T{U}) = 0$ since $\V{\Delta}_{d,ij}\V{1}\V{c}\Tra = 0$ for all $(i,j) \in \E_d$.

Now suppose that $R_d(\T{U})$ is zero. Take an arbitrary pair $(i,j)$ with $i < j$. By Assumption 4.1, there exists a path $i \rightarrow k \rightarrow \cdots \rightarrow l \rightarrow j$ along which the weights are positive. Let $w$ denote the smallest weight along this path, namely $w = \min \{\VE{w}{d,ik}, \ldots, \VE{w}{d,lj}\}$. By the triangle inequality
\begin{eqnarray*}
\lVert \T{U} \times_d \V{\Delta}_{d,ij} \rVert_{\text{F}} & \leq & \lVert \T{U} \times_d \V{\Delta}_{d,ik} \rVert_{\text{F}}
+ \cdots + \lVert \T{U} \times_d \V{\Delta}_{d,lj} \rVert_{\text{F}}.
\end{eqnarray*}
We can then conclude that
\begin{eqnarray*}
w\lVert \T{U} \times_d \V{\Delta}_{d,ij} \rVert_{\text{F}} & \leq & R_d(\T{U}) \amp = \amp 0.
\end{eqnarray*}
It follows that $\V{e}_i\Tra\Mz{U}{d} = \V{e}_j\Tra\Mz{U}{d}$, since $w$ is positive. Since the pair $(i,j)$ is arbitrary, it follows that all the rows of $\Mz{U}{d}$ are identical or in other words, $\Mz{U}{d} = \V{1}\V{c}\Tra$ for some $\V{c} \in \Real^{n_{-d}}$. \hfill $\blacksquare$

\subsection{Proof of \Prop{coalesce}}
\label{sec:coalesce}

We will show that there is a $\gamma_{\max}$ such that for all $\gamma \geq \gamma_{\max}$, the grand mean tensor $\Tbar{X}$ is the unique global minimizer to the primal objective \Eqn{primal_objective}. We will certify that $\Tbar{X}$ is the solution to the primal problem by showing that the optimal value of a dual problem, which lower bounds the primal, equals $F_\gamma(\Tbar{X})$.

Note that the Lagrangian dual given in \Eqn{dual_expanded} is a tight lower bound on $F_\gamma(\T{U})$.
\begin{eqnarray*}
\underset{\V{\lambda} \in C_\gamma}{\max}\;
-\frac{1}{2} \lVert \M{A}\Tra\V{\lambda} \rVert_2^2 + \langle \V{\lambda}, \M{A}\V{x} \rangle.
\end{eqnarray*}

For sufficiently large $\gamma$, the solution to the dual maximization problem coincides with the solution to the unconstrained maximization problem
\begin{eqnarray*}
\underset{\V{\lambda}}{\max}\;
-\frac{1}{2} \lVert \M{A}\Tra\V{\lambda} \rVert_2^2 + \langle \V{\lambda}, \M{A}\V{x} \rangle,
\end{eqnarray*}
whose solution is $\V{\lambda}^\star = \left (\M{A}\M{A}\Tra \right )^\dagger\M{A}\V{x}$.  Plugging $\V{\lambda}^\star$ into the dual objective gives an optimal value of
\begin{eqnarray*}
\frac{1}{2} \lVert \M{A}\Tra\left (\M{A}\M{A}\Tra \right )^\dagger \M{A} \V{x} \rVert_2^2 \amp = \amp
\frac{1}{2} \lVert \V{x} - \bigg [\M{I} - \M{A}\Tra\left (\M{A}\M{A}\Tra \right )^\dagger \M{A} \bigg ]\V{x} \rVert_2^2.
\end{eqnarray*}
Note that  $\left [\M{I} - \M{A}\Tra\left (\M{A}\M{A}\Tra \right )^\dagger \M{A} \right ]$
is the projection onto the orthogonal complement of the column space of $\M{A}\Tra$, which is equivalent to the null space or kernel of $\M{A}$, denoted Ker$(\M{A})$. We will show below that Ker($\M{A}$) is the span of the all ones vector. Consequently,

\begin{eqnarray*}
\left [\M{I} - \M{A}\Tra\left (\M{A}\M{A}\Tra \right )^\dagger \M{A} \right ]\V{x} & = & \frac{1}{n}\langle \V{x}, \V{1} \rangle \V{1}.
\end{eqnarray*}
Note that the smallest $\gamma$ such that $\V{\lambda}^\star \in C_\gamma$ is an upper bound on $\gamma_{\max}$. 

We now argue that Ker($\M{A}$) is the span of $\V{1} \in \Real^{n}$.
We rely on the following fact: If $\M{\Phi}_d$ is an incidence matrix of a connected graph with $n_d$ vertices, then the rank of $\M{\Phi}_d$ is $n_d-1$
\citep[Theorem 7.2]{Deo1974}. According to \As{connectedness}, the mode-$d$ graphs are connected; it follows that $\M{\Phi}_d \in \{-1,0,1\}^{\lvert \mathcal{E}_d \rvert \times n_d}$ has rank $n_d-1$. It follows then that Ker($\M{\Phi}_d$) has dimension one. Furthermore, since each row of $\M{\Phi}_d$ has one $1$ and one $-1$, it follows that $\V{1} \in$ Ker($\M{\Phi}_d$) $\subset \Real^{n_d}$. A vector $\V{z} \in $Ker($\M{A}$) if and only if $\V{z} \in$ Ker($\M{A}_d$) for all $d$.

Recall that the rank of the Kronecker product $\M{A} \Kron \M{B}$ is the product of the ranks of the matrices $\M{A}$ and $\M{B}$. This rank property of Kronecker products of matrices implies that the dimension of Ker($\M{A}_d$) equals $n_{-d}$. Let $\V{b}_i = \V{1}_{n_D} \Kron \cdots \Kron \V{1}_{n_{d+1}} \Kron \V{e}_i \Kron \V{1}_{n_{d-1}} \Kron \cdots \Kron \V{1}_{n_1}$ where $\V{1}_p \in \Real^p$ is the vector of all ones and $\V{e}_i \in \Real^{n_d}$ is the $i$th standard basis vector. Then that the set of vectors $\mathcal{B} = \{\V{b}_1, \V{b}_2, \ldots, \V{b}_{n_{d}}\}$ forms a basis for Ker($\M{A}_d$).

Take an arbitrary element from Ker($\M{A}_d$), namely a vector of the form $\V{1}_{n'} \Kron \V{a} \Kron \V{1}_{n''}$, where $n' = \prod_{j=d+1}^{D} n_j$ and $n'' = \prod_{j = 1}^{d-1}$. We will show that in order for $\V{1}_{n'} \Kron \V{a} \Kron \V{1}_{d''} \in$ Ker($\M{i} \Kron \M{\Phi}_d$), $\V{a}$ must be a multiple of $\V{1}_{n_d}$. Consider the relevant matrix-vector product 
\begin{eqnarray*}
\M{A}_d\bigg (\V{1}_{n_D} \Kron \V{a} \Kron \V{1}_{n_1}\bigg) & = &
(\V{1}_{n_D} \Kron \cdots \Kron \V{1}_{n_{d+1}}\Kron \M{\Phi}_d\V{a} \Kron \V{1}_{n_{d-1}} \Kron \cdots \Kron \V{1}_{n_1} ).
\end{eqnarray*}
Therefore, $\M{A}_d\bigg(\V{1}_{n'} \Kron \V{a} \Kron \V{1}_{n''}\bigg) = \V{0}$ if and only if $\M{\Phi}_d\V{a} = \V{0}$. But
the only way for $\M{\Phi}_d\V{a}$ to be zero is for $\V{a} = c\V{1}_{n_d}$ for some $c \in \Real$. Thus, Ker($\M{A}$) is the span of $\V{1}_n$. \hfill $\blacksquare$

\subsection{Proof of \Prop{non_expansive}}
\label{sec:non_expansive}

Note that $\That{U}$ is the proximal mapping of the closed, convex function
\begin{eqnarray*}
\sum_{d=1}^D R_d(\T{U})
\end{eqnarray*}
Then $\That{U}$ is firmly nonexpansive in $\T{X}$ \citep[Lemma 2.4]{ComWaj2005}. Finally, firmly nonexpansive mappings are nonexpansive, which completes the proof. \hfill $\blacksquare$

\section{Proof of \Thm{final_error}}\phantomsection
\label{sec:proof_final_error}

We first prove some auxiliary lemmas before proving our prediction error result. 

\subsection{Auxiliary Lemmas}
\label{sec:other_lemma}
The following lemma considers the concentration of a random quadratic form $\V{y}\Tra \M{B}  \V{y}$ for a $M$-concentrated random vector $\V{y}$ and a deterministic matrix $\M{B}$ \citep{VuWang2015}. It can be viewed as a generalization of the standard Hanson and Wright inequality for the quadratic forms of independent sub-Gaussian random variables \citep{HW71}. 

\begin{lemma}
\label{lem:VuWang2015} 
Let $\V{y} \in \Real^n$ be a $M$-concentrated random vector, see \Def{error}. Then there are constants $C, C^{'} >0$ such that for any matrix $\M{B} \in \Real^{n\times n}$,
\begin{eqnarray*}
\mathbb P\Big( \lvert \V{y}\Tra \M{B}  \V{y} - \tr(\M{B}) \lvert \ge t \Big) & \le & C \log(n) \exp\left\{ - C^{'}M^{-2} \min\left[\frac{t^2}{\lVert \M{B} \rVert_{\text{F}}^2 \log(n)}, \frac{t}{\lVert \M{B} \rVert_2}  \right]  \right\}.
\end{eqnarray*}
\end{lemma}

The next lemma studies the properties of the matrix $\M{A}_{d,ij}$, defined in \Eqn{Adij}, in the penalty function. Denote $\M{S}_d$ as the matrix constructed by concatenating $\M{A}_{d,ij}, i<j$ vertically. That is, 
\begin{eqnarray}
\label{eq:Sk}
\M{S}_d & = & \begin{pmatrix}
\M{A}_{d,12}\Tra & \M{A}_{d,13}\Tra & \ldots & \M{A}_{d,n_d-1 n_d}\Tra
\end{pmatrix}\Tra \in \Real^{[{n_d \choose 2} n_{-d}] \times n}.
\end{eqnarray}

\begin{lemma}
\label{lem:propertyS} 
For each $d = 1, \ldots, D$, the rank of the matrix $\M{S}_d$ is $(n_d-1)n_{-d}$. Denote $\sigma_{\min}(\M{S}_d)$ and $\sigma_{\max}(\M{S}_d)$ as the minimum non-zero singular value and maximum singular value of $\M{S}_d$, respectively. We have $\sigma_{\min}(\M{S}_d) = \sigma_{\max}(\M{S}_d) = \sqrt{n_d}$. 
\end{lemma}

The proof of \Lem{propertyS} follows from Lemma 1 in \cite{Tan2015} and is omitted. According to \Lem{propertyS}, we can construct a singular value decomposition of $\M{S}_d = \M{U}_d \M{\Lambda}_d \M{V}_d\Tra$, where $\M{U}_d \in \Real^{[{n_d \choose 2} n_{-d}] \times (n_d-1)n_{-d}}$, $\M{\Lambda}_d \in \Real^{(n_d-1)n_{-d} \times (n_d-1)n_{-d}}$, and $\M{V}_d \in \Real^{ n \times (n_d-1)n_{-d}}$. Denote 
\begin{eqnarray}
\label{eq:Gk}
\M{G}_{d} & = & \M{U}_d \M{\Lambda}_d \in  \Real^{[{n_d \choose 2} n_{-d}] \times (n_d-1)n_{-d}},
\end{eqnarray}
and its pseudo-inverse as $\M{G}_{d}^{\dagger} \in  \Real^{(n_d-1)n_{-d} \times [{n_d \choose 2} n_{-d}] }$. The following lemma studies the properties of $\M{G}_{d}$ and $\M{G}_{d}^{\dagger}$, for each $d=1, \ldots, D$.

\begin{lemma}
\label{lem:propertyG} 
For each $d = 1, \ldots, D$, the rank of the matrix $\M{G}_d$ is $(n_d-1)n_{-d}$. The minimal non-zero singular value and maximal singular value of $\M{G}_d$ are $\sigma_{\min}(\M{G}_d) = \sigma_{\max}(\M{G}_d) = \sqrt{n_d}$. Moreover, $\sigma_{\min}(\M{G}_d^{\dagger}) = \sigma_{\max}(\M{G}_d^{\dagger}) = 1/ \sqrt{n_d}$.
\end{lemma}
\Lem{propertyG} follows directly from the conclusions in \Lem{propertyS}.

\subsection{Proof of Main Theorem}\phantomsection
\label{sec:proof_theorem}

We first reformulate our optimization problem via a decomposition approach to simplify the theoretical analysis. Such strategy was developed in \cite{LiuYuan2013} and has been successfully applied in \cite{Tan2015, WanZhaSun2016}. 

Denote $\gamma_d = \gamma / n_d$. Our convex tensor co-clustering method is equivalent to solving
\begin{eqnarray}
\label{eq:obj}
\Vhat{u} & = & \underset{\V{u}}{\arg\min} \left\{ \frac{1}{2} \lVert \V{x} - \V{u} \rVert_{2}^2 +  \sum_{d=1}^D \gamma_d \sum_{(i,j) \in \mathcal{E}_d}  \lVert \M{A}_{d,ij} \V{u} \rVert_2 \right\}.
\end{eqnarray}

According to the definition of $\M{S}_d$ in \Eqn{Sk}, we define the penalty function $R(\cdot)$ such that 
\begin{eqnarray*}
R(\M{S}_d \V{u}) & = & \sum_{(i,j) \in \mathcal{E}_d} \lVert \M{A}_{d,ij} \V{u} \rVert_2.
\end{eqnarray*}

According to the singular value decomposition of $\M{S}_d = \M{U}_d \M{\Lambda}_d \M{V}_d\Tra$, there exists a matrix $\M{W}_d \in \Real^{n \times n_{-d}}$ such that $\tilde{\M{V}}_d = [\M{W}_d, \M{V}_d] \in \Real^{n \times n}$ is an orthogonal matrix and $\M{W}_d\Tra \M{V}_d = \M{0}$. Let  $\V{\alpha}_d = \M{W}_d\Tra \V{u} \in \Real^{n_{-d}}$ and $\V{\beta}_d = \M{V}_d\Tra \V{u} \in \Real^{n}$. Clearly, we have 
\begin{eqnarray}
\label{eq:alphabeta}
\M{W}_d \V{\alpha}_d + \M{V}_d\V{\beta}_d & = & \M{W}_d\M{W}_d\Tra \V{u} + \M{V}_d\M{V}_d\Tra \V{u} \amp = \amp \tilde{\M{V}}_d\tilde{\M{V}}_d\Tra\V{u} \amp = \amp \V{u},
\end{eqnarray}
for any $d=1, \ldots, D$. This fact together with the definition of $\M{G}_{d} = \M{U}_d \M{\Lambda}_d$ in \Eqn{Gk} imply that solving our convex tensor clustering in \Eqn{obj} is equivalent to solving
\begin{equation}
\label{eq:newobj}
\min_{\V{\alpha}_d, \V{\beta}_d, d = 1,\ldots, D}  \sum_{d=1}^D \left\{ \frac{1}{2D} \lVert \V{x} - \M{W}_d \V{\alpha}_d + \M{V}_d\V{\beta}_d \rVert_{2}^2 +  \gamma_d R(\M{G}_{d}\V{\beta}_d) \right\}
\end{equation}

Denote the solution of \Eqn{newobj} as $\hat{\V{\alpha}}_d, \hat{\V{\beta}}_d$, $d=1,\ldots, D$, which corresponds to the estimator $\Vhat{u}$ in \Eqn{obj} according to \Eqn{alphabeta}. Similarly, we denote the true parameters as $\V{\alpha}_d^*, \V{\beta}_d^*$ that corresponds to $\V{u}^*$ defined in \As{model}. Our goal is to derive the upper bound of $\lVert \Vhat{u} - \V{u}^* \rVert_2^2$ by above reparametrization. Since $\hat{\V{\alpha}}_d, \hat{\V{\beta}}_d$, $d=1, \ldots, D$ minimizes the objective function in \Eqn{newobj}, we have
\begin{eqnarray*}
&& \sum_{d=1}^D \left\{ \frac{1}{2D} \lVert \V{x} - \M{W}_d \hat{\V{\alpha}}_d + \M{V}_d\hat{\V{\beta}}_d \rVert_{2}^2 +  \gamma_d R(\M{G}_{d}\hat{\V{\beta}}_d) \right\}\\ 
&&~~ \le \amp  \sum_{d=1}^D \left\{ \frac{1}{2D} \lVert \V{x} - \M{W}_d \V{\alpha}_d^* + \M{V}_d\V{\beta}_d^* \rVert_{2}^2 +  \gamma_d R(\M{G}_{d}\V{\beta}_d^*) \right\}.
\end{eqnarray*}
Note that $\lVert \V{x} - \Vhat{u} \rVert_{2}^2 - \lVert \V{x} - \V{u}^* \rVert_{2}^2 = \lVert \Vhat{u} \rVert_{2}^2 - \lVert \V{u}^* \rVert_{2}^2 - 2 \V{x}\Tra(\Vhat{u} - \V{u}^*) = \lVert \Vhat{u} - \V{u}^* \rVert_{2}^2 + 2 \V{\epsilon}\Tra(\Vhat{u} - \V{u}^*)$, where the last equality is due to the model assumption $\V{x} = \V{u}^* + \V{\epsilon}$. Therefore, we have
\begin{eqnarray}
\label{eq:finalbound}
&&\frac{1}{2}\lVert \Vhat{u} - \V{u}^* \rVert_2^2 + \sum_{d=1}^D\gamma_d R(\M{G}_{d}\hat{\V{\beta}}_d) \amp \le \amp \frac{1}{2D} \sum_{d=1}^D \V{\epsilon}\Tra( \V{u}^* - \Vhat{u})  + \sum_{d=1}^D\gamma_d R(\M{G}_{d}\V{\beta}^*_d)\nonumber\\
&&~~\le \amp \frac{1}{2D} \sum_{d=1}^D \underbrace{ \left| \V{\epsilon}\Tra[\M{W}_d (\V{\alpha}_d^* - \hat{\V{\alpha}}_d)  +  \M{V}_d(\V{\beta}_d^* - \hat{\V{\beta}}_d)]\right| }_{f(\hat{\V{\alpha}}_d, \hat{\V{\beta}}_d)}  + \sum_{d=1}^D\gamma_d R(\M{G}_{d}\V{\beta}^*_d). 
\end{eqnarray}

Next we derive the bound for $f(\hat{\V{\alpha}}_d, \hat{\V{\beta}}_d)$. Note that the optimization over $\V{\alpha}_d$ in \Eqn{newobj} has a closed-form since the penalty term is independent of $\V{\alpha}_d$. In particular, by setting the derivative of $\lVert \V{x} - \M{W}_d \V{\alpha}_d + \M{V}_d\V{\beta}_d \rVert_{2}^2$ with respect to $\V{\alpha}_d$ to be zero, we obtain that $\V{\alpha}_d = \M{W}_d\Tra(\V{x} - \M{V}_d\V{\beta}_d)$. This implies that
\begin{eqnarray}
\label{eq:alphahat}
\hat{\V{\alpha}}_d & = & \M{W}_d\Tra(\V{x} - \M{V}_d\hat{\V{\beta}}_d) \nonumber \\
& = & \M{W}_d\Tra(\M{W}_d \V{\alpha}_d^* + \M{V}_d\V{\beta}_d^* + \V{\epsilon} - \M{V}_d\hat{\V{\beta}}_d) \\
& = & \V{\alpha}_d^* + \M{W}_d\Tra\V{\epsilon}, \nonumber
\end{eqnarray}
where the second equality is due to $\V{x} = \V{u}^* + \V{\epsilon}$ and the last equality is due to the fact that $\M{W}_d\Tra \M{V}_d = 0$ and $\M{W}_d\Tra\M{W}_d = \M{I}$. According to \Eqn{alphahat}, we have
\begin{eqnarray}
\label{eq:f}
f(\hat{\V{\alpha}}_d, \hat{\V{\beta}}_d) & = & \left|  \V{\epsilon}\Tra\M{W}_d\M{W}_d\Tra\V{\epsilon} +  \V{\epsilon}\Tra \M{V}_d(\V{\beta}_d^* - \hat{\V{\beta}}_d)]  \right| \nonumber \\
& \le &   \underbrace{\left|  \V{\epsilon}\Tra\M{W}_d\M{W}_d\Tra\V{\epsilon} \right|}_{(I)} +    \underbrace{\left|\V{\epsilon}\Tra \M{V}_d(\V{\beta}_d^* - \hat{\V{\beta}}_d)]  \right|}_{(II)}.
\end{eqnarray}

{\bf Bound $(I)$:} We apply the concentration inequality in \Lem{VuWang2015} to bound $(I)$. It remains to compute $\lVert \M{W}_d\M{W}_d\Tra\rVert_{\text{F}}^2$ and $\lVert \M{W}_d\M{W}_d\Tra \rVert_2$. By construction, $\M{W}_d\M{W}_d\Tra \in \Real^{n  \times n }$ is a projection matrix since $\tilde{\M{V}}_d\tilde{\M{V}}_d\Tra = \M{W}_d\M{W}_d\Tra + \M{V}_d\M{V}_d\Tra = \M{I}$. Therefore, the rank of $\M{W}_d\M{W}_d\Tra$ is $\prod_{j\ne d} n_j$, $\lVert \M{W}_d\M{W}_d\Tra\rVert_{\text{F}}^2 = \prod_{j\ne d} n_j, \lVert \M{W}_d\M{W}_d\Tra \rVert_2 = 1$, and $\tr(\M{W}_d\M{W}_d\Tra) = \prod_{j\ne d} n_j$. 

Denote $n = \prod_{d=1}^D n_d$. By \Lem{VuWang2015} and \As{model}, we have 
\begin{eqnarray*}
\mathbb P\Big( \V{\epsilon}\Tra\M{W}_d\M{W}_d\Tra\V{\epsilon} \ge t +  n_{-d} \Big) & \le & C \log(n) \exp\left\{ - C^{'}M^{-2} \min\left[\frac{t^2}{\log(n) n_{-d} }, t  \right]  \right\}.
\end{eqnarray*}
Setting $t = \sqrt{ n_{-d} \log(n)^2}$, we have
\begin{eqnarray}
\label{eq:bound_I}
\mathbb P\left( \V{\epsilon}\Tra\M{W}_d\M{W}_d\Tra\V{\epsilon} \ge \log(n)\sqrt{n_{-d}} + n_{-d} \right) & \le & C \exp\left\{ \log\log(n) - C^{'}M^{-2}\log(n)  \right\}, 
\end{eqnarray}
where the right hand side converges to zero as the dimension $n = \prod_{d=1}^D n_d \rightarrow \infty$. Note that our error $\V{\epsilon}$ in \As{model} is assumed to be a $M$-concentrated random variable. If we assume a stronger condition such that $\V{\epsilon}$ is a vector with iid sub-Gaussian, we can obtain a upper bound $\sqrt{ \log(n)n_{-d}} +  n_{-d}$ according to the Hanson and Wright inequality \citep{HW71}. Therefore, in spite of the relaxation in the error assumption, our bound in \Eqn{bound_I} is only up to a log-term larger.

{\bf Bound $(II)$:} By definitions of $\M{G}_{d}$ in \Eqn{Gk} and $\M{G}_{d}^{\dagger}$, we have $\M{G}_{d}^{\dagger}\M{G}_{d} = \M{I}$. Furthermore, let $\M{G}_{d, ij}^{\dagger}$ refer to the column of $\M{G}_{d}^{\dagger}$ that corresponds to the index $(i,j)$, and let $\M{G}_{d, ij}$ refer to the row of $\M{G}_{d}$ that corresponds to the index $(i,j)$. We have
\begin{eqnarray*}
(II) & = & \left|\V{\epsilon}\Tra \M{V}_d(\V{\beta}_d^* - \hat{\V{\beta}}_d) \right| \amp = \amp \left|\V{\epsilon}\Tra\M{V}_d \M{G}_{d}^{\dagger}\M{G}_{d}(\V{\beta}_d^* - \hat{\V{\beta}}_d)  \right| \amp = \amp  \left| \sum_{i<j}\V{\epsilon}\Tra \M{V}_d\M{G}_{d,ij}^{\dagger}\M{G}_{d,ij}(\V{\beta}_d^* - \hat{\V{\beta}}_d)  \right|\\
& \le &  \sum_{i<j} \lVert \V{\epsilon}\Tra\M{V}_d \M{G}_{d,ij}^{\dagger} \rVert_2 \lVert \M{G}_{d,ij}(\V{\beta}_d^* - \hat{\V{\beta}}_d) \rVert_2 \amp \le \amp  \underbrace{\max_{i<j} \lVert \V{\epsilon}\Tra\M{V}_d \M{G}_{d,ij}^{\dagger} \rVert_2}_{II_1} \cdot \sum_{i<j} \lVert \M{G}_{d,ij}(\V{\beta}_d^* - \hat{\V{\beta}}_d) \rVert_2
\end{eqnarray*}

{\bf Bound $II_1$:} By construction, $\V{\epsilon}\Tra\M{V}_d \M{G}_{d,ij}^{\dagger} \in \Real^{n_{-d}}$. We have 
\begin{eqnarray*}
\lVert \V{\epsilon}\Tra\M{V}_d \M{G}_{d,ij}^{\dagger} \rVert_2 & \le & \sqrt{n_{-d}} \, \lVert \V{\epsilon}\Tra\M{V}_d\M{G}_{d,ij}^{\dagger} \rVert_{\infty},
\end{eqnarray*}
and hence
\begin{eqnarray*}
\max_{i<j}\, \lVert \V{\epsilon}\Tra\M{V}_d \M{G}_{d,ij}^{\dagger} \rVert_2 & \le & \sqrt{n_{-d}} \max_{i<j}\, \lVert \V{\epsilon}\Tra\M{V}_d \M{G}_{d,ij}^{\dagger} \rVert_{\infty} \amp = \amp \sqrt{n_{-d}} \, \lVert \V{\epsilon}\Tra\M{V}_d \M{G}_{d}^{\dagger} \rVert_{\infty}
\end{eqnarray*}
Let $\eta_j = \V{e}_j\Tra \M{G}_{d}^{\dagger \top}\M{V}_d\Tra \V{\epsilon} \in \Real$, where $\V{e}_j  \in \Real^{{n_d \choose 2} n_{-d}}$ is the basis vector with the $j$th entry one and the rest zeros. According to \Lem{propertyG} and the property of $\M{V}_d$ which consists of singular vectors, we have $\sigma_{\max}(\M{V}_d) = 1$ and $\sigma_{\max}(\M{G}_{d}^{\dagger}) = 1/\sqrt{n_d}$. Therefore, we have $\eta_j$ is a $M/\sqrt{n_d}$-concentrated random variable with mean zero. According to the definition of concentrated random variable in \Def{error}, we have
\begin{eqnarray*}
\mathbb P\left(|\eta_j | \ge t_1 \right) & \le & C_1 \exp\left(- \frac{C_2 n_d t_1^2}{M^2}\right).
\end{eqnarray*}
Therefore, by union bound, we have
\begin{eqnarray*}
\mathbb P\left( \max_j \lvert \eta_j \rvert \ge t_1 \right) \le C_1 {n_d \choose 2} \left( n_{-d} \right)\exp\left(- \frac{C_2 n_d t_1^2}{M^2}\right).
\end{eqnarray*}
By setting $t_1 = \sqrt{\log(n) \log \bigg [{n_d \choose 2} n_{-d} \bigg] / n_d}$, we have
\begin{eqnarray*}
\mathbb P\left(\lVert \V{\epsilon}\Tra\M{V}_d \M{G}_{d}^{\dagger} \rVert_{\infty} \ge \sqrt{\log(n) \log\left[{n_d \choose 2} n_{-d}\right] / n_d} \right) & \le & \frac{C_3}{n},
\end{eqnarray*}
for some constant $C_3>0$. Hence with probability at least $1 - C_3/n$, we have
\begin{eqnarray}
\label{eq:bound_II1}
II_1 & \le & \sqrt{n_{-d}\log(n) \log\left [{n_d \choose 2} n_{-d}\right] / n_d}.
\end{eqnarray}

Plugging the results in \Eqn{bound_I} and \Eqn{bound_II1} into \Eqn{f}, we obtain that, for each $d =1, \ldots, D$
\begin{eqnarray*}
f(\hat{\V{\alpha}}_d, \hat{\V{\beta}}_d) & \le & \log(n)\sqrt{ n_{-d}} +  n_{-d} + \sqrt{n_{-d}\log(n) \log\left[{n_d \choose 2} n_{-d}\right] / n_d} \sum_{i<j} \lVert \M{G}_{d,ij}(\V{\beta}_d^* - \hat{\V{\beta}}_d) \rVert_2.
\end{eqnarray*}
Therefore, \As{tuning} on the tuning parameter $\gamma_d$ implies that
\begin{eqnarray*}
f(\hat{\V{\alpha}}_d, \hat{\V{\beta}}_d) & \le & \log(n)\sqrt{ n_{-d}} +  n_{-d} + D \gamma_d \sum_{d=1}^D\gamma_d \sum_{i<j} \lVert \M{G}_{d,ij}(\V{\beta}_d^* - \hat{\V{\beta}}_d) \rVert_2,
\end{eqnarray*}
by noting that $\log({n_d \choose 2} n_{-d}) \le \log(n_d^2 n_{-d}) \le 2\log( n )$. This combines with the inequality in \Eqn{finalbound} lead to
\begin{eqnarray}
\label{eq:bound_u}
\frac{1}{2} \lVert \Vhat{u} - \V{u}^* \rVert_2^2 & \le & \frac{1}{2D}\sum_{d=1}^D \bigg[ \log(n)\sqrt{ n_{-d}} +  n_{-d} \bigg] + \frac{3}{2} \sum_{d=1}^D \gamma_d R(\M{S}_d \V{u}^*).
\end{eqnarray}

According to the cluster structure assumption in \As{model}, there are $k_d$ clusters along the $d$th mode of the tensor. Therefore, along each mode the true parameter $\T{U}^*$ only has a few different slices. Denote $\T{U}^*_{\cdots i \cdots}$ as the $i$-th mode-$d$ subarray. Formally, we have
\begin{eqnarray}
\label{eq:bound_r}
R(\M{S}_d \V{u}^*) & = & \sum_{(i,j), i < j, i,j=1,\ldots, n_d} \lVert \M{A}_{d,ij} \V{u} \rVert_2 \nonumber \\
&=& \sum_{(i,j), i < j, i,j=1,\ldots, n_d} \lVert  \T{U}^*_{\cdots i \cdots} -  \T{U}^*_{\cdots j \cdots} \rVert_{\text{F}} \amp \le \amp 4C_0^2 {n_d \choose 2} \sqrt{\prod_{j\ne d} k_j},
\end{eqnarray}
where $C_0$ is a constant upper bound for the entries of $\T{U}^*$. Combining the inequalities in \Eqn{bound_u} and \Eqn{bound_r} with the condition on $\gamma_d$ given in \As{tuning} implies that
\begin{eqnarray*}
\frac{1}{2} \lVert \Vhat{u} - \V{u}^* \rVert_2^2 
&\le& \frac{1}{2D}\sum_{d=1}^D \left( \log(n)\sqrt{ n_{-d}} +  n_{-d} \right) + \frac{3}{2} \sum_{d=1}^D  \frac{2c_0 \log(n)\sqrt{n}}{Dn_d} 4C_0^2 {n_d \choose 2} \sqrt{\prod_{j\ne d} k_j}.
\end{eqnarray*}
Dividing both sides by $n$ gives to the prediction error bound in \Eqn{final_error}. This ends the proof of \Thm{final_error}.  \hfill $\blacksquare$

\section{Derivation of Lagrangian Dual}\phantomsection
\label{sec:dual_derivation}

Let $\T{U} \times_d \M{A}$ denote the multiplication of $\T{U}$ along mode $d$ by the matrix $\M{A}$. Recall that for a tensor $\T{U} \in \Real^{n_1 \times \cdots \times n_d}$ and a matrix $\M{A} \in \Real^{L \times n_d}$
\begin{eqnarray*}
\vec (\T{U} \times_d \M{A}) 
& = & (\M{I}_{n_D} \Kron \cdots \Kron \M{I}_{n_{d+1}}\Kron \M{A} \Kron \M{I}_{n_{d-1}} \Kron \cdots \Kron \M{I}_{n_1} )\V{u},
\end{eqnarray*}
where $\V{u} = \vec(\T{u}) = \vec (\Mz{U}{1})$, namely the column-major vectorization of the mode-1 matricization of the tensor $\T{u}$. So,
Note that $\T{Y} = \T{U} \times_d \M{A}$ is equivalent to $\Mz{Y}{d} = \M{A}\Mz{U}{d}$. We rewrite the penalty function $R_d$ as follows. 

\begin{eqnarray*}
R_d(\T{U}) & = & \sum_{l \in \mathcal{E}_d}w_{d,l} \lVert \T{U} \times_d \V{\Delta}_{d,l} \rVert_{\text{F}}
\amp = \amp \sum_{l \in \mathcal{E}_d}w_{d,l} \lVert \vec(\T{U} \times_d \V{\Delta}_{d,l}) \rVert_{2} 
\amp = \amp \sum_{l \in \mathcal{E}_d}w_{d,l} \lVert \M{A}_{d,l}\V{u} \rVert_{2},
\end{eqnarray*}
where $\M{A}_{d,l} = (\M{I}_{n_D} \Kron \cdots \Kron \M{I}_{n_{d+1}}\Kron \V{\Delta}_{d,l}\Kron \M{I}_{n_{d-1}} \Kron \cdots \Kron \M{I}_{n_1} )$.

We now write down the Lagrangian:
\begin{eqnarray*}
\mathcal{L}(\V{u}, \V{v}, \V{\lambda})
& = & \frac{1}{2} \lVert \V{x} - \V{U} \rVert_{2}^2 +  \sum_{d=1}^D \sum_{l \in \mathcal{E}_d} \bigg\{\gamma \VE{w}{d,l}\lVert \V{v}_{d,l} \rVert_2 + \langle \V{\lambda}_{d,l}, \M{A}_{d,l} \V{u} - \V{v}_{d,l} \rangle \bigg \} \\
& = &  \bigg \{\frac{1}{2} \lVert \V{x} - \V{U} \rVert_{2}^2 + \sum_{d=1}^D \langle \M{A}_d\Tra\V{\lambda}_d, \V{u} \rangle \bigg \}
 -  
\sum_{d=1}^D \sum_{l \in \mathcal{E}_d} \bigg \{\langle \V{\lambda}_{d,l}, \V{v}_{d,l} \rangle -   \gamma\VE{w}{d,l}\lVert \V{v}_{d,l} \rVert_2 \bigg \} \\
 & = & \bigg \{\frac{1}{2} \lVert \V{x} - \V{U} \rVert_{2}^2 +  \langle \M{A}\Tra\V{\lambda}, \V{u} \rangle \bigg\}
 -  
\sum_{d=1}^D \sum_{l \in \mathcal{E}_d} \bigg \{\langle \V{\lambda}_{d,l}, \V{v}_{d,l} \rangle -   \gamma\VE{w}{d,l}\lVert \V{v}_{d,l} \rVert_2 \bigg \}.
\end{eqnarray*}

The Lagrangian dual objective is given by $G(\V{\lambda})$ by minimizing the Lagrangian $\mathcal{L}(\V{u}, \V{v}, \V{\lambda})$ over the primal variables $\V{u}$ and $\V{v}$, namely
\begin{eqnarray}
\label{eq:dual_expanded}
G(\V{\lambda}) & = &  \underset{\V{u},\V{v}}{\min}\; \mathcal{L}(\V{u}, \V{v}, \V{\lambda}) \nonumber \\
& = &  \underset{\V{u}}{\min}  \bigg \{\frac{1}{2} \lVert \V{x} - \V{U} \rVert_{2}^2 +  \langle \M{A}\Tra\V{\lambda}, \V{u} \rangle \bigg \}
 -  
\sum_{d=1}^D \sum_{l \in \mathcal{E}_d}
  \underset{\V{v}_{d,l}}{\max}\bigg \{
\langle \V{\lambda}_{d,l}, \V{v}_{d,l} \rangle -   \gamma\VE{w}{d,l}\lVert \V{v}_{d,l} \rVert_2 \bigg \} \nonumber \\
& = &
\frac{1}{2} \lVert \V{x} \rVert_2^2- \frac{1}{2} \lVert \V{x} -  \M{A}\Tra\V{\lambda} \rVert_{2}^2 
- \sum_{d=1}^D \sum_{l \in \mathcal{E}_d} \iota_{C_{d,l}}(\V{\lambda}_{d,l}),
\end{eqnarray}
where $\iota_{C_{d,l}}$ is the indicator function of the closed convex set $C_{d,l} = \{\V{z} : \lVert \V{z} \rVert_2 \leq \gamma w_{d,l} \}$.

The last equality in \Eqn{dual_expanded} follows from the fact that the Fenchel conjugate of a norm is the indicator function of the unit dual norm ball. Recall that the Fenchel conjugate $f^\star$ of a function $f$ is given by
\begin{eqnarray*}
f^\star(\V{\lambda}) & = & \underset{\V{v}}{\sup} \bigg \{ \langle \V{\lambda}, \V{v} \rangle - f(\V{v}) \bigg \}.
\end{eqnarray*}
Let $B = \{\V{\lambda} : \lVert \V{\lambda} \rVert_2 \leq 1 \}$ denote the unit $\ell_2$-norm ball. Since the $\ell_2$-norm is self dual, we arrive at the identity
\begin{eqnarray*}
\iota_B(\V{\lambda}) & = & \underset{\V{v}}{\sup} \bigg \{ \langle \V{\lambda, \V{v}} \rangle - \lVert \V{v} \rVert_2 \bigg\}.
\end{eqnarray*}

\section{Projected Gradient Applied to the Lagrangian Dual}\phantomsection
\label{sec:projected_gradient}

Note that the dual problem \Eqn{dual_problem} has the form 
\begin{equation}
\label{eq:constrained_smooth}
\begin{split}
\text{minimize}&\; g(\V{\lambda}) \\
\text{subject to}&\; \V{\lambda} \in C,
\end{split}
\end{equation}
where $g(\V{\lambda})$ is a convex and Lipschitz-differentiable function and the constraint set $C$ is a closed convex set, which implies that every point $\V{\lambda}$ possesses a unique orthogonal projection, $\mathcal{P}_C(\V{\lambda}) = \arg\min_{\V{\theta}\in C} \lVert \V{\theta} - \V{\lambda} \rVert_2$, onto $C$. When $\mathcal{P}_C(\V{\lambda})$ can be computed analytically, a simple and effective iterative algorithm for solving problems like \Eqn{constrained_smooth} is the projected gradient descent algorithm, a special case of proximal gradient descent algorithm \citep{ComWaj2005, combettes2011proximal}. Recall that projected gradient descent alternates between taking a gradient step and projecting onto the set $C$. Thus, at the $m$th iteration, we perform the following update
\begin{eqnarray}
\label{eq:projected_gradient}
\Vn{\lambda}{m} & = & \mathcal{P}_C\left (\Vn{\lambda}{m-1} - \eta \nabla g(\V{\lambda}) \right),
\end{eqnarray}
where $\eta$ is a step-length parameter.

Applying the update rule in \Eqn{projected_gradient} to the dual problem \Eqn{dual_problem}, we obtain the following rule for computing the $m$th iteration
\begin{eqnarray*}
\Vn{u}{m}  &= & \V{x} - \M{A}\Tra\Vn{\lambda}{m-1} \\
\Vn{\lambda}{m} & = & \mathcal{P}_C\left (\Vn{\lambda}{m-1} + \eta \M{A}\Vn{u}{m} \right).
\end{eqnarray*}
Note that, at the $m$th iteration, the gradient of the least squares objective in \Eqn{dual_problem} is given by $-\M{A}\Vn{u}{m}$. Thus, we automatically update our CoCo estimator $\Vn{u}{m}$ as part of our gradient calculation. Finally, we note that the projection onto the set $C$ consists of independent projections onto the sets $C_{d,l}$ that can be carried out in parallel.

\subsection{Per-Iteration and Storage Costs}

The gradient update is dominated by the matrix-vector multiplications $\M{A}\Tra\V{\lambda}$ and $\M{A}\V{u}$. Although $\M{A}$ is a $\sum_{d=1}^D \lvert \E_d \rvert n_{-d}$-by-$n$ matrix it has only $2 \sum_{d=1}^D \lvert \E_d \rvert n_{-d}$ non-zero elements. Thus, computing the gradient step requires $\mathcal{O}(\sum_{d=1}^D \lvert \E_d \rvert n_{-d})$ flops. Projecting onto the set $C$ also requires $\mathcal{O}(\sum_{d=1}^D \lvert \E_d \rvert n_{-d})$ flops since projecting onto the set $C_{d,l}$ requires $\mathcal{O}(n_{-d})$ flops. Thus, the per-iteration cost is $\mathcal{O}(\sum_{d=1}^D \lvert \E_d \rvert n_{-d})$ flops. The storage cost is dominated by storing the dual variable $\V{\lambda}$, which has $\sum_{d=1}^D \lvert \E_d \rvert n_{-d}$ elements. At first glance these storage and per-iteration costs may seem prohibitive, as $\lvert \E_d \rvert$ can be as large as $\mathcal{O}(n_d^2)$ for a fully connected mode-$d$ graph. Shrinking together all combinations of pairs of mode-$d$ subarrays, however, typically produces poor clustering results in comparison to shrinking together mode-$d$ subarrays that are nearest-neighbors as observed in prior work in convex clustering \citep{Chen2015, Chi2015} and convex biclustering \citep{ChiAllenBaraniuk2017}. Consequently, we employ sparse weights. Specifically, we keep positive weights between approximately nearest-neighbor mode-$d$ subarrays so that $\lvert \E_d \rvert$ is $\mathcal{O}(n_d)$. By using these sparse weights, the per-iteration and storage costs scale more reasonably as $\mathcal{O}(Dn)$, namely linearly in either the number of dimensions $D$ or in the number of elements $n$. Details on our weights choices are elaborated in \Sec{weights}.

\subsection{Convergence}
\label{sec:convergence}

The sequence of dual iterates $\Vn{\lambda}{m}$ is guaranteed to converge to a solution $\Vhat{\lambda}$ of \Eqn{dual_problem} provided that the step-size parameter $\eta$ is less than twice the reciprocal of the spectral radius of the matrix $\M{A}\Tra\M{A}$ \citep[Theorem 3.4]{ComWaj2005}. Consequently, the sequence of primal iterates $\Vn{u}{m}$ is guaranteed to converge to the CoCo estimator $\Vhat{u}$. We note that under the same step-size conditions, convergence of the sequence $\Vn{u}{m}$ can also be guaranteed by observing that the projected gradient algorithm applied to the dual problem \Eqn{dual_problem} is an example of the alternating minimization algorithm \citep[Proposition 2]{Tseng1991}.

\subsection{Monitoring Convergence via the Duality Gap}

Recall that we can bound the suboptimality of the $m$th iterate, $F_\gamma(\Vn{u}{m}) - F_\gamma(\Vhat{u})$, by the duality gap
$F_\gamma(\Vn{u}{m}) - G(\Vn{\lambda}{m})$, which can be expressed solely in terms of the $m$th iterate of the primal variable $\Vn{u}{m}$, namely
\begin{eqnarray*}
\label{eq:duality_gap}
F_\gamma(\Vn{u}{m}) - G(\Vn{\lambda}{m}) \amp = \amp \lVert \Vn{u}{m} \rVert_2^2 - \langle \V{x}, \Vn{u}{m} \rangle + \gamma \sum_{d=1}^D\sum_{l \in \E_d} w_{d,l} \lVert \M{A}_{d,l}\Vn{u}{m}\rVert_2.
\end{eqnarray*}
For any optimal dual solution $\Vhat{\lambda}$, the gap vanishes, namely $F_\gamma(\Vhat{u}) = G(\Vhat{\lambda})$. Note that computing the duality gap incurs minimal additional cost as $\Vn{u}{m}$ and $\M{A}_{d,l}\Vn{u}{m}$ are already computed as part of the gradient step. In short, including a duality gap computation will not change the $\mathcal{O}(Dn)$ per-iteration cost of the projected gradient algorithm. In practice, we can terminate the algorithm once the duality gap falls below some small tolerance.

\subsection{Computing Mode-$d$ Difference Variables}
\label{sec:compute_v}

In \Sec{partition}, we explained how clustering assignments along the $d$th mode are made using the mode-$d$ difference variables $\V{v}_{d,l} = \T{U} \times_d \V{\Delta}_{d,l}$. In practice we must deal with the fact that the $\Vhat{u}$ recovered by computing $\V{x} - \M{A}\Tra\Vhat{\lambda}$ may exhibit a nearly but not exactly checkerbox structure due to limitations in numerical precision.  This creates a practical issue as a small but non-zero difference variable will lead to an incorrect clustering assignment. Addressing this issue, however, is simple. The projected gradient algorithm used to compute CoCo is a natural generalization of the projected gradient algorithm used in \cite{Chi2015} for convex clustering. Consequently, we can use the obvious adaptation of the procedure for computing the differences variables in convex clustering. The following brief technical discussion is expanded in more detail in \cite{Chi2015}. 

The key fact that we use is that the projected gradient algorithm is equivalent to the alternating minimization algorithm (AMA) applied to the following augmented Lagrangian function
\begin{equation*}
\mathcal{L}_\eta(\V{u},\V{v},\V{\lambda})  =  \frac{1}{2} \lVert \V{x} - \V{u} \rVert_2^2 + \sum_{d = 1}^D\sum_{l \in \mathcal{E}_d} \left [ \gamma w_{d,l} \lVert\V{v}_{d,l} \rVert_2 + \langle \V{\lambda}_{d,l}, \V{v}_{d,l} - \M{A}_{d,l}\V{u} \rangle +\frac{\eta}{2} \lVert \V{v}_{d,l} - \M{A}_{d,l}\V{u} \rVert_2^2 \right].
\end{equation*}
The mode-$d$ difference vector $\V{v}_{d,l}$ is determined by the proximal map
\begin{equation}
\label{eq:update_v}
\begin{split}
\V{v}_{d,l} & \amp = \amp \underset{\V{v}_{d,l}}{\arg\min} \; \frac{1}{2}  \left [\lVert\V{v}_{d,l} - \M{A}_{d,l}\V{u} - \eta^{-1}\V{\lambda}_{d,l} \rVert_2^2 + \frac{\gamma w_{d,l}}{\eta} \lVert \V{v}_{d,l} \rVert_2 \right ] \\
 & \amp = \amp \prox_{\sigma_{d,l} \lVert \cdot \rVert_2}\left(\M{A}_{d,l}\V{u} - \eta^{-1}\V{\lambda}_{d,l}\right),
\end{split}
\end{equation}
where $\sigma_{d,l} = \gamma w_{d,l}/\eta$. Because the proximal mapping can produce mode-$d$ difference variables that are {\em exactly} zero, the procedure for computing $\V{v}_{d,l}$ in \Eqn{update_v} is immune to the numerical precision issues that hinder the direct computation $\That{U} \times_d \V{\Delta}_{d,l}$.

\section{Details on Denoising with the Tucker Decomposition for Setting Weights}\phantomsection
\label{sec:weight_extra_discussion}

Employing the Tucker decomposition introduces another tuning parameter, namely the rank of the decomposition.  When applicable, a user can leverage problem-specific knowledge to select the rank for the decomposition.  Nonetheless, the availability of an automatic approach is desirable to handle cases when such knowledge is unavailable.  Selecting the rank in a tensor decomposition, however, is an open question \citep{KoldaBader2009, YokotaLee2017}. During initial experiments, a few different methods for selecting the Tucker decomposition rank from the literature were compared: an $L$-curve approach that attempts to strike a balance between the decomposition's relative error and compression ratio, as implemented by the {\tt mlrankest} function in the {\tt Tensorlab} \textsc{Matlab} toolbox \citep{tensorlab}, minimum description length \citep{rissanen1978modeling, YokotaLee2017}, and the recently-proposed SCORE algorithm \citep{YokotaLee2017}. Out of these, the SCORE algorithm produced the best average CoCo estimator performance.  The SCORE algorithm itself includes a tuning parameter, $\hat{\rho}$, and \cite{YokotaLee2017} suggest setting $\hat{\rho} \in [10^{-4}, 10^{-2}]$.  We considered $\hat{\rho} \in \{10^{-4}, 10^{-3}, 10^{-2}\}$ and found $10^{-3}$ to perform the best, which also matches the value used in the experiments by \cite{YokotaLee2017}. 

We also developed a simple yet effective heuristic for choosing the rank where we set the Tucker rank for the $d$th mode to be the floor of $\sqrt{n_d}/2$.   Two principles motivating the heuristic are that the rank of the decomposition should be both small relative to and also in proportion to the length of the modes. Both the SCORE algorithm and our heuristic were employed in our simulations described in \Sec{simulations} as a robustness check to ensure our CoCo estimator's performance does not crucially depend on the choice of the rank.

The basic Tucker decomposition computation is accomplished by the higher order SVD (HOSVD) method \citep{delathauwer2000multilinear} which computes for each mdoe $k$ the $r_k$ leading left singular values of the mode-$k$ matricization and stores them as a factor matrix $\M{U}_k$. The HOSVD then computes the core tensor by contracting the data tensor $\T{X} \times_k \M{U}_k$. Thus, the main cost is computing $D$ SVDs. This is an illustrative calculation, however, and more efficient alternatives exist \citep{Vannieuwenhoven2012, MinsterSaibabaKilmer2020}.

\section{CPD+$k$-means}\phantomsection
\label{sec:cpdkmeans}

We describe in greater detail the CPD+$k$-means method for co-clustering a $D$-way tensor $\T{X} \in \Real^{n_1 \times \cdots \times n_D}$. The method consists of two steps

\begin{itemize}
    \item[Step 1.] Compute a rank-$R$ CP decomposition
    \begin{eqnarray*}
	    \T{X} & \approx & \sum_{i = 1}^R \Vn{a}{1}_i  \circ \Vn{a}{2}_i \circ \cdots \circ \Vn{a}{D}_i,
    \end{eqnarray*}
where $\circ$ represents the outer product and $\Vn{a}{d}_i$ is the $i$th column of the $d$th factor matrix $\Mn{A}{d} \in \Real^{n_d \times R}$.
    \item[Step 2.] For each factor matrices $\Mn{A}{d}$, apply $k$-means clustering on the $n_d$ rows of $\Mn{A}{d}$. Note that the $D$ applications of $k$-means are done independently for each mode-$d$ factor matrix $\Mn{A}{d}$. 
\end{itemize}

\noindent {\bf Tuning parameters:} There are two sets of tuning parameters: (i) the rank parameter $R$, used in Step 1, and (ii) the $D$ cluster number parameters for each factor matrix, used in Step 2. To choose the rank parameter $R$, we create a candidate set of ranks 
$\mathcal{R}_{\text{candidate}} \subset \{1, 2, 3, \ldots\}$ and select $R^\star \in \mathcal{R}_{\text{candidate}}$ using the tuning procedure in \cite{sun2015provable}. We then compute a CP decomposition using the selected rank $R^\star$ and obtain the factor matrices $\Mn{A}{d}$ for $d = 1, \ldots, D$. To choose the $D$ cluster number parameters, we create $D$ candidate sets of cluster numbers $\mathcal{K}^{d}_{\text{candidate}} \subset \{1, 2, 3, \ldots, n_d\}$ and select $k_d^\star \in \mathcal{K}^{d}_{\text{candidate}}$ for $d = 1, \ldots, D$ using the gap statistic procedure in \cite{TibshiraniWaltherHastie2001}. We use the $D$ clustering results from running $k$-means on the rows of each of the $\Mn{A}{d}$ using $k_d^\star$.

\section{Additional Simulations on Rectangular Tensors}\phantomsection
\label{sec:simulations_rectangular_extra}

\begin{figure}[t]
    \centering
    \subfloat[Adjusted Rand Index, Mode 1]{\label{fig:checkerCube_nonCube_101050_202050_ARI_mode1}
    \includegraphics[scale=0.45]{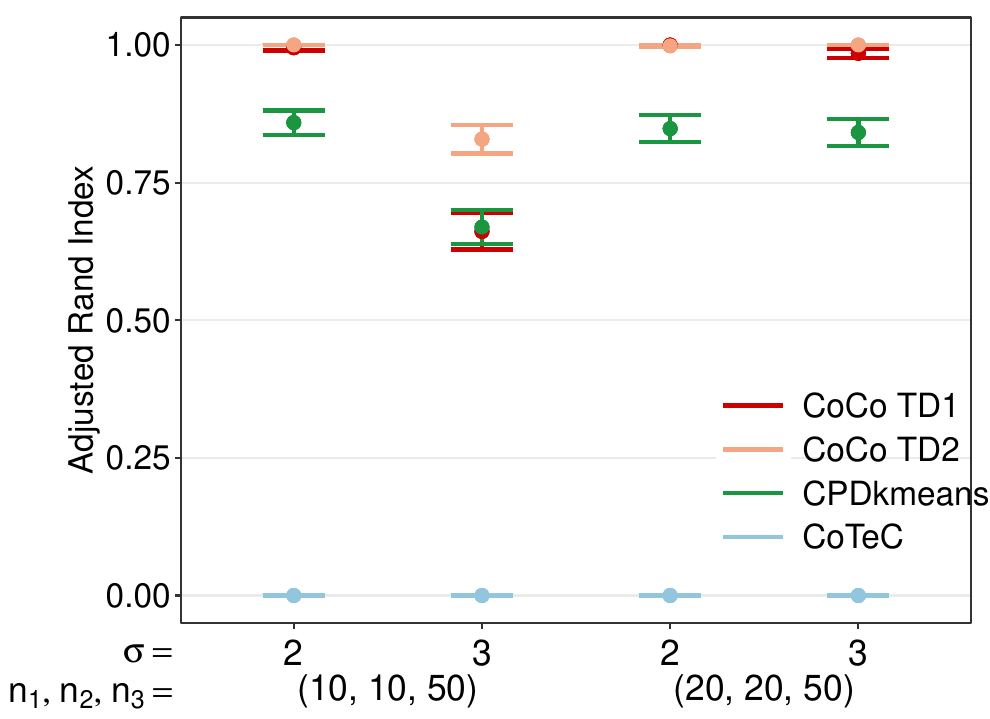}}
    \subfloat[Adjusted Rand Index, Mode 2]{\label{fig:checkerCube_nonCube_101050_202050_ARI_mode2}
    \includegraphics[scale=0.45]{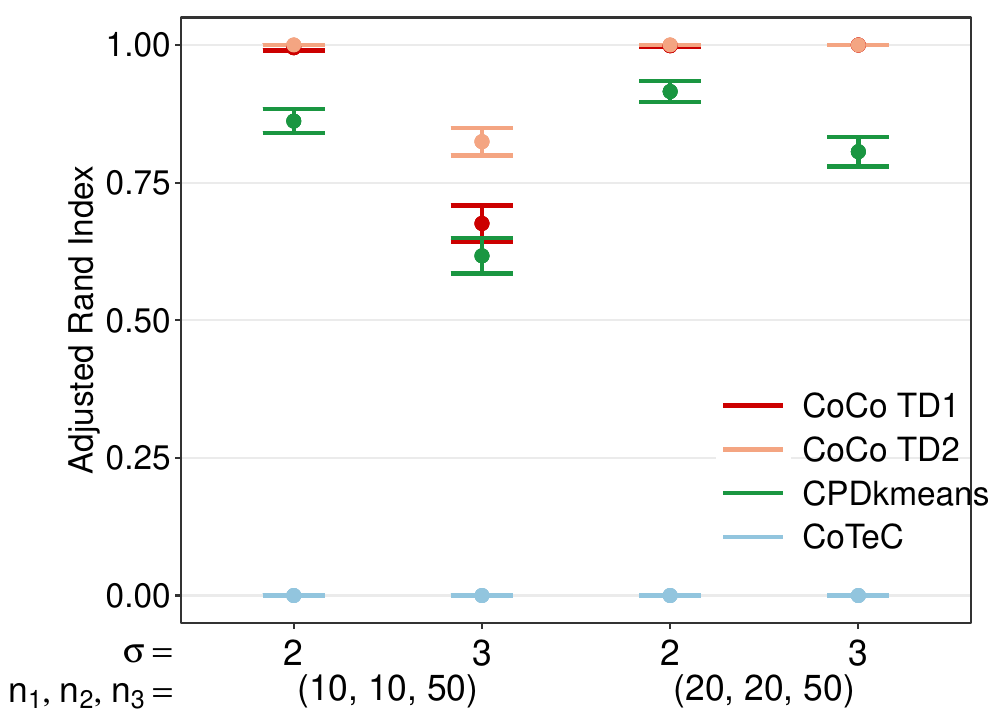}} \\
    \subfloat[Adjusted Rand Index, Mode 3]{\label{fig:checkerCube_nonCube_101050_202050_ARI_mode3}
    \includegraphics[scale=0.45]{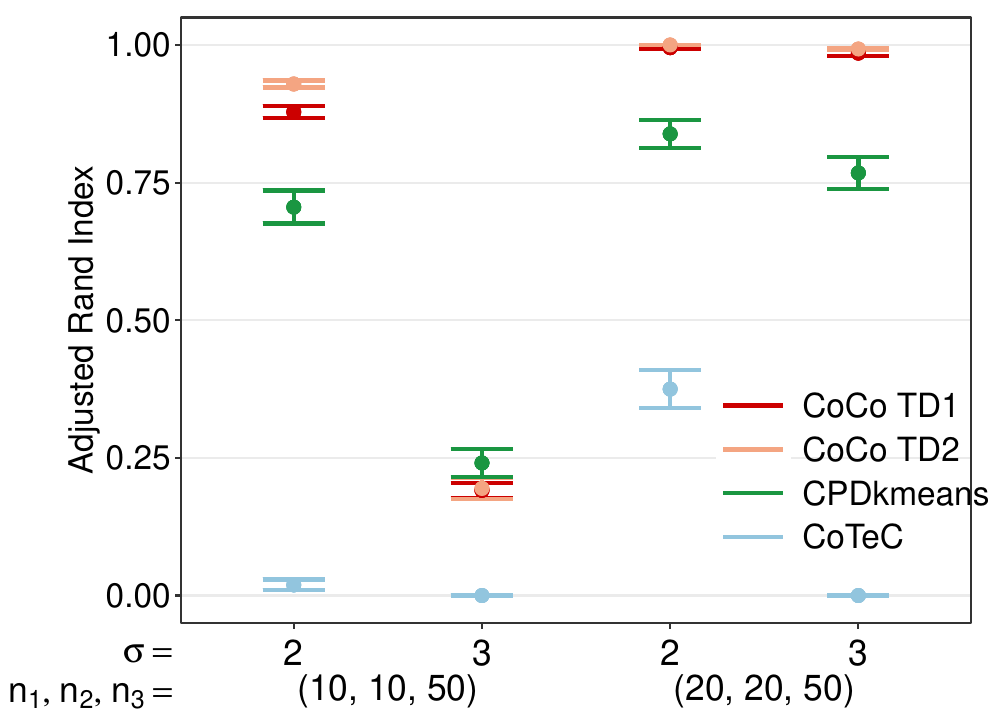}}
    \subfloat[Adjusted Rand Index, Co-clusters]{\label{fig:checkerCube_nonCube_101050_202050_ARI_triclust}
    \includegraphics[scale=0.45]{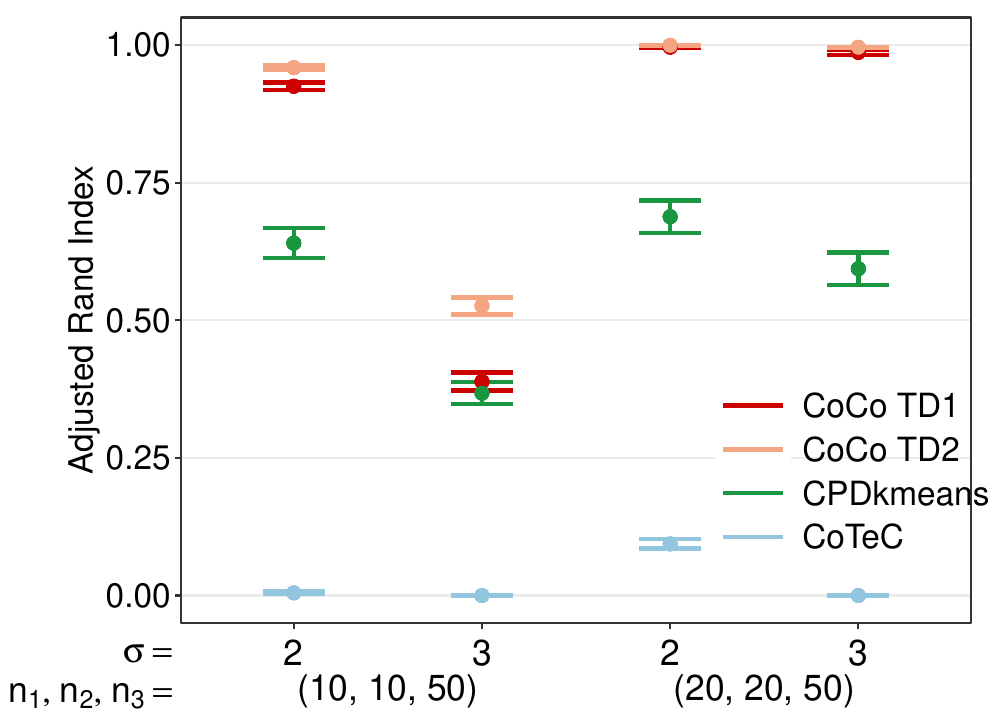}}
    \caption{Checkerbox Simulation Results:  Impact of Tensor Shape.  \small Two balanced clusters per mode with two levels of homoskedastic noise for a tensor with two short modes and one longer mode. \small Average adjusted rand index plus/minus one standard error for different noise levels and mode lengths.} \label{fig:nonCube_101050_202050}
\end{figure}

The first rectangular tensor is one in which there are two short modes ($n_1 = n_2 = 10$) and one relatively longer mode ($n_3 = 50$). \Fig{nonCube_101050_202050} presents the clustering results for this tensor shape.

At a lower noise level ($\sigma = 2$), CoCo performs very well and outperforms CPD+$k$-means and CoTeC in terms of both single-mode clustering and co-clustering.  When the noise level is bumped up ($\sigma = 3$), both methods experience a noticeable drop off in their performance and now perform more similarly.  Interestingly, CoCo's single-mode clustering results are better along the two shorter modes (modes 1 and 2), which is not what we expected.  This provides some evidence that the performance along a mode depends on both the length of that mode as well as the lengths of the other modes.  When the length of the shorter modes are increased slightly (from $n_d = 10$ to $n_d= 20$ for $d =1, 2$), CoCo has near-perfect performance while CPD+$k$-means performs roughly the same as before.  Thus, CoCo struggles with this tensor shape only when the short modes are really short (only 10 observations).

\begin{figure}[t]
    \centering
    \subfloat[Adjusted Rand Index, Mode 1]{\label{fig:checkerCube_nonCube_1050100_2050100_ARI_mode1}
    \includegraphics[scale=0.45]{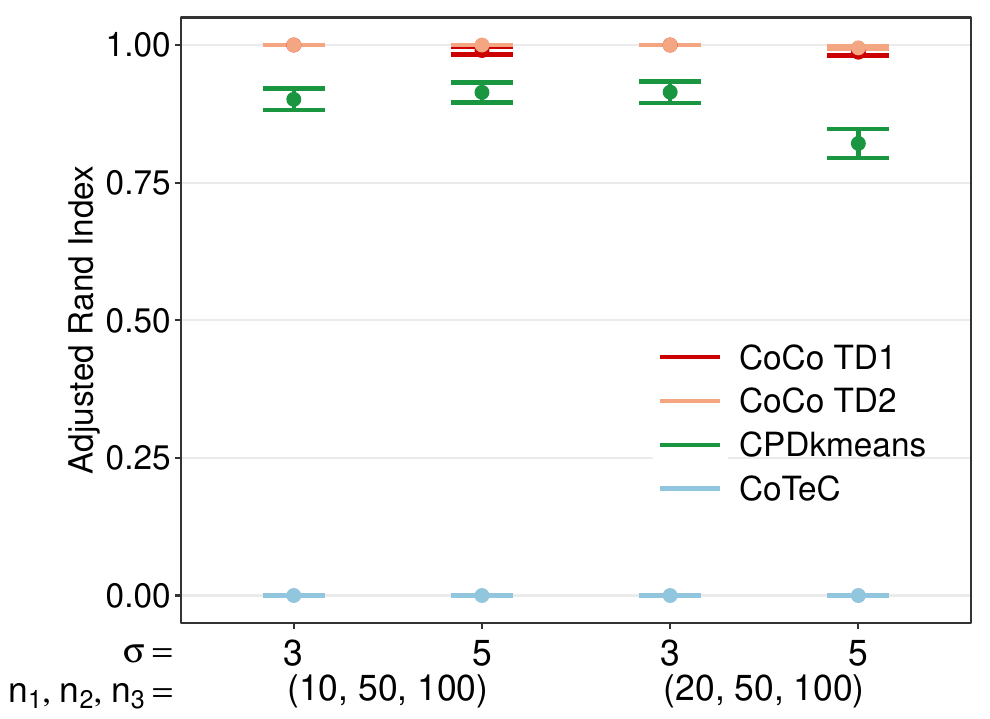}}
    \subfloat[Adjusted Rand Index, Mode 2]{\label{fig:checkerCube_nonCube_1050100_2050100_ARI_mode2}
    \includegraphics[scale=0.45]{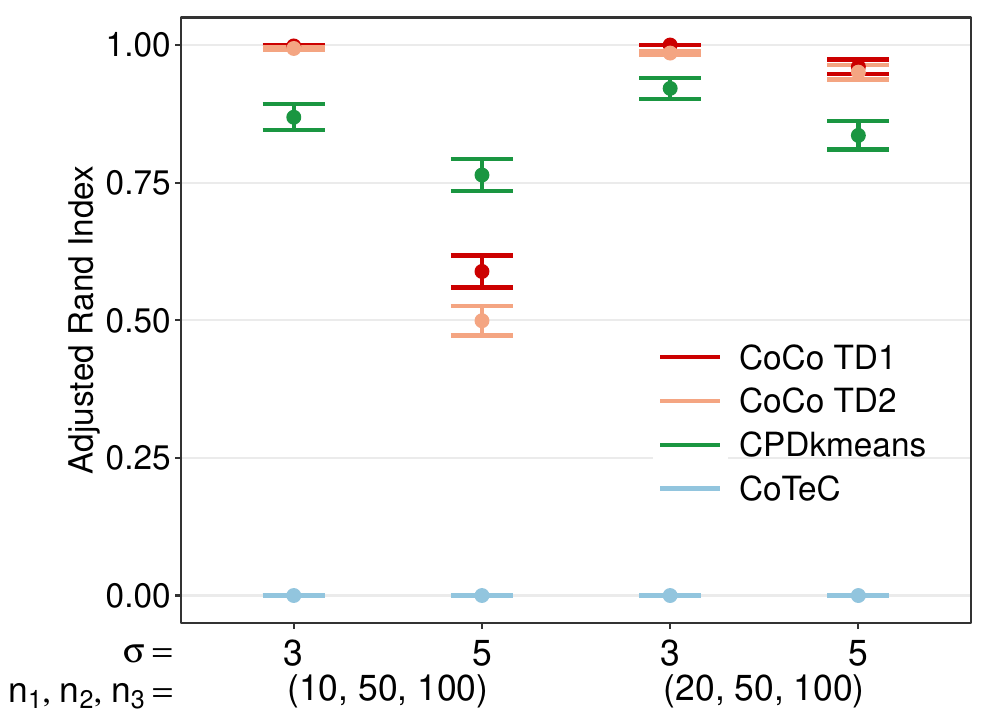}} \\
    \subfloat[Adjusted Rand Index, Mode 3]{\label{fig:checkerCube_nonCube_1050100_2050100_ARI_mode3}
    \includegraphics[scale=0.45]{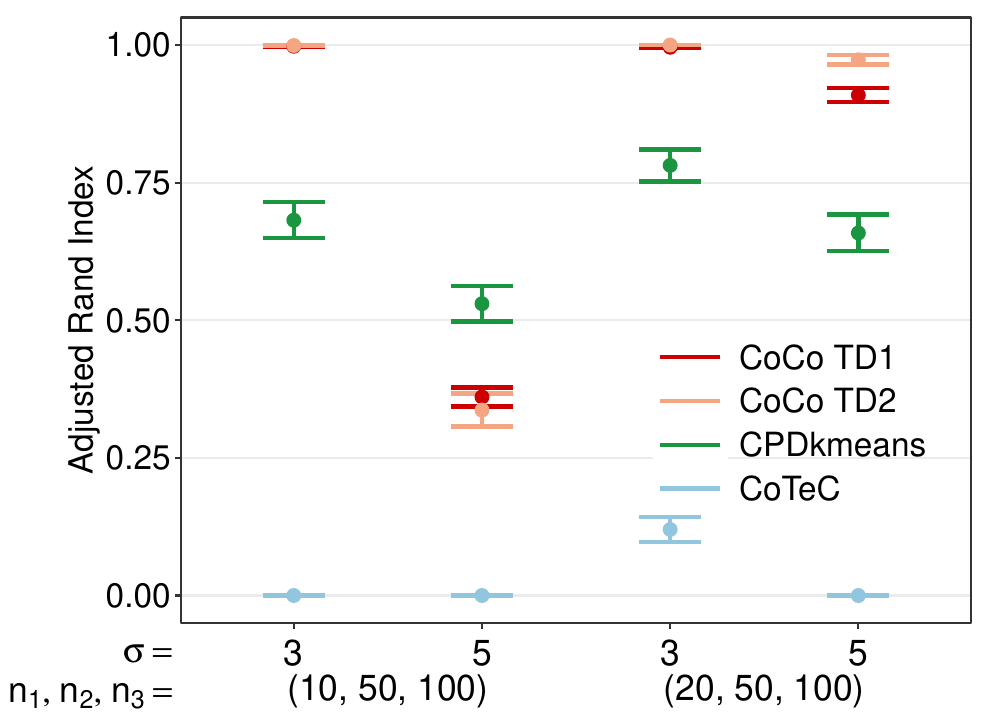}}
    \subfloat[Adjusted Rand Index, Co-clusters]{\label{fig:checkerCube_nonCube_1050100_2050100_ARI_triclust}
    \includegraphics[scale=0.45]{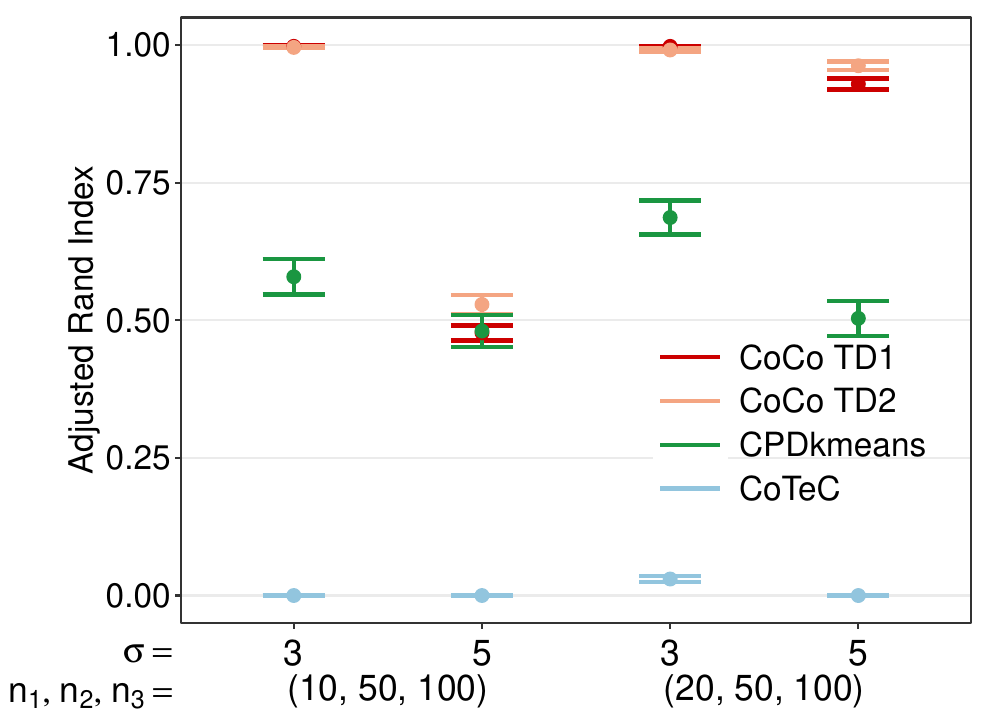}}
    \caption{Checkerbox Simulation Results:  Impact of Tensor Shape.  \small Two balanced clusters per mode with two levels of homoskedastic noise for a tensor with short, medium, and long mode lengths.  Average adjusted rand index plus/minus one standard error for different noise levels and mode lengths. } \label{fig:nonCube_1050100_2050100}
\end{figure}

To further investigate the mode-by-mode performance with rectangular tensors, we also apply the clustering methods to a ``Goldilocks'' tensor with mode lengths that are short, medium, and long.   This setting was again motivated by the results from the previous two tensor shapes to see how the performance is impacted when the size of a longer mode is increased.  The ARI results for this tensor shape are given in \Fig{checkerCube_nonCube_1050100_2050100_ARI_triclust}, and they are consistent with what was observed previously.  When the short mode has only 10 observations, CoCo initially performs very well until the noise reaches a certain level.  At this point, its performance for the longer modes declines sharply and actually performs worse than CPD+$k$-means, and this pattern is more pronounced for the longest mode ($n_3 = 100$).  The overall co-clustering performance for both methods remains similar, however.  As before, CoCo does not experience as much of a decrease when the shortest mode is made slightly longer ($n_1 = 20$), and does noticeably better than CPD+$k$-means for the most part.

Overall, from clustering these different tensor shapes we see that CoCo still generally performs very well and better than CPD+$k$-means.  The main issue it encounters is when at least one mode is very short ($n_d = 10$).  CoCo performs very well a lower noise levels but has a sharp decline in performance once the noise reaches a certain level.  Unexpectedly, the decline in single-mode performance is worse for the longer modes.  However, even when this happens, CoCo's overall co-clustering performance is still comparable to CPD+$k$-means.  Additionally, this pattern is much less striking when the length of the shortest mode is increased slightly.   

\bibliographystyle{plain}
\bibliography{CoCo}

\end{document}